\documentclass[11pt,a4paper]{article}
\pdfoutput=1
\usepackage{jheppub}
\usepackage{tensor}
\usepackage{physics}
\usepackage{cancel}
\usepackage{amsmath,amsfonts}
\usepackage[english]{babel}
\usepackage{tikz}
\usepackage[normalem]{ulem}
\usepackage{subcaption}

\usetikzlibrary{3d}
\usetikzlibrary{decorations.pathmorphing}

\usepackage{jheppub}
\usepackage{amsthm}
\usepackage{slashed}
\usepackage[utf8]{inputenc}
\usepackage[T1]{fontenc}
\usepackage{mathtools}
\usepackage{float}
\usepackage{empheq}
\usepackage{bm}

\def\nt{\notag}
	
\def\wt{\widetilde}

\def\R{\mathbb{R}}

\def\cD{\mathcal{D}}

\def\cF{\mathcal{F}}

\def\cH{\mathcal{H}}
\def\cI{\mathcal{I}}

\def\cM {\mathcal{M}}
\def\cN{\mathcal{N}}

\def\p{\partial}

\def\/{\over}
\def\ov{\over}

\def\rn{\rangle}
\def\ln{\langle}

\def\t{\theta}
\def\s{\sigma}

\def\ve{\varepsilon}
\def\vphi{\varphi}
\def\a{\alpha}
\def\b{\beta}
\def\d{\delta}

\def\g {\gamma}
\def\la {\lambda}

\def\z{\zeta}

\def\l{\ell}
\def\mn{{\mu\nu}}

\def\n {\nabla}

\def\S{\Sigma}

\def\ra{\rightarrow}

\def\Tr{\mathrm{Tr}}

\def\r{\mathrm}

\def\_{\hspace{2cm}}
\def\'{\:\:}
\def\-{\\\notag}
\def\={&=&}

\newcommand\be{\begin{equation}}
\newcommand\ee{\end{equation}}

\newcommand{\bea}{\begin{eqnarray}}
\newcommand{\eea}{\end{eqnarray}}

\newcommand{\bpm}{\begin{pmatrix}}
\newcommand{\epm}{\end{pmatrix}}

\newcommand{\bit}{\begin{itemize}}
\newcommand{\eit}{\end{itemize}}

\newcommand{\ben}{\begin{enumerate}}
\newcommand{\een}{\end{enumerate}}

\newcommand\bsp{\begin{split}}
\newcommand\esp{\end{split}}

\def\le{\left}
\def\ri{\right}

\def\wg{\wedge}
\def\l{\ell}

\def\qq{\qquad}

\def\cos{\r{cos}}
\def\sin{\r{sin}}
\def\tan{\r{tan}}
\def\cosh{\r{cosh}}
\def\sinh{\r{sinh}}

\def\lph{\tfrac{\l_\phi}{2}}

\definecolor{bluemathematica}{rgb}{0.37, 0.51, 0.71}
\definecolor{orangemathematica}{rgb}{0.92, 0.39, 0.21}

\subheader{\begin{flushright}
\end{flushright}}

\newcommand{\subf}[2]{%
	{\small\begin{tabular}[t]{@{}c@{}}
			#1\\#2
		\end{tabular}}%
	}

\def\tx{{\tilde{x}}}
\def\ty{{\tilde{y}}}
\def\tz{{\tilde{z}}}
\def\tit{{\tilde{t}}}

\title{Gravitation in flat spacetime from entanglement }

\author[a]{Victor Godet}
\author[b]{and Charles Marteau}
\emailAdd{v.z.godet@uva.nl}
\emailAdd{marteau.charles.75@gmail.com}
\affiliation[a]{\small Institute for Theoretical Physics, University of Amsterdam, 1090 GL Amsterdam, Netherlands}
\affiliation[b]{\small Centre de Physique Théorique, CNRS, Institut Polytechnique de Paris, France}

\abstract{We explore holographic entanglement entropy for  Minkowski spacetime in three and four dimensions. Under some general assumptions on the putative holographic dual, the entanglement entropy associated to a special class of subregions can be computed using an analog of the Ryu-Takayanagi formula. We refine the existing prescription in three dimensions and propose a generalization to four dimensions. Under reasonable assumptions on the holographic stress tensor, we show that the first law of entanglement is equivalent to the gravitational equations of motion in the bulk, linearized around Minkowski spacetime.   }

\begin{document}

\maketitle

\section{Introduction}

The AdS/CFT correspondence has been a fruitful avenue to understand quantum gravity in asymptotically AdS spacetimes. A question of interest is whether the holographic principle makes sense in more general spacetimes, such as our own universe. Some proposals have been made for de Sitter \cite{Strominger:2001pn}, Kerr \cite{Guica:2008mu} or warped AdS \cite{Anninos:2008fx, Detournay:2012pc}. The asymptotically flat case is particularly interesting because it can be obtained as a flat limit of AdS \cite{Bagchi:2012xr, Barnich:2012aw}.
Other approaches to flat space holography exist, such as applying AdS/CFT on hyperbolic foliations of Minkowski spacetime \cite{deBoer:2003vf} or using the recently discovered equivalence between BMS Ward identities and Weinberg's soft theorems \cite{Strominger:2013jfa}.

The flat space limit of AdS is an ultra-relativistic limit, or Carrollian limit, of the dual field theory. Already at the level of the symmetries, one can show that the conformal Carroll group is the BMS group \cite{Duval:2014uva}, which is the symmetry group of asymptotically flat gravity \cite{Barnich:2010eb}. More precisely, the conformal Carroll group associated with  the future boundary, i.e. null infinity $\mathcal{I}^+$, is isomorphic to BMS$_3$ when $\mathcal{I}^+=\mathbb{R}\times S^1$ and to BMS$_4$ when $\mathcal{I}^+=\mathbb{R}\times S^2$. Therefore, the putative dual theory should enjoy a Carrollian symmetry.
Recent works have been able to match the gravitational dynamics with ultra-relativistic conservation laws \cite{Ciambelli:2018ojf,Ciambelli:2018wre}. This suggests that the holographic duals of asymptotically flat spacetimes should be Carrollian CFTs \cite{Bagchi:2019xfx}.

An important insight from AdS/CFT is the role of entanglement in the emergence of the bulk spacetime from the field theory degrees of freedom. The Ryu-Takayanagi prescription \cite{Ryu:2006bv}, and its covariant generalization \cite{Hubeny:2007xt}, have lead to a more precise understanding of bulk reconstruction \cite{Almheiri:2014lwa, Dong:2016eik} and a landmark result was the derivation of the gravitational equation, linearized around AdS, from the first law of entanglement in the CFT \cite{Lashkari:2013koa, Faulkner:2013ica,Faulkner:2017tkh}. This suggests that linearized gravity can be understood as the thermodynamics of entanglement. Jacobson's earlier result \cite{Jacobson:1995ab}, and its more recent refinements \cite{Jacobson:2015hqa, Jacobson:2018ahi}, suggest that this connection is very general and goes beyond asymptotically AdS spacetimes. In this paper, we show that a similar result holds for flat space holography in three and four dimensions, under some general assumptions that allow us to use an analog of the Ryu-Takayanagi prescription.

Entanglement entropies in 3d Minkowski spacetime were considered in \cite{Bagchi:2014iea} and were matched with computations in conjectured dual theories. We will follow the geometrical picture proposed in \cite{Jiang:2017ecm}, where the authors used a generalization of the CHM transformation \cite{Casini:2011kv}, to propose an RT prescription for flat spacetime. This requires some assumptions on the putative dual theory which are given in full details below. Under the same working assumptions, we refine their 3d prescription to include perturbations and propose a generalization to 4d.

This paper is organized as follows. In Sec.\ \ref{assumptions} we detail our working assumptions on flat holography. This allows us to use an analog of the Ryu-Takayanagi prescription in Minkowski spacetimes. We review and generalize the existing 3d prescription in Sec.\ \ref{3drev} to include perturbations. In Sec.\ \ref{3dproof} we prove that the gravitational equations, linearized around 3d Minkowski, follow from the first law of entanglement.\footnote{Before submitting our paper, we learned that another group is currently pursuing similar ideas \cite{toappear}.}  In Sec.\ \ref{holst} we perform a flat limit of AdS$_3$, also considered in \cite{Barnich:2012aw, Campoleoni:2018ltl}, to identify the holographic stress tensor associated of 3d Minkowski, a necessary ingredient for the proof. In Sec.\ \ref{4dproof} we generalize the RT prescription to 4d Minkowski and prove that the first law of entanglement is equivalent to the gravitational equations of motion. Our proof is valid for general theories of gravity. 

\newpage
\section{Working assumptions on flat holography} \label{assumptions} 

Holography in asymptotically flat spacetimes is not well understood. The putative dual field theory should be defined  on null surfaces and it is not clear how one should understand objects such as local operators or path integrals.
Therefore, to obtain a well-defined equivalent of the Ryu-Takayanagi prescription, we need some general assumptions on holography in flat spacetime which are listed below:
\bit
\item (Assumption 1) There exists a quantum system living on the future boundary $\cI^+$, such that we can associate a Hilbert space $\cH$ to any slice $\S$ of constant retarded time $u$. To any bulk configuration on $\S$, we can associate a state in $\cH$. For the purpose of this work, we could also weaken this assumption by taking the bulk configurations to be only linear perturbations of Minkowski.
\item (Assumption 2) For a subregion $A$ of $\p \S$ among a special class, we can associate a density matrix $\rho_A$. If the Hilbert space factorizes on subregions, we expect that $\rho_A = \r{Tr}_{\bar{A}}|0\rn\ln0|$ where $\bar{A}$ is the complement of $A$ on the slice and $|0\rn$ is the Minkowski vacuum. We allow $\rho_A$ to be only defined on some subspace $\cH_\text{code}$ of $\cH$.
\eit
The domain of dependence $\cD$ of $A$ is defined to be the union of all the images of $A$ under translation along the $u$ direction. This is simply the ultra-relativistic limit of the Lorentzian domain of dependence. Indeed, in this limit, the width of the lightcone vanishes (see Fig.\ \ref{fig:modflow} for an illustration). Following \cite{Jiang:2017ecm}, we define a generalized Rindler transformation to be a symmetry transformation on $\cI^+$ which maps $\cD$ to a spacetime which has a thermal circle.\footnote{This means that one coordinate of the new spacetime should have an imaginary identification $x\sim x+ i \b$.} The  generator $\z_A$ of the thermal identification, which is called the modular flow generator, is required to annihilate the vacuum and leave $\cD$ and $\p\cD$ invariant. A Rindler transformation is a generalization of the CHM conformal transformation \cite{Casini:2011kv}.
\bit
\item (Assumption 3) If we can find a Rindler transformation, the density matrix can be written as $\rho_A = U^{-1} e^{-K_A} U $ where $K_A$ is the operator that generate translations along the thermal circle and $U$ is a unitary operator acting on the Hilbert space which implements the symmetry transformation. For this definition to make sense, $K_A$ needs to be bounded from below in $\cH_\text{code}$.
\eit
From the knowledge of the boundary modular flow $\z_A$, one can find a bulk modular flow $\xi_A$. It is the Killing vector field of Minkowski spacetime which asymptotes to $\z_A$. 
\bit
\item (Assumption 4) The expectation value $\delta\ln K_A\rn$ for a linear perturbation of the vacuum is computed by the Iyer-Wald energy $\d E_A^\text{grav}$ associated to the Killing vector $\xi_A$ of the corresponding bulk configuration on $\S$.
\item (Assumption 5) The von Neumann entropy $S_A = -\r{Tr}\,\rho_A\log\rho_A$ is computed by the area\footnote{Or the adequate functional for other theories than Einstein gravity.} of the special bulk surface $\wt{A}$ that is preserved by the bulk modular flow $\xi_A$ and is homologous to $A$. This is the analog of the Ryu-Takayanagi (RT) prescription and $\wt{A}$ will be called the RT surface.
\eit
These assumptions can be derived for holographic CFTs with AdS duals. There, the special class of entangling regions are spatial balls in the boundary CFT. Also, Assumptions 3 and 5 were obtained in \cite{Casini:2011kv} and Assumption 4 is a consequence of the AdS/CFT holographic dictionary. The RT prescription for more general entangling regions was derived in \cite{Lewkowycz:2013nqa, Dong:2016hjy}. 

In this work, we want to consider the implications of the above assumptions for flat holography. In particular, we will investigate the consequences of the first law of entanglement $\d S_A = \d\ln K_A\rn$ which is valid for any quantum system where these objects can be defined. Paralleling the AdS story \cite{Faulkner:2013ica}, we will show that the linearized gravitational equations of motion are equivalent to the first law. We believe that although the microscopic theory is not well understood, this approach can provide valuable insights about holography in non-AdS spacetimes.

The results that we have proven can also be phrased purely in classical gravity. We have shown that for linearized perturbations of Minkowski spacetime, the gravitational equations of motion are \emph{equivalent} to the first law
\be
\d S_A^\text{grav}= \d E_A^\text{grav},
\ee
for a set of boundary regions $A$ among a special class, and where $S^{\text{grav}}_A$ is the gravitational entropy of the surface $\wt{A}$ defined to be the surface homologous to $A$ and fixed by the Killing vector field $\xi_A$. The existence of a holographic theory such that $\d S_A^\text{grav}= \d S_A $ and $\d E_A^\text{grav} = \d\ln K_A\rn$ provides a microscopic realization and an interpretation in term of entanglement which renders the first law automatic.

\section{Ryu-Takayanagi prescription in 3d Minkowski}\label{3drev}
{}
We consider three-dimensional flat spacetime in Bondi gauge
\be
ds^2 = -du^2 - 2dudr + r^2 d\phi^2,
\ee
where $u = t-r$.  The boundary is the null infinity $\cI^+$ (at $r=\infty$) and the boundary metric is degenerate:
\be
ds^2 = 0 \times du^2 + d\phi^2.
\ee
Let's pick a region $A$ on $\cI^+$. We would like to compute the entanglement entropy associated to $A$ in a putative holographic theory living on $\cI^+$. This can be computed with an analog of the Ryu-Takayanagi formula, which was proposed in \cite{Jiang:2017ecm}.  In this section, we will review and refine this prescription.

\subsection{Review of the 3d prescription}\label{3dreview}

In \cite{Jiang:2017ecm}, the authors proposed an RT prescription for 3d Minkowski spacetime by using a "generalized Rindler method". This consists of finding a transformation, which satisfies the same properties as the Casini-Huerta-Myers conformal mapping \cite{Casini:2011kv}. One should look for a symmetry transformation which maps the domain of dependence $\cD$ of a subregion $A$ to a Rindler spacetime characterized by a thermal identification. The modular flow generator, which is the generator of the thermal identification, is required to annihilate the vacuum and to leave $\cD$ and $\p\cD$ invariant. 

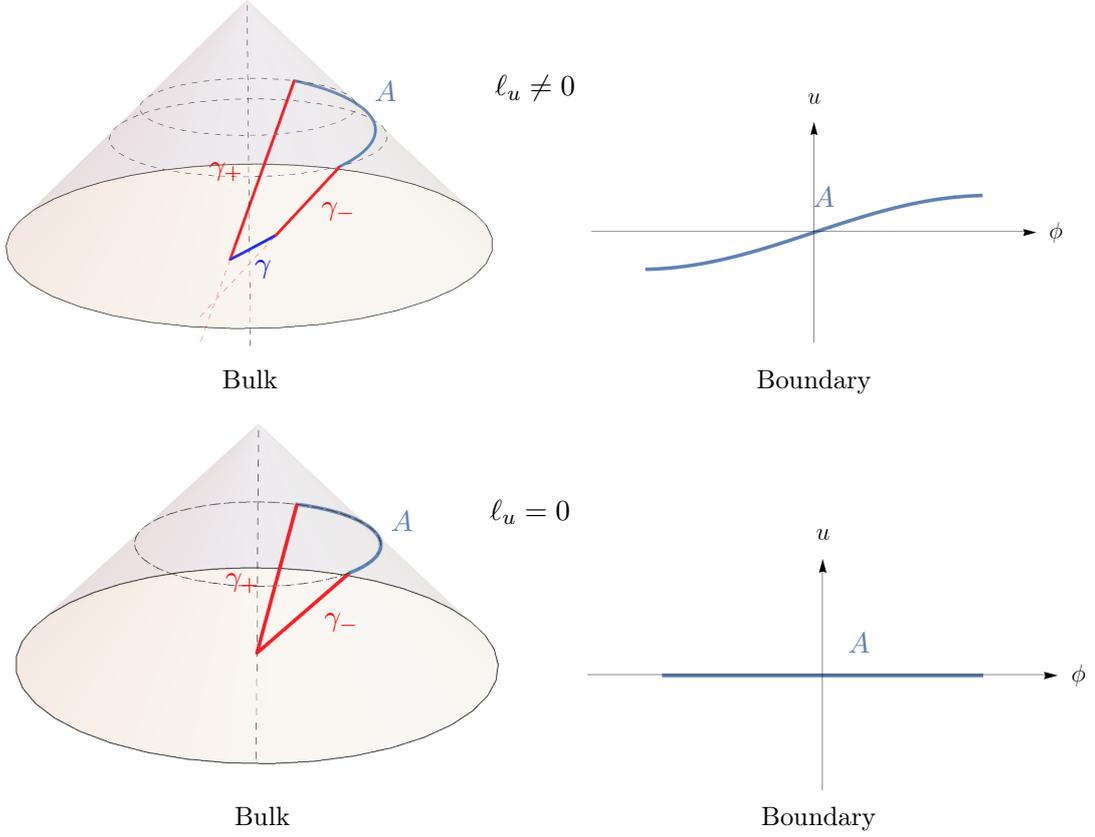
\begin{figure}
\centering
	\begin{tabular}{cc}
		\subf{\includegraphics[width=7cm]{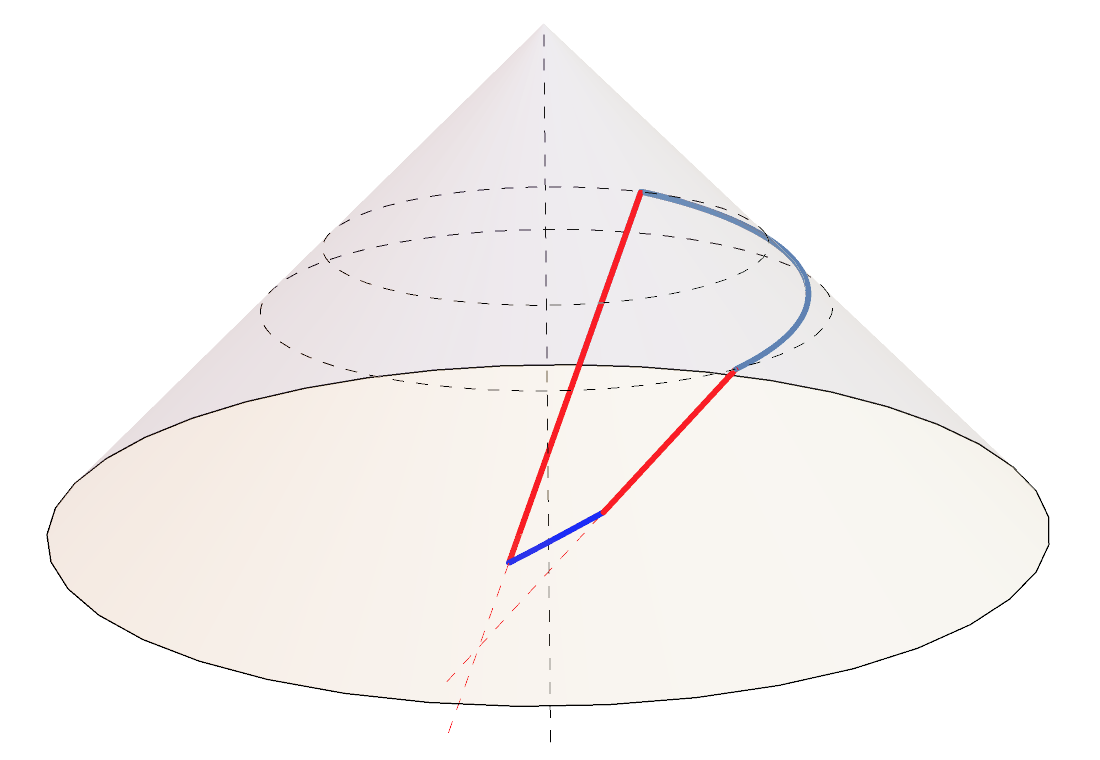}}{Bulk} &
		\subf{\includegraphics[width=7cm]{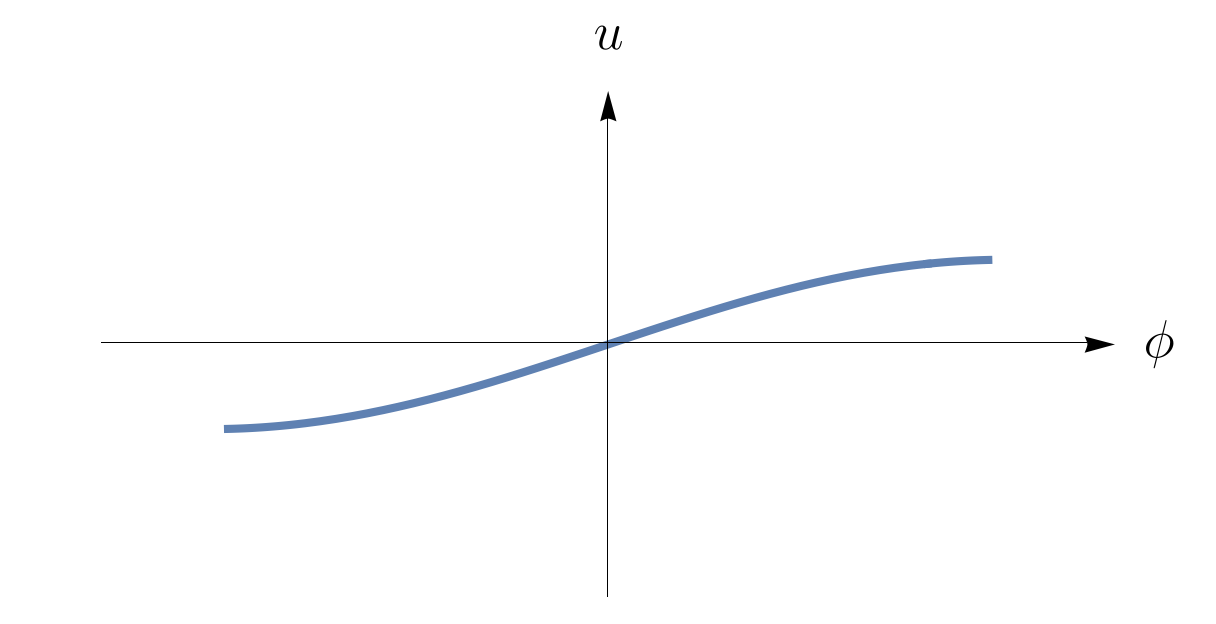}}{Boundary}
			\end{tabular}
\put(-225,40){{$\l_u\neq0$}}
\put(-332, 10){{\color{red}{$\g_+$}}}
\put(-290, -5){{\color{red}{$\g_-$}}}
\put(-315, -27){{\color{blue}{$\g$}}}
\put(-270, 38){{\color{bluemathematica}{$A$}}}
\put(-106, -2){{\color{bluemathematica}{$A$}}}

\centering
	\begin{tabular}{cc}
		\subf{\includegraphics[width=7.6cm]{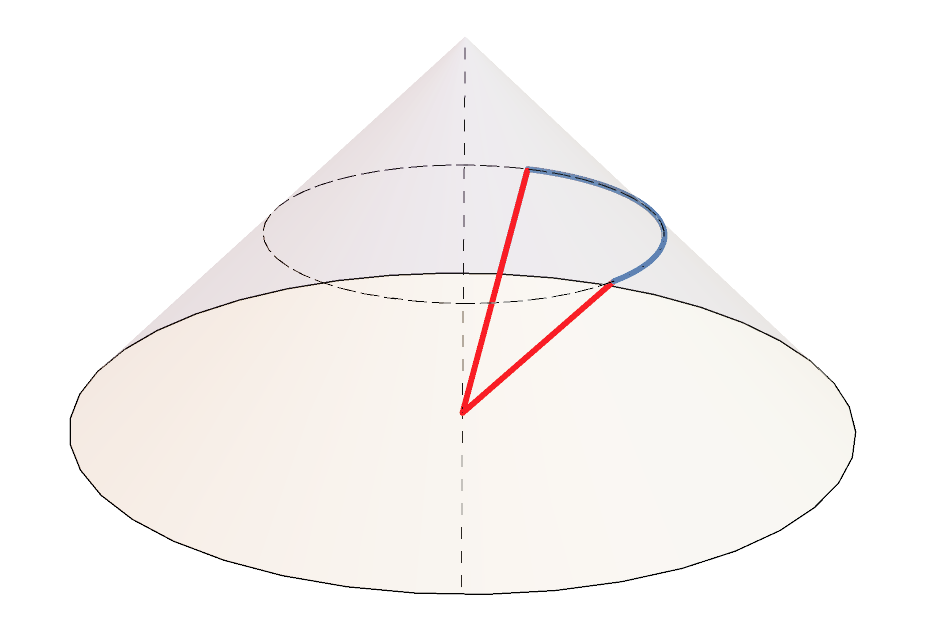}}{Bulk} &
		\subf{\includegraphics[width=7cm]{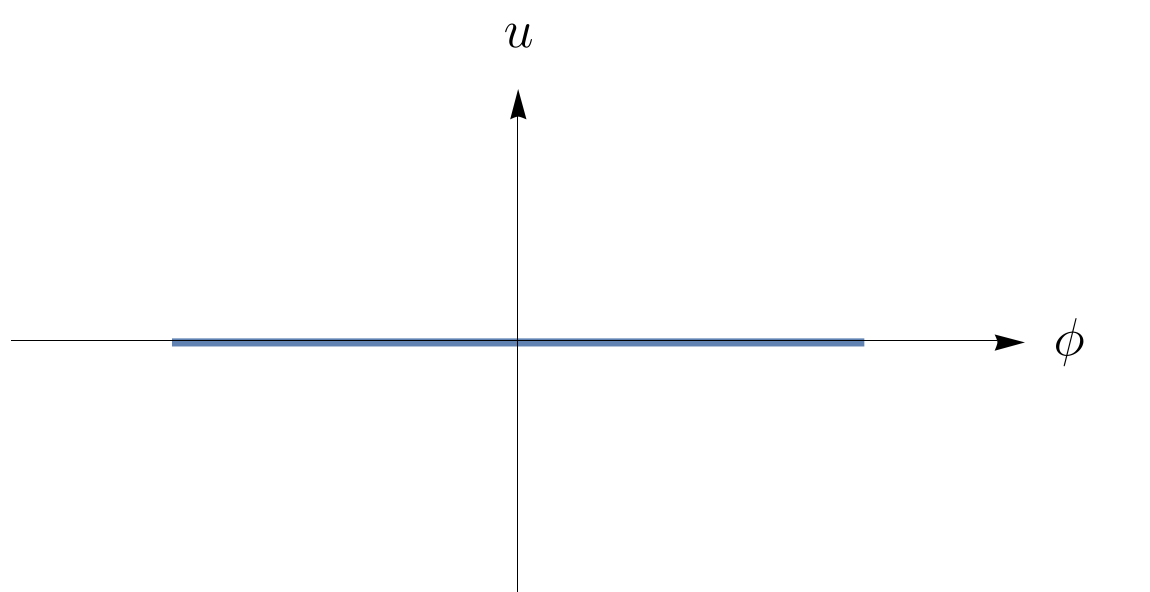}}{\hspace{-0.8cm}Boundary} 
	\end{tabular}
\put(-240,40){{$\l_u=0$}}
\put(-339, 15){{\color{red}{$\g_+$}}}
\put(-302, 0){{\color{red}{$\g_-$}}}
\put(-277, 36){{\color{bluemathematica}{$A$}}}
\put(-106, -10){{\color{bluemathematica}{$A$}}}

\vspace{0.3cm}
\caption{Examples of Ryu-Takayanagi surfaces in 3d Minkowski spacetime}\label{fig:3dprescription}\label{fig:3dpresc}
\end{figure}

Let's consider an interval $A$ on the boundary, it is characterized by its sizes $\l_u$ and $\l_\phi$ in the $u$ and $\phi$ directions. The authors of  \cite{Jiang:2017ecm} were able to find a Rindler transformation for $A$ and to derive a boundary modular flow. Then, the Rindler transformation was extended into the bulk by finding a suitable change of coordinates. The bulk image of the transformation is a flat space cosmological solution \cite{Cornalba:2002fi}, which is the flat space analog of the hyperbolic black hole in AdS$_3$. This maps the entanglement entropy into thermal entropy, which is computed geometrically from the area of the horizon of the flat space cosmological solution. This leads to the following picture: the RT surface is the union of three curves
\be
\wt{A} = \g_+ \cup \g \cup \g_-,
\ee
where $\g_\pm$ are two light rays emanating from the two extremities $\p A$ of the interval and $\g$ is a bulk curve connecting $\g_+$ and $\g_-$. In Einstein gravity, the entanglement entropy is then obtained as
\be\label{3dprescriptionEinstein}
S_A = {\text{Length}(\g)\/4 G} \,.
\ee
We illustrate this procedure in Fig.\ \ref{fig:3dpresc}. This prescription is consistent with computations in conjectured dual theories \cite{Bagchi:2014iea}. This RT surface was also shown in \cite{Hijano:2017eii} to correspond to an extremal surface. See also \cite{Wen:2018mev} for a discussion on the replica trick in this context. 

We would like to consider more general theories of gravity and derive a first law. In a more general context, the RT configuration is the same but the entanglement entropy is given by Wald's functional
\be\label{3dprescription}
S_A =  \int_{\wt{A}} {\rm \textbf{Q}}[\xi_A]
\ee 
where $\xi_A$ is the bulk modular flow reviewed below. As we will show, it is important to integrate over $\wt{A}$ here, instead of just $\g$, if we want to have a first law. In Einstein gravity, \eqref{3dprescription} reduces to \eqref{3dprescriptionEinstein} because Wald's functional vanishes when integrated on $\g_+$ and $\g_-$.

\begin{figure}
\centering
	\begin{tabular}{cc}
		\subf{\includegraphics[width=7cm]{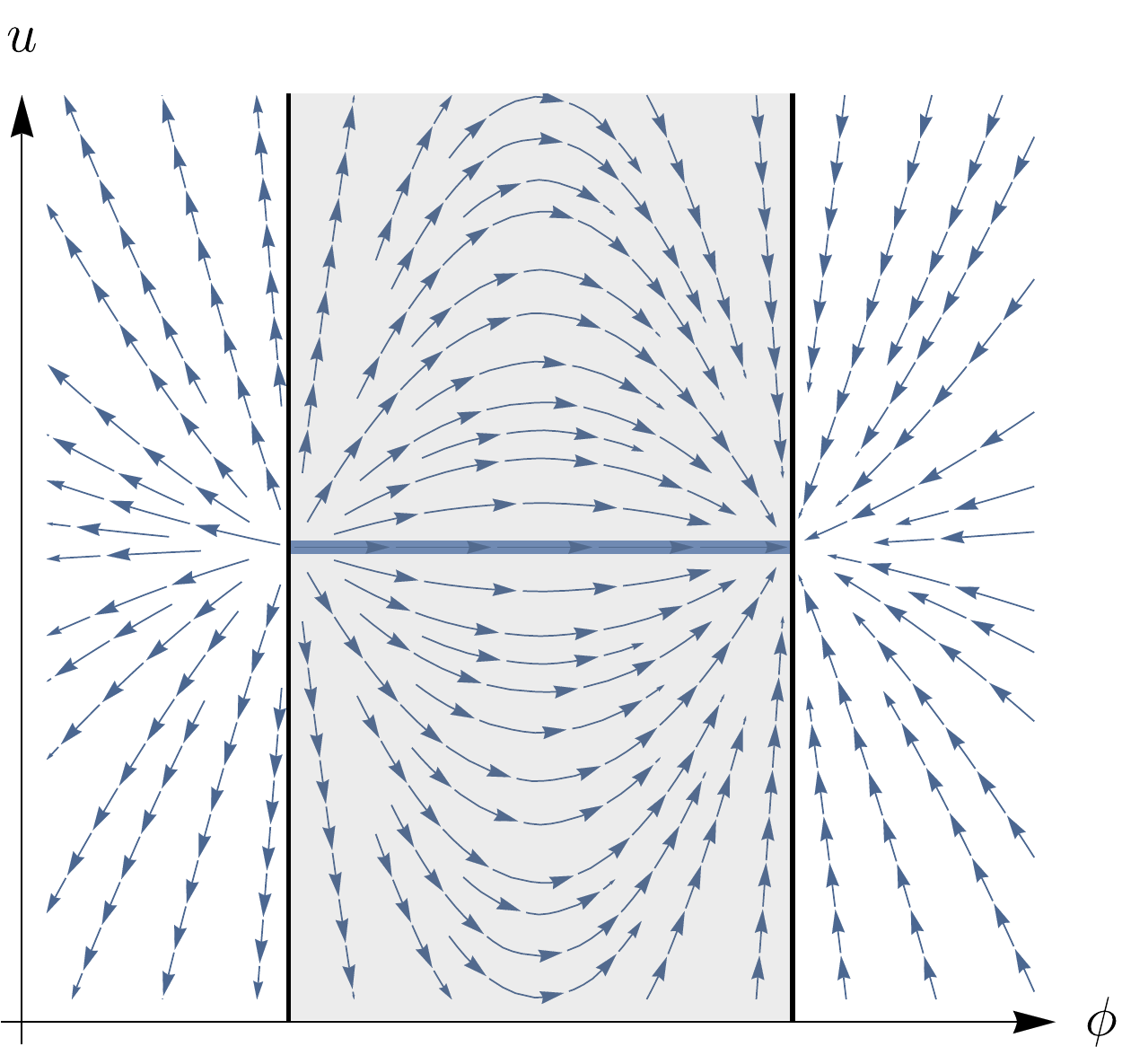}}{Modular flow in $(u,\phi)$} &
		\subf{\includegraphics[width=7cm]{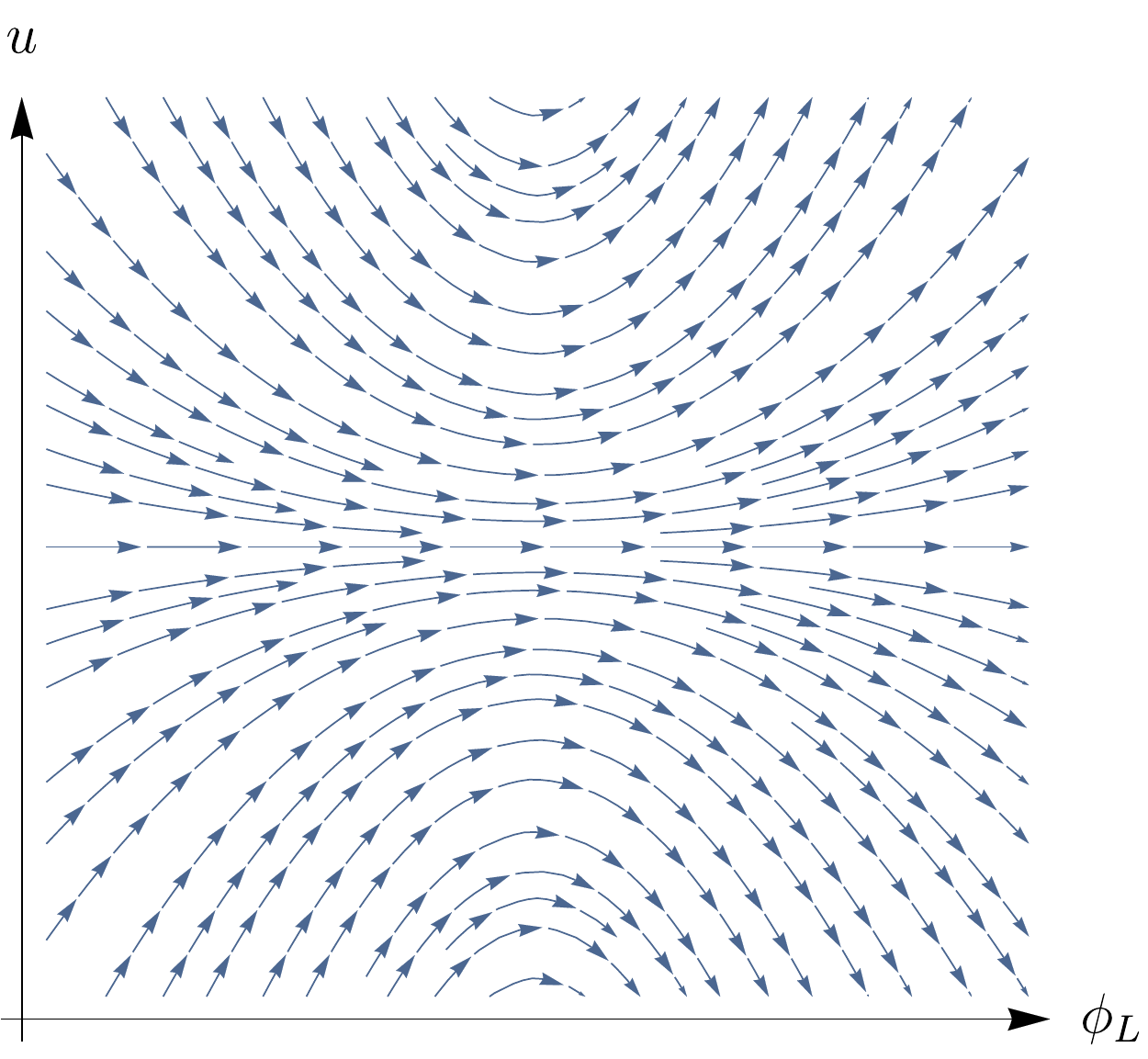}}{Modular flow in $(u,\phi_L)$ with $\phi_L = i\phi$} 
			\end{tabular}
	\put(-243,110){{$\l_u=0$}}
\put(-324,94){{$\cD$}}

\vspace{0.7cm}
\centering
	\begin{tabular}{cc}
		\subf{\includegraphics[width=7cm]{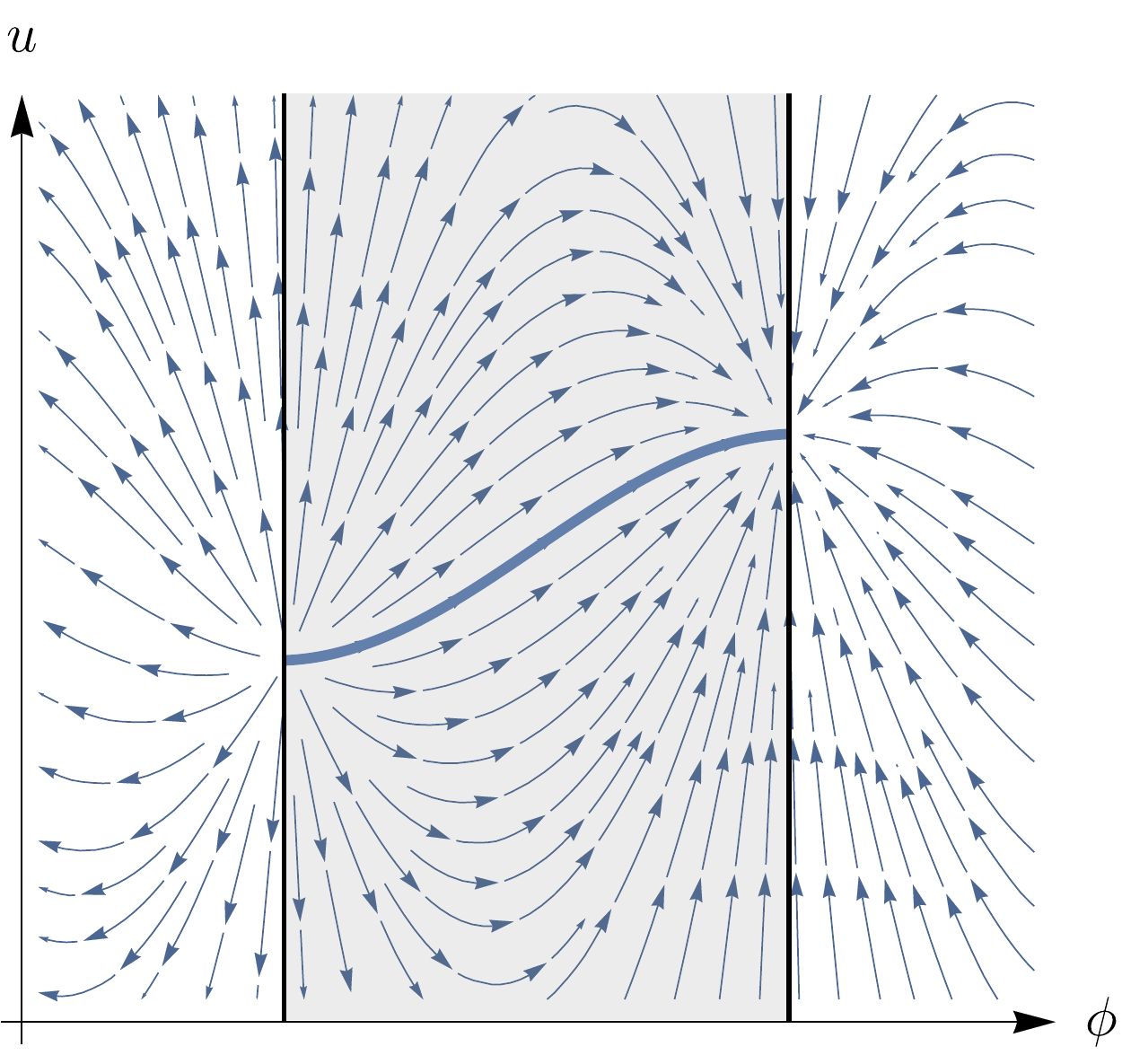}}{Modular flow  in $(u,\phi)$} &
		\subf{\includegraphics[width=7cm]{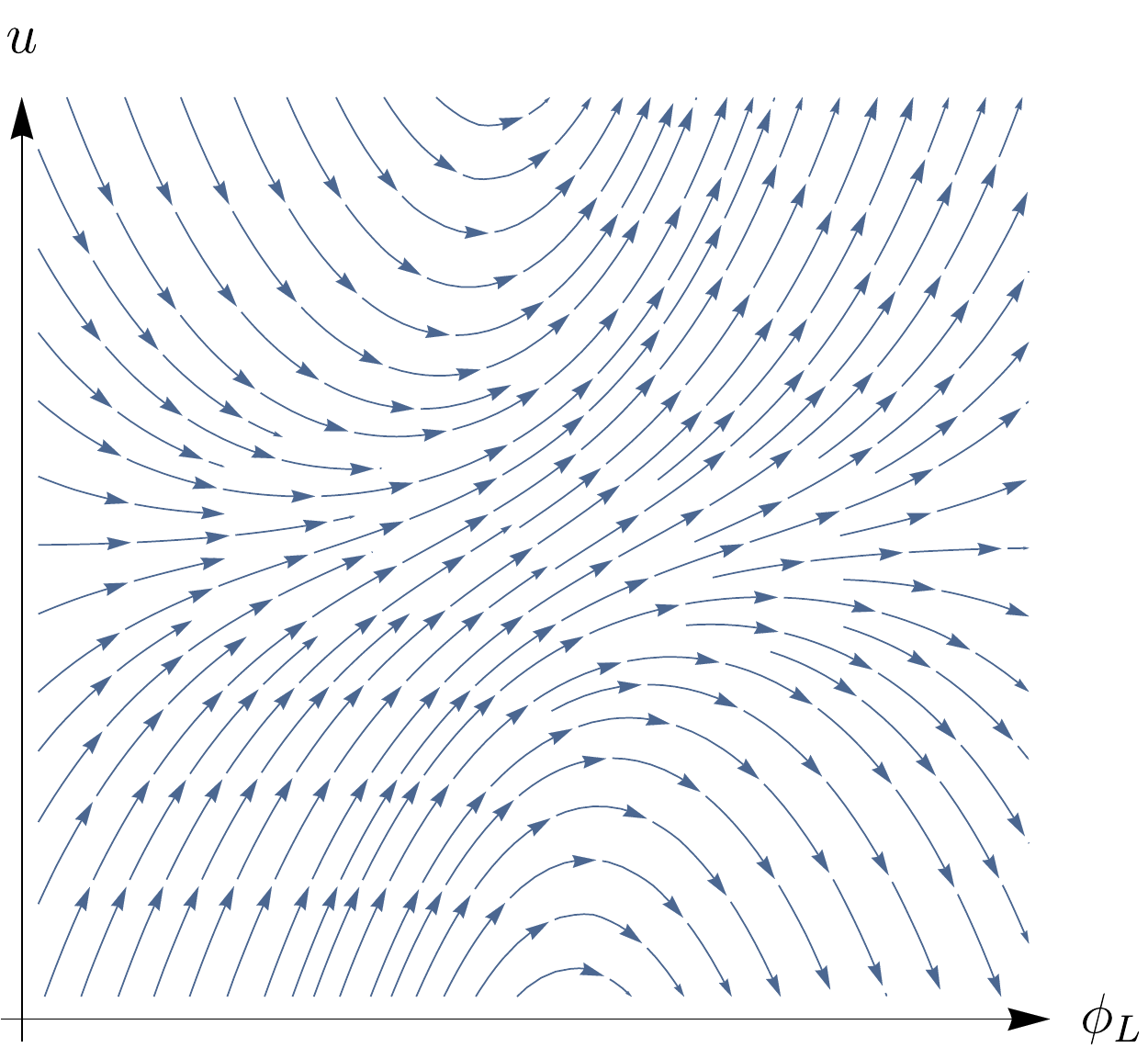}}{Modular flow in $(u,\phi_L)$ with $\phi_L = i\phi$} \\
	\end{tabular}
	\put(-243,110){{$\l_u\neq 0$}}
	\put(-324,94){{$\cD$}}

	\vspace{0.3cm}

\caption{Boundary modular flow for 3d Minkowski. The left pictures represents the modular flow with the entangling region $A$ (in blue) and its domain of dependence $\cD$ (shaded) for $\l_u = 0$ and $\l_u\neq 0$. The right picture is the Wick rotated version with $\phi_L=i\phi$, where we see that the modular flow circles around a point at infinity. In contrast with the corresponding AdS/CFT picture (which is Fig.\ 2 in \cite{Faulkner:2017tkh}), the modular flow does not "transport" the entangling region $A$ but is parallel to it. This suggests that the density matrix $\rho_A$ is more naturally associated with the domain of dependence $\cD$, as argued by \cite{Czech:2012bh} in the AdS/CFT context. Since they have the same domain of dependence, this suggests that the case $\l_u=0$ is really equivalent to the case $\l_u\neq 0$, as we will explain in Sec.\ \ref{3dgenpresc}. } \label{fig:modflow}

\end{figure}

\paragraph{Generalized Rindler method.} We are now going to review how the generalized Rindler method is implemented in \cite{Jiang:2017ecm}. The Rindler transformation in the 2d boundary theory is
\bea\label{rindler3d}
u \= {\r{sin}(\lph)\,\/\r{cosh}\,\rho + \r{cos}(\lph)}\left(\tau+\frac{\ell_u}{2\,\sin(\frac{\ell_\phi}{2})}\sinh\, \rho\right) \,, \-
\phi \=  \r{arctan}\le({\r{sin}(\lph)\,\r{sinh}\,\rho\/1+\r{cos}(\lph)\,\r{cosh}\,\rho}\ri) \,.
\eea
The thermal identification is given by $\rho\sim \rho+2\pi i$. The boundary modular flow is the thermal generator $2\pi \p_\rho$ which is 
\be\label{3dmodflow}
\z_A= {2\pi \/\r{sin}(\lph)} \le[\le(- u \,\r{sin}\,\phi\,+\frac{\ell_u \,\cos\,\phi}{2\,\tan\,(\frac{\ell_\phi}{2})} -\frac{\ell_u}{2\,\sin\,(\frac{\ell_\phi}{2})}\ri)  \p_u+  \le(\r{cos}\,\phi-\r{cos}(\lph)\ri) \p_\phi \ri].
\ee
This modular flow generates a transformation of BMS$_3$ since it can be written as
\be
\zeta_A=(u\,Y'(\phi)+T(\phi))\p_u+Y(\phi)\p_\phi,
\ee
where $Y(\phi)$ corresponds to a superrotation and $T(\phi)$ to a supertranslation. It is depicted together with its Wick rotated version in Fig.\ \ref{fig:modflow}. A simple shape for the region $A$ when $\l_u\neq 0 $ is a portion of sinusoid with equation
\be
u = {\l_u\/2\,\r{sin}(\lph)}\r{sin}\,\phi~,
\ee
although the precise shape doesn't matter in the computation of the entanglement entropy. The bulk modular flow can be found by looking for a Killing vector of 3d Minkowski which asymptotes to $\z_A$. It takes the form
\bea\label{3dbulkmodularflow}
\xi_A \=   {2\pi\/\r{sin}(\lph)} \le[ \le( u\,\r{sin}\,\phi + {\l_u\/2 \,\r{tan}(\lph)}\r{cos}\,\phi - {\l_u\/2 \,\r{sin}(\lph)}\ri)\p_u \ri. \-
&& \hspace{1.5cm}\le.+ \le( \r{cos}(\lph)-\r{cos}\,\phi - {u\/r} \r{cos}\,\phi + {\l_u\/2 \,\r{tan}(\lph)} {\r{sin}\,\phi\/ r} \ri)\p_\phi \ri. \-
&& \hspace{1.5cm}\le. - \le( (u+r)\,\r{sin}\,\phi + {\l_u\/2\,\r{tan}(\lph)} \r{cos}\,\phi\ri)\p_r\ri]~.
\eea 
The bulk modular flow $\xi_A$ vanishes on the curve $\g$. It doesn't vanish on the two light rays $\g_\pm$ but is tangent to them. This is enough to guarantee the existence of a first law, as explained in Sec.\ \ref{First Law}.

\paragraph{Entanglement entropy as Rindler entropy.}\label{RindlerHorizon} To understand better the bulk picture described above, it is useful to go to Cartesian coordinates $(t,x,y)$ defined as
\begin{equation}
t=u+r, \qq x=r\,\cos\,\phi, \qq y=r\,\sin\,\phi~.
\end{equation}
In these coordinates, the bulk modular flow becomes
\be\label{3dmodflow}
\xi _A= {2\pi\/\r{sin}(\lph)} \le[ \le( y+{\l_u\/2\,\r{sin}(\lph)} \ri)\p_t + \le(y\,\r{cos}(\lph) + {\l_u\/2\,\r{tan}(\lph)} \ri) \p_x + \le(t - x\,\r{cos}(\lph)\ri) \p_y\ri],
\ee
which is simply a boost, as can be seen by defining  new Cartesian coordinates 
\be\label{3drindlercoord}
\tilde{t} = {t\/\r{sin}(\lph)} -\r{cot}(\lph) \,x, \qq \tilde{x} = {x\/\r{sin}(\lph)}- \r{cot}(\lph) \,t ,\qq \tilde{y} = y + {\l_u\/2\,\r{sin}(\lph)}.
\ee
In these coordinates, the modular flow is simply
\be
\xi_A = 2\pi \le(\tilde{y}\, \p_{\tilde{t}} + \tilde{t}\, \p_{\tilde{y}} \ri)~.
\ee
In App.\ \ref{appbulkrindler}, we confirm that the Rindler thermal circle is the same as the one appearing in the generalized Rindler transform \eqref{rindler3d}.\footnote{One should remember that in the upper wedge, the Rindler time is spacelike, which is consistent with the boundary picture,  see Fig.\ \ref{fig:modflow}.} This geometry should be seen as the analog of the hyperbolic black hole in AdS.

\begin{figure}
	\centering
	\includegraphics[width=10cm]{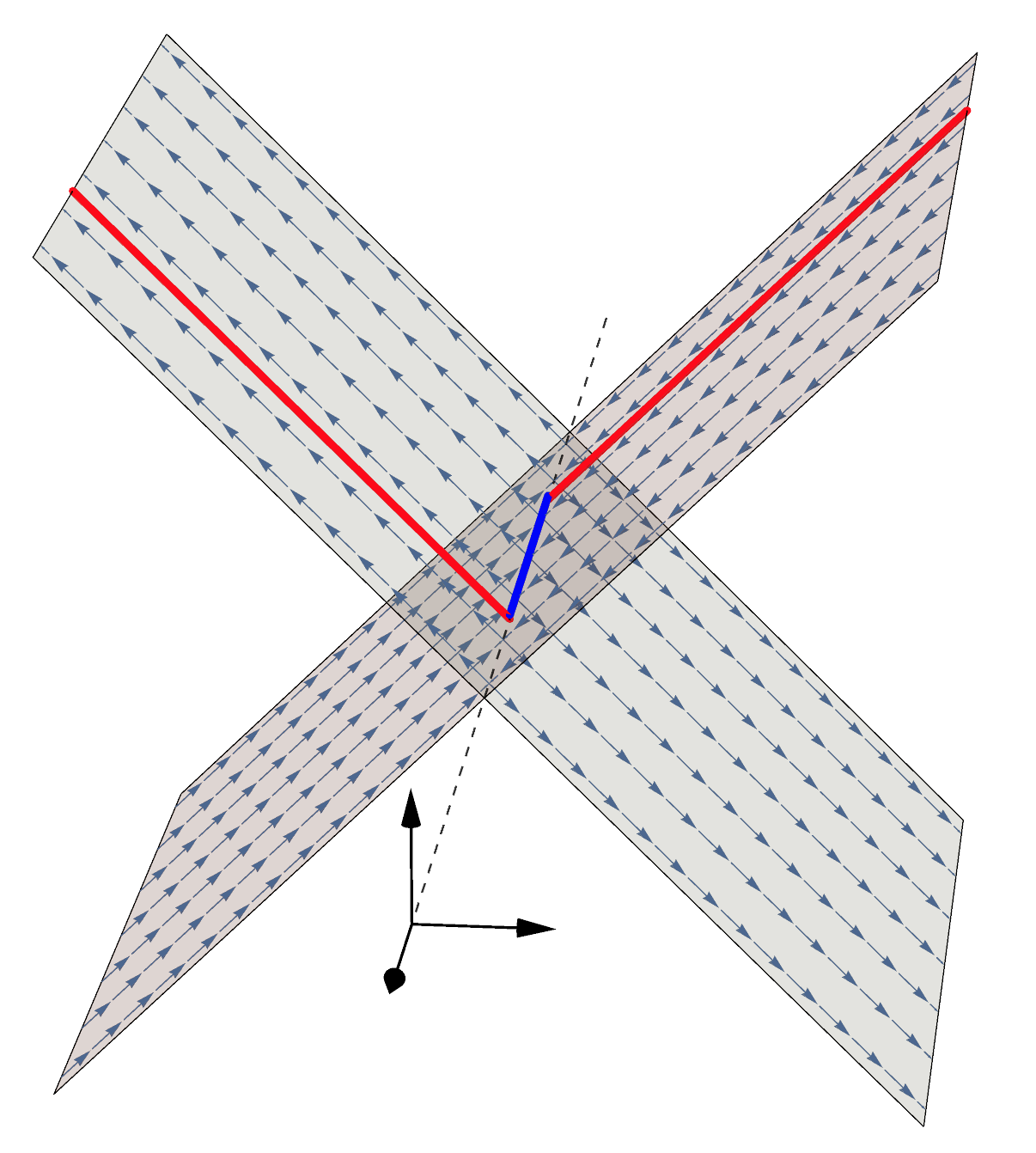}
		\put(-182,32){$\tx$}
		\put(-182,85){$\tit$}
		\put(-128,57){$\ty$}
		\put(-100, 242){{\color{red}{$\g_+$}}}
\put(-200, 210){{\color{red}{$\g_-$}}}
\put(-156, 180){{\color{blue}{$\g$}}}
	\caption{Ryu-Takayanagi surface in coordinates $(\tit,\tx,\ty)$ in which the bulk modular flow is a boost. It is given by $\wt{A} = \g_-\cup\g \cup\g_+$. The surface $\g$ lies on the Rindler bifurcation surface (the dashed line) and the light rays $\g_+$ and $\g_-$ are tangent to the modular flow. }\label{3drindler}
\end{figure}

We will now review the explicit RT prescription of \cite{Jiang:2017ecm} but in Cartesian coordinates where the description becomes simpler. This will be important in discussing the more general prescription in Sec.\ \ref{3dgenpresc} and the 4d generalization in Sec.\ \ref{4dproof}. As depicted in Fig.\ \ref{fig:3dpresc}, we consider two bulk light rays that go to the two extremity points of $A$ on $\cI^+$. There is an ambiguity in choosing such light rays, as discussed in Sec.\ \ref{3dgenpresc}. The prescription adopted in \cite{Jiang:2017ecm} is to impose that these two light rays pass through the spatial origin $r=0$, which is natural given a choice of Bondi coordinates. A parametrization of these two light rays is
\be
\g_+ : \quad\begin{cases} t = - {\l_u\/2} + s \\ x = s\,\r{cos}(\lph) \\ y = - s\,\r{sin}(\lph) \end{cases},\qq \g_- : \quad\begin{cases} t = {\l_u\/2} + s \\ x = s\,\r{cos}(\lph) \\ y =  s\,\r{sin}(\lph) \end{cases}~.
\ee
In the limit $r \to+\infty$, we have 
\begin{eqnarray}
\g_+:\quad \begin{cases} u\to  {\l_u\/2}, \\  \phi \to  {\l_\phi\/2} \end{cases},\qq \g_-:\quad \begin{cases} u\to - {\l_u\/2}, \\  \phi \to - {\l_\phi\/2} \end{cases}~,
\end{eqnarray}
so that they intersect the two extremities of $A$ on $\cI^+$ as required. The bulk modular flow vanishes on the Rindler bifurcation surface 
\be
\tit = \ty =0~.
\ee
The curve $\g$ should be located where the  bulk modular flow vanishes. Therefore, it has to lie on the bifurcation surface. To determine which portion it covers, we should look for the intersection of $\g_\pm$ with the bifurcation surface which gives two points $P_+$ and $P_-$ with coordinates
\begin{eqnarray}
P_\pm:\quad \tilde{t}=\tilde{y}=0, \quad \tilde{x}=\pm\frac{\ell_u}{2\,\sin^2(\frac{\ell_\phi}{2})}.
\end{eqnarray}
The curve $\g$ is then the segment $[P_-P_+]$. The resulting RT surface becomes
\be
\wt{A} = \g_+\cup \g_-\cup \g~,
\ee
where it is understood that we only consider the portions of $\g_\pm$ that connect $\g$ to $A$. From the general prescription \eqref{3dprescription}, the entanglement entropy of the region $A$ is be given by the integral of Wald's functional on $\wt{A}$. For Einstein gravity, this reduces to the length of $\g$ and this leads to
\begin{equation}\label{SAeinstein}
S_A=\frac{\ell_u}{4G}\,\r{cot}(\lph) \qq \text{(Einstein gravity)}~.
\end{equation}
We illustrate this prescription in Fig.\ \ref{3drindler} in the coordinates \eqref{3drindlercoord} where the modular flow is a boost. A success of the prescription of \cite{Jiang:2017ecm} is that this reproduces the entanglement entropies obtained through field theoretic methods in \cite{Bagchi:2014iea}. We can now understand what is going to happen when we will perturb the bulk geometry: the portion of the bifurcation surface in consideration will satisfy a first law on-shell (this is true for any Killing horizon) that will map, through the assumptions we have made earlier, to a first law of entanglement of a putative dual field theory. This is explained in details in Sec.\ \ref{First Law}.

 \paragraph{More RT surfaces.} The authors of \cite{Jiang:2017ecm} derived a prescription to compute the entanglement entropies for a particular set of boundary  regions. The prescription is summarized in Fig.\ \ref{fig:3dprescription} with two qualitatively different cases $\l_u = 0$ or $\l_u\neq 0$. There is a simple way to generate the RT surfaces associated to more general regions on $\cI^+$. This can be done by acting with bulk isometries on the initial configurations. In Minkowski spacetime, we should act with elements of the Poincaré group. Their actions on $\cI^+$ are given by BMS$_3$ transformations which transform $A$ into a new region $A'$. This new region will be a more complicated curve. The corresponding RT surface $\wt{A}'$ is simply obtained as the image of $\wt{A}$ under the bulk isometry. These transformed RT surfaces are depicted in Fig.\ \ref{fig:newRT} and play a crucial role in the proof of the linearized gravitational equations of motion from the first law of entanglement.

\subsection{General 3d prescription} \label{3dgenpresc}

We will explain an important ambiguity in the RT prescription of \cite{Jiang:2017ecm}, which we reviewed above, corresponding to the choice of how the light rays reach infinity. This ambiguity was also considered in \cite{Hijano:2017eii}. As a result, we will show that additional RT configurations are possible.

\paragraph{Infalling light sheaf.}

This ambiguity is most apparent when we consider the following fact: the case $\l_u\neq 0 $ can actually be obtained from the case $\l_u = 0$ by acting with the bulk translation
\be
y\ra  y +{\l_u\/2\,\r{sin}(\lph)}~.
\ee
This is apparent from the formula of the bulk modular flow \eqref{3dmodflow}: the modular flow for $\l_u\neq0$ is simply the image of the bulk modular flow for $\l_u=0$ under this translation. On the boundary, this translation becomes 
\begin{equation}
u\rightarrow u+{\l_u\/2\,\r{sin}(\lph)}\sin\,\phi~,
\end{equation}
and maps the boundary interval with $\ell_u=0$ to the one with $\ell_u\neq 0$, see Fig.\ \ref{fig:modflow}. This fact is puzzling because it implies that the configuration with $\l_u= 0$ and the configuration with $\l_u\neq 0$ are physically equivalent, as they are related by a bulk translation (which should be a true symmetry of the Minkowski vacuum). However, the entanglement entropies computed earlier are not the same for $\l_u = 0$ and $\l_u \neq0$, as seen for \eqref{SAeinstein}.

In fact, this arises because the RT prescription depends on a choice of how the light rays arrive at infinity, or a choice of \emph{infalling light sheaf}. For a given point on $\cI^+$ with coordinates $(u,\phi)$, there are many inequivalent bulk light rays that go to this point, differing by bulk translations. We define an infalling light sheaf to be a set of light rays whose intersection with $\cI^+$ is $\p A$. The RT prescription will depend on the choice of such a light sheaf and acting with a bulk translation will modify this choice. To obtain a good RT prescription, we must require that the light sheaf satisfies the following two conditions: 
\ben
\item Each light ray in the light sheaf must intersect the Rindler bifurcation surface.
\item The bulk modular flow must be tangent to the light sheaf.
\een
The first condition is necessary to be able to define an RT surface (which should contain a portion of the Rindler bifurcation surface) while the second condition ensures the existence of a well-defined first law as we will show in the next section. 

Heuristically, the choice of a light sheaf amounts to a choice of cutoff surface at infinity. In more mundane language, we are just saying that the entanglement entropy is cutoff dependent (even though it is finite). It is difficult to be more precise about what we mean by "cutoff" because the dual theory is not well-understood. We believe that this ambiguity reflects some properties of the UV structure of the dual theory.

\paragraph{Generalized 3d prescription.}

In 3d, the boundary $\p A$ consists of two points $B_+$ and $B_-$. Hence, the choice of infalling light sheaf is the choice of two light rays $\g_+$ and $\g_-$ that arrive at these points and satisfy the two conditions stated above. An explicit parametrization of this light sheaf can be given as
\bea\label{3dlightsheet}
\g_+:\quad \begin{cases} t={\l_u\/2}+ s + Y_+\, \r{sin}(\lph) \\ 
x = s\,\r{cos}(\lph) \\ 
y = s\,\r{sin}\,(\lph) + Y_+ \end{cases},\qq
\g_-: \quad \begin{cases} t=- {\l_u\/2}+ s- Y_-\, \r{sin}(\lph) \\ 
x = s\,\r{cos}(\lph) \\ 
y =- s\,\r{sin}\,(\lph) + Y_- \end{cases}
\eea
where $s \in \R$ is a parameter on the light ray and $Y_+, Y_-$ are arbitrary constants. The light rays $\g_\pm$ arrive on $\cI_+$ respectively at the points $B_\pm$. As required, they intersect the bifurcation surface $\tilde{y}=\tilde{t}=0$ and are tangent to the bulk modular flow. Note that we have also used the freedom of reparametrization of $s$ to reduce the number of independent parameters. At the end, we obtain a family of light sheaf parametrized by two arbitrary constants $Y_+$ and $Y_-$. The light rays $\g_\pm$ intersect the bifurcation surface at $\tilde{x} = \tilde{x}_\pm$ with
\bea
\tilde{x}_+ = -{\l_u\/2\,\r{tan}(\lph)} - Y_+\,\r{cos}(\lph) ,\qq \tilde{x}_- = {\l_u\/2\,\r{tan}(\lph)} + Y_-\,\r{cos}(\lph) ~.
\eea
The length of $\g$ is therefore given by the separation in $\tilde{x}$ which leads to the entropy
\be
S_A = {1\/4G}\le| \l_u \,\text{cot}(\lph) + (Y_++Y_-)\,\r{cos}(\lph)\ri|~.
\ee
The case $Y_+= Y_-=0$ corresponds to the prescription adopted of \cite{Jiang:2017ecm} described above. This prescription can also be obtained by requiring that the light rays intersect the line $r=0$, which makes this prescription natural given a choice of Bondi coordinates.  Another simple choice is 
\be
Y_+ = Y_-=-{\l_u\/2\,\r{sin}(\lph)}~.
\ee
In this case, the two light rays $\g_+$ and $\g_-$ intersect at the point 
\be\label{interpointgenpresc}
\tilde{t}= \tilde{x}=0,\qq \tilde{y} = -{\l_u\/2\,\r{sin}(\lph)}~.
\ee
This gives a vanishing entropy and it corresponds to the case where we have applied a bulk translation to go from the $\l_u=0$ configuration shown in Fig.\ \ref{fig:3dprescription} to a configuration with $\l_u\neq 0$ in which the light rays $\g_+$ and $\g_-$ still meet. We can see that the intersection point \eqref{interpointgenpresc} is indeed precisely the image of the origin by this translation. We would like to emphasize that there are no reason to favor one prescription or the other. Instead, we believe that we are free to choose any light sheaf satisfying the two conditions described above, and we interpret this choice as reflecting a choice of regulator in the putative dual theory.

\subsection{First law of entanglement}\label{First Law}

In quantum mechanics, the first law of entanglement is a general property of the von Neumann entropy, which holds whenever we have a well-defined density matrix. It states that under a variation $\rho \to \rho + \d\rho$, we have
\be
\d S = \d \ln K \rn ,
\ee
where $S = - \Tr\,\rho\log\rho$ and $K = - \log\rho$. The proof uses simple manipulations on density matrices and is given in \cite{Faulkner:2013ica}. When $\rho$ is the density matrix associated to the boundary region $A$, we will denote $\d S_A$ the entropy variation and $\d E_A = \d \ln K\rn $ the energy variation. The first law of entanglement states that
\be
\d S_A =\d E_A \,.
\ee
We would like to compute the corresponding gravitational quantities $\d S_A^\text{grav}$ and $\d E_A^\text{grav}$ under a general perturbation of the metric. Following the general prescription discussed above, we consider the RT surface $\wt{A}=\g_+ \cup \g \cup\g_-$ where $\g_\pm$ are given in \eqref{3dlightsheet}. In Einstein gravity, the gravitational entropy associated to the RT surface $\wt{A}$ is nothing but its area in Planck units. The variation of the entropy is then computed from the variation of the area of $\wt{A}$. We want to allow for general theories of gravity so we introduce Wald's Noether charge ${\rm \bf Q}[\xi_A]$ associated to the Killing vector field $\xi_A$. The variation of the gravitational entropy is then given by
\be
\d S^\text{grav}_A = \int_{\wt{A}} \d {\rm\bf Q}[\xi_A].
\ee
The gravitational energy is defined as the boundary term appearing in the expression of the canonical energy of the region $\S$ such that $\partial \S=A\cup \tilde{A}$. It has the expression
\be
\d E^\text{grav}_A=\int_\S \le(\d {\bf Q}[\xi_A] - \xi_A\cdot{\bm\Theta}(\d\phi)\ri)~ ,
\ee
where ${\bm\Theta}$ is the presymplectic form. Paralleling the AdS story \cite{Faulkner:2013ica}, let's define the form
\be\label{defchi}
{\bm\chi} = \d {\bf Q}[\xi_A] - \xi_A\cdot{\bm\Theta}(\d\phi)\,,
\ee
we will show that ${\bm\chi}$ satisfies the same properties as its AdS counterpart. The bulk modular flow $\xi_A$ vanishes on $\g$. It doesn't vanish on $\g^\pm$ where it is tangent, nonetheless, the integral of $\xi_A\cdot{\bm\Theta}(\d\phi)$ on $\g^\pm$ vanishes because  $\xi_A \cdot (\xi_A\cdot{\bm\Theta}(\d\phi)) = 0$ since ${\bm \Theta}$ is a 2-form. This shows that  $
\int_{\wt{A}} \xi_A\cdot{\bm\Theta}(\d\phi) = 0$ and that we have
\be
\d S_A^\text{grav} = \int_{\wt{A}} {\bm\chi}\,.
\ee
Using similar manipulations as in Sec.\ 5.1 of \cite{Faulkner:2013ica}, we can also show that 
\be
\d E^\text{grav}_A  = \int_A {\bm\chi}\, ,
\ee
and that
\be\label{dchiform}
d{\bm\chi} = -2\xi_A^a \d E_{ab} {\bm\ve}^b \, ,
\ee
where $\d E_{ab} $ are the equations of motion. Therefore, the gravitational entropy and energy satisfy a first law for on-shell perturbations
\be
\d S^\text{grav}_A = \d E^\text{grav}_A \,,
\ee
which follows from the fact that
\be\label{firstlawonshell}
\d E^\text{grav}_A - \d S^\text{grav}_A = \int_A{\bm\chi} - \int_{\wt{A}} {\bm\chi} = \int_\S d{\bm\chi} =0\,.
\ee
The goal of our paper is to show that the converse also holds: the first law of entanglement for all the regions $A$ (among a special class) implies the gravitational equations of motion.

\paragraph{Einstein gravity.}

For pure Einstein gravity, we have
\bea
{\bm \Theta}(\d g) \= {1\/16\pi G} (\n_b \d g^{ab} - \n^a \d g_b^{~b}) \,,\qq{\bm {\mathrm  Q}}[\xi] = -{1\/16\pi G} \n^a \xi^b {\bm \ve}_{ab}\,.
\eea
The expression for ${\bm\chi}$ reads
\bea\label{chiexplicit}
{\bm\chi}(\d g) \= \d {\bf Q}[\xi_A](\d g)- {\xi_A}\cdot{\bm\Theta}(\d g)\-
 \={1\/16\pi G} {\bm \ve}_{ab}\le( \d g^{ac} \n_c \xi^b_A - {1\/2} \d g_c^{~c} \n^a \xi^b_A+ \n^b \d g^a_{~c} \xi^c_A - \n_c \d g^{ac}\xi^b_A + \n^a \d g^c_{~c} \xi^b_A\ri) \,.
\eea
We now consider a small perturbation of the metric around Minkowski 
\begin{equation}
g_{ab}=\eta_{ab}+\lambda h_{ab},
\end{equation}
such that $\d g_{ab} =\la h_{ab}$, where $\lambda$ is small. For instance, one can consider a perturbation in Bondi gauge (see Sec. \ref{Flat Limit} for a complete description),
\be\label{generalpert3d}
h_{ab} dx^a dx^b = \le(\frac{V}{r} -2 \b\ri) du^2 - 4 \b dudr - 2 r^2 U du dr + 2r^2 \vphi \,d\phi^2\,,
\ee
where $V$, $\beta$, $U$ are functions of all coordinates, while $\varphi$ depends only on $u$ and $r$. 
The linearized Einstein equation are obtained for small $\lambda$:
\be
R_{ab} - {1\/2 }R g_{ab} = \d E_{ab}(h) \la + O(\la^2)\,.
\ee
Using \eqref{chiexplicit}, we have computed ${\bm\chi}$ explicitly and checked that indeed
\be
d{\bm\chi} = -2\xi^a \d E_{ab} {\bm\ve}^b\,.
\ee
Note that this formula follows from the general derivation given in \cite{Guica:2008mu}. It ensures the validity of the first law for on-shell perturbations. A simple class of asymptotically flat on-shell perturbations is 
\be
ds^2=\eta_{ab} dx^a dx^b +\la\left(\Theta(\phi)\, du^2+2\le(\Xi(\phi)+\frac{u}{2}\partial_{\phi}\Theta(\phi)\ri)du d\phi\right),
\label{pert3d}
\ee
where $\Theta$ and $\Xi$ are arbitrary functions of $\phi$. They were found in \cite{Barnich:2010eb} and we show how to obtain them in Sec.\ \ref{Flat Limit}. We focus on an interval $A$ on the slice $u=0$ (taking $\ell_u=0$) and with width $\l_\phi$. We compute explicitly the energy variation
\be
\d E_A = \int_A {\bm\chi} = {1\/ 4 \,\r{sin}(\lph) }\int_{-\lph}^{\lph}d\phi\,\le(\r{cos}\,\phi-\r{cos}(\lph)\ri)\,\Xi(\phi)\,.
\ee
Note that this can be written in term of the modular flow \eqref{3dmodflow} as
\be\label{dEA}
\d E_A = {1\/8\pi }\int_A d\phi \, \z_A^\phi\, \Xi(\phi)\,.
\ee
We conclude that this perturbation should be accompanied by a variation of the entropy for the first law to be satisfied. 

\paragraph{Refined prescription.} In \cite{Jiang:2017ecm}, the RT prescription was proposed only for Minkowski spacetime. For linearized perturbations at first order, the RT surface $\wt{A}$ is unchanged so we expect to be able to use the same prescription for perturbed Einstein gravity:
\be\label{naivepresc}
S_A = {\r{Length}(\wt{A})\/4 G}\,,
\ee
where the length is computed in the perturbed geometry. For the perturbation \eqref{pert3d}, it is easy to see that $\gamma_+$ and $\gamma_-$ are still light rays that intersect at the origin and, since $\tilde{A}$ is the union of them, the prescription would imply that $\d S_A = 0$.\footnote{We are using here the light sheaf prescription  where we impose that the light rays pass through the origin $r=0$. This is the prescription used in \cite{Jiang:2017ecm}.} This contradicts the first law of entanglement because $\d E_A\neq 0$. The resolution of this problem comes from the corner in $\wt{A}$ between $\g_+$ and $\g_-$. We should regulate it by considering a smooth curve $\wt{A}_\text{reg}$ arbitrarily close to $\wt{A} = \g_+\cup \g_-$. In other words, the corner has a non-trivial contribution to the integral.\footnote{There is a similar problem with the origin in polar coordinates. For example, we have $\int_{S^1_\ve} d\t=2\pi$ for a circle $S^1_\ve$ of radius $\ve$. Stokes theorem implies that this integral doesn't depend on $\ve$. In the limit $\ve\to0$ though, $S^1_\ve$ reduces to a point which suggests that the integral should be set to zero. This is incorrect because $d\t$ is not defined at the origin.} The correct prescription is then

\be
S_A = \int_{\wt{A}_\text{reg}} {\rm\bf Q}[\xi_A] = \lim_{\ve\to0}\int_{\wt{A}_\ve} {\rm\bf Q}[\xi_A]\,,
\label{refinedpresc}
\ee
where $\wt{A}_\ve$ is a smooth curve that regulates the corner in $\wt{A} = \g_+\cup \g_-$ and converges to $\wt{A}$ when $\ve\to 0$. From the fact that $d{\bm\chi} = 0$ on-shell and that $\wt{A}_\ve$ is a smooth curve homologous to $A$, we have
\be
\int_{\wt{A}_\ve} {\bm\chi} = \int_A {\bm\chi} = \d E_A~,
\ee
which would not be necessarily true if $\wt{A}_\ve$ had corners. From the definition \eqref{defchi} of ${\bm\chi}$, we can see that
\be
\d S_A = \lim_{\ve\to0} \int_{\wt{A}_\ve} \le( {\bm\chi} + \xi_A\cdot{\bm\Theta}\ri).
\ee
In the limit where $\ve\to 0$, the integral of $\xi_A\cdot{\bm\Theta}$ vanishes because $\xi_A$ is tangent to $\g_\pm$ and vanishes at the corner $\g_+\cap\g_-$ (while $\bm\Theta$ is finite at the corner). Therefore, we have checked the validity of the first law of entanglement for the RT prescription, 
\be
\d S_A = \d E_A\,.
\ee
Note that for Einstein gravity, \eqref{refinedpresc} doesn't reduce to the length of $\wt{A}_\text{reg}$ because ${\rm\bf Q}[\xi_A]$ computes only the length of the surface on which $\xi_A$ vanishes. In particular, $S_A$ can become negative for some choices of perturbations. We comment on this in Sec.\ \ref{constraints}.

\subsection{Positivity constraints}\label{constraints}

Let's consider the interval $A$ with $\l_u=0$ and use the prescription in which the light rays intersect at the origin, see Fig.\ \ref{fig:3dprescription}. In Einstein gravity, the entanglement entropy $S_A$ vanishes. This implies that the state $\rho_A $ is pure. This is unlike any standard quantum field theory, where the vacuum entanglement entropy has a universal divergence. This suggests some form of ultralocality as discussed in \cite{Wall:2011hj}: the vacuum factorizes between subregions of a constant $u$ slice of $\cI^+$. A perturbation will then create a nonzero entropy
\be
S_A= \d S_A = \d E_A\, .
\ee
From the explicit expression of \eqref{dEA}, we can see that this expression can become negative. This is in tension with the fact that von Neumann entropies are always positive. This gives a constraint on perturbations of the form \eqref{pert3d} that can be described within a quantum system on $\cI^+$ satisfying our assumptions. Imposing that 
\be\label{Xipos}
S_A=\d E_A \geq0
\ee
gives a constraint on $\Xi(\phi)$ according to \eqref{dEA}. To understand this better, let's restrict the Hilbert space $\cH$ that contains only the perturbations \eqref{pert3d} of 3d Minkowski. The condition \eqref{Xipos} implies that we should  restrict to the subspace $\cH_\text{code} \subset \cH$ on which $\d\ln K_A\rn \geq 0 $. This implies that the operator $K_A$ is bounded from below on $\cH_\text{code}$ and hence, that the density operator $e^{-K_A}$ is well-defined there. As a result, positivity of the entropy gives a constraint on the perturbations that can be described within a quantum system satisfying our assumptions. This is similar to the constraints on AdS perturbations coming from quantum information inequalities \cite{Lashkari:2014kda, Lashkari:2015hha, Lashkari:2016idm}.

\paragraph{Sign ambiguity.} The generalized Rindler method doesn't fix the sign of the modular flow. If a path integral formulation can eventually be given, the sign would be fixed from the choice of the vacuum state. Choosing the new modular flow $\z_A'=-\z_A$, with new modular Hamiltonian $K_A'=-K_A$, the condition $S_A\geq 0 $ selects a different subspace  $\cH_\text{code}' \subset \cH$: the subspace on which $K'_A$ is a positive operator. This ensures that for the modular flow $\z_A'$, we have a density operator $e^{-K_A'}$ which is well-defined on $\cH_\text{code}'$. Hence, changing the sign of the modular flow amounts to selecting a different subspace on which $\rho_A$ is well-defined.

\begin{figure}
\centering
	\begin{tabular}{cc}
		\subf{\includegraphics[width=7cm]{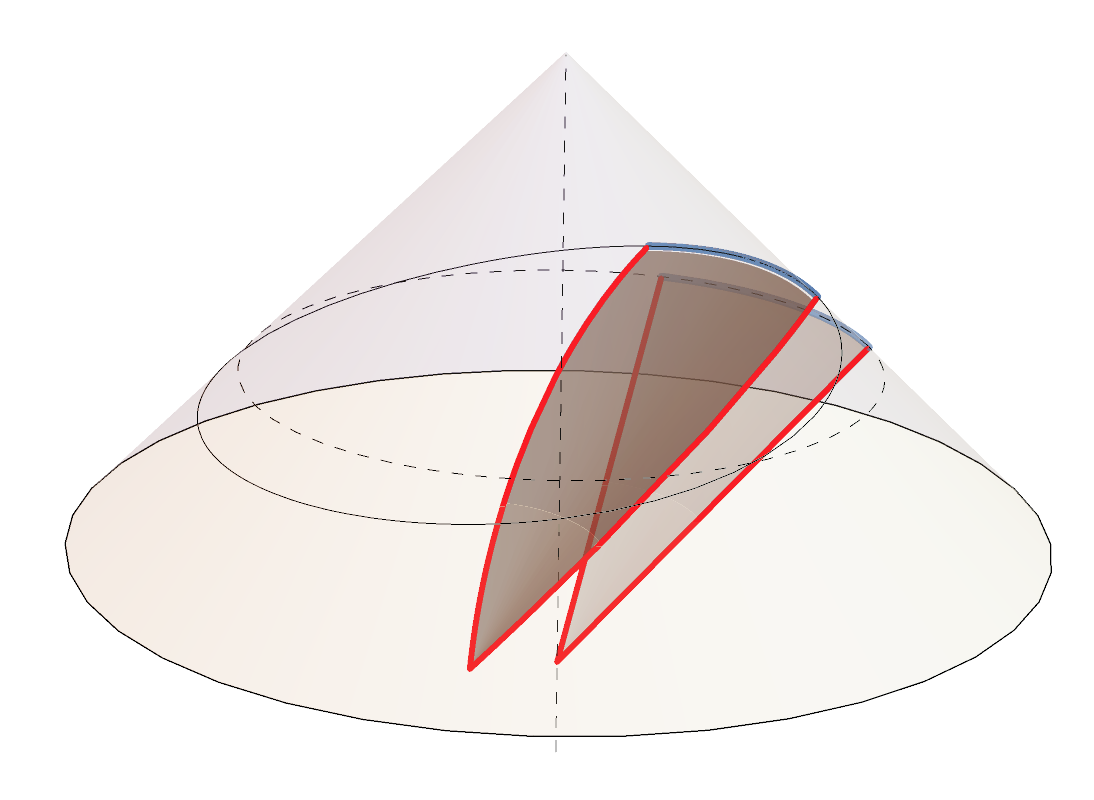}}{Translation in the bulk} &
		\subf{\includegraphics[width=7cm]{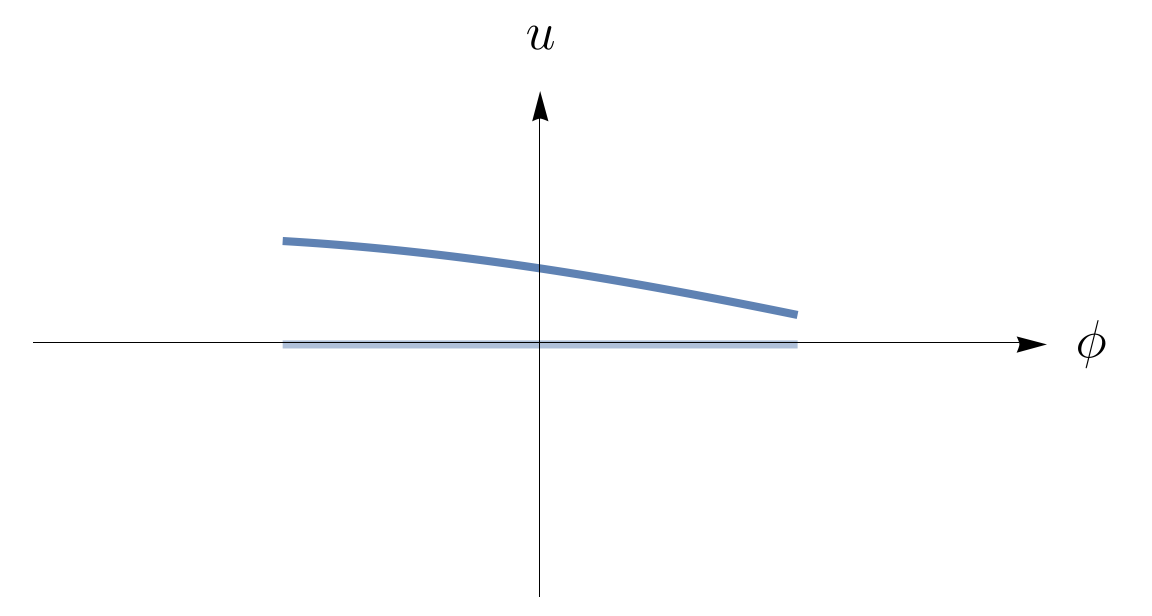}}{Translation in the boundary}
			\end{tabular}
\put(-335, 10){{\color{red}{$\wt{A}'$}}}
\put(-270, -10){{\color{red}{$\wt{A}$}}}
\put(-285, 45){{\color{bluemathematica}{$A'$}}}
\put(-262, 32){{\color{bluemathematica}{$A$}}}
\put(-106, 5){{\color{bluemathematica}{$A'$}}}
\put(-125, -25){{\color{bluemathematica}{$A$}}}

\centering
	\begin{tabular}{cc}
		\subf{\includegraphics[width=7cm]{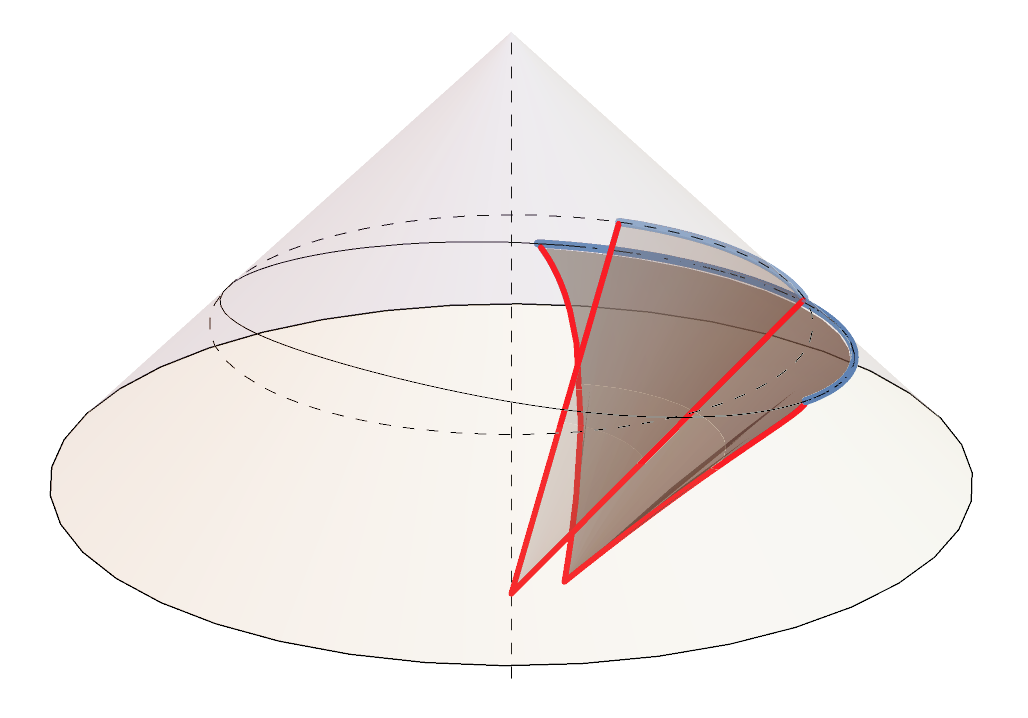}}{Boost in the bulk} &
		\subf{\includegraphics[width=7cm]{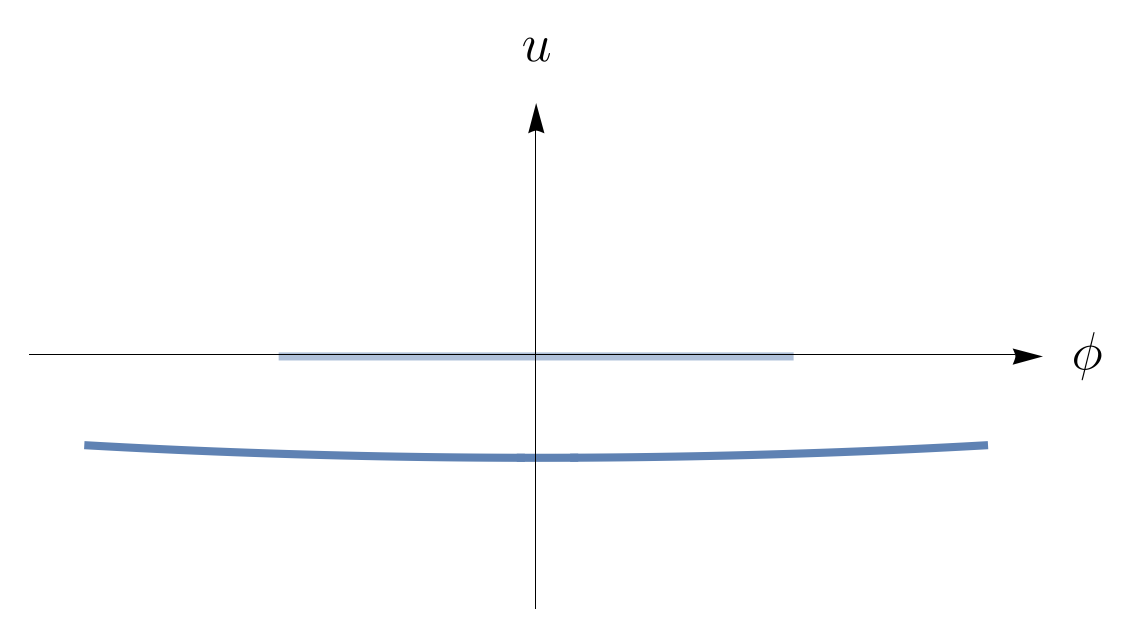}}{Boost in the boundary} 
	\end{tabular}
\put(-325, 15){{\color{red}{$\wt{A}'$}}}
\put(-330, -25){{\color{red}{$\wt{A}$}}}
\put(-273, 36){{\color{bluemathematica}{$A$}}}
\put(-250, 20){{\color{bluemathematica}{$A'$}}}
\put(-125, -2){{\color{bluemathematica}{$A$}}}
\put(-106, -40){{\color{bluemathematica}{$A'$}}}

\vspace{0.3cm}
\caption{Examples of new RT surfaces obtained by bulk isometries acting on the reference configuration for $\ell_u=0$.}\label{fig:newRT}
\end{figure}
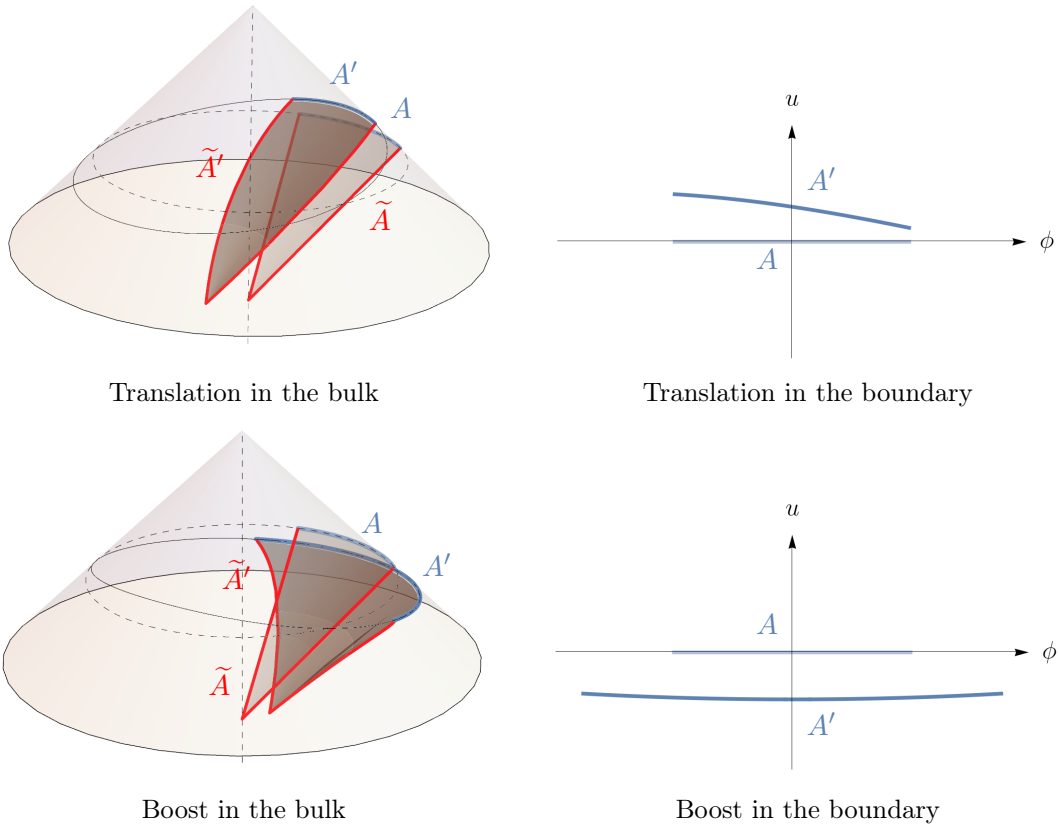

\section{Flat 3d gravity from entanglement}\label{3dproof}

In this section, we show that the first law of entanglement implies the gravitational equations of motion, linearized around three-dimensional Minkowski spacetime. Our proof is valid for any theory of gravity, including higher-derivative terms. The generalization to four dimensions is treated in the Sec.\ \ref{4dproof}.

\subsection{General strategy}\label{genstrat}

Let's consider a general off-shell perturbation of 3d Minkowski. The one-form ${\bm\chi}$ satisfies
\be\label{dchideltaE}
d{\bm\chi} = -2 \xi^a \d E_{ab} {\bm\ve}^b,
\ee
where $\d E_{ab}$ are the equations of motion for the perturbations and  ${\bm\ve}_a = {1\/2}\ve_{abc} dx^b\wg dx^c$.\footnote{$\ve_{abc}$ is a totally antisymmetric tensor such that $\ve_{u r \phi}=\sqrt{-g}$.} As explained in \eqref{firstlawonshell}, the first law of entanglement implies that for all surfaces $\S$ bounded by $A$ and $\wt{A}$, we have
\be
\int_\S d{\bm\chi}=0\,.
\ee
We would like to show that this implies that $\d E_{ab} = 0$. This is reasonable because we have a large number of such surfaces $\S$. The derivation will be similar to the AdS case \cite{Faulkner:2013ica} although the RT surfaces are more involved here. Bulk isometries will play a crucial role.

The strategy is to start with some reference configuration. By varying the parameters of this configuration, we will obtain constraints on the gravitational equations $\d E_{ab}$. We will then act on this configuration with bulk isometries to obtain new constraints. This amounts to probing the perturbation with new RT surfaces, obtained by applying a bulk isometry to the reference configuration. The new constraint is obtained by replacing $\d E_{ab}$ by its image under the transformation. The logic can be phrased as follows: the first law of entanglement gives the equation
\be
\int_\S  \xi^a \d E_{ab} (x){\bm\ve}^b =0  \,.
\label{1}
\ee
We can consider a new configuration $\wt\S$ obtained by performing a bulk isometry $x\to\tilde{x}$. The associated bulk modular flow $\tilde\xi^a $ and volume form ${\bm\tilde\ve}^b$ can be obtained by applying the transformation to $\xi^a$ and ${\bm\ve}^b$, which gives
\be
\int_{\wt\S}   \tilde{\xi}^{a} \d E_{ab}(\tilde{x}) \tilde{\bm\ve}^{b} =0\,. 
\ee
We are probing the same perturbation $\d E_{ab}$ with a different RT surface and we emphasize that $\d E_{ab}(\tilde{x})$ is now evaluated on the new RT surface $\tilde{\S}$. Now, we can change variables in the integral using the inverse bulk isometry $x\to x'$. This gives
\be\label{newconstraint}
\int_\S  \xi^c \le( {\p \tilde{x}^a \/ \p x^c}{\p \tilde{x}^b \/ \p x^d}\d E_{ab} (\tilde{x}(x))\ri) {\bm\ve}^d =0 \,.
\ee
This shows that if \eqref{1} allows us to prove that some functional of the equations of motion vanishes: 
\be
\cF\left[\d E_{ab}(x)\right]=0,
\ee 
then we immediately have that the same functional but applied to the transformed equations of motion vanishes:
\be 
\cF\left[ {\p \tilde{x}^c \/ \p x^a}{\p \tilde{x}^d \/ \p x^b}\d E_{cd} (\tilde{x}(x))\right]=0.
\ee
This procedure is made mathematically precise in App.\ \ref{precisionstrat}.

\subsection{Linearized gravitational equations}\label{sec:3dproof}

We now describe the proof of the gravitational equations, linearized around 3d Minkowski spacetime. Although the proof is conceptually similar to the AdS case derived in \cite{Faulkner:2013ica}, it is rather more challenging in flat space. In particular, we will have to use different RT prescriptions as discussed in Sec.\ \ref{3dgenpresc}. Bulk isometries will also play an important role in generating enough constraints on the perturbation. 

\paragraph{Reference configuration.} The reference configuration is an interval $A$ with $\l_u=0$ at $u=0$ and with length $\l_\phi$ centered at $\phi=0$. We can parametrize the interval $A$ by
\be\label{boundaryAref}
A :\quad  u=0 ,\quad \phi \in [-\lph,\lph]\,.
\ee
The RT surface $\wt{A}$ consists of two semi-infinite light rays starting at the origin and ending at the extremities $\p A$, as in Fig.\ \ref{fig:3dprescription}. The surface $\S$ at $u=0$ which is bounded by $A$ and $\wt{A}$ can be parametrized by $r$ and $\phi$ with
\be
\S : \quad u = 0 ,\quad  r\geq 0 ,\quad \phi \in [-\lph,\lph]\,. 
\ee
The bulk modular flow \eqref{3dbulkmodularflow} evaluated on $\S$ reduces to
\bea\label{3dbulkmodref}
\xi_A \=   {2\pi \/\r{sin}(\lph)}  \le( r \,\r{sin}\,\phi\,\p_r  +  (\r{cos}\,\phi-\r{cos}(\lph)) \p_\phi\ri).
\eea 
Let's write explicitly the equation \eqref{dchideltaE}. In Bondi coordinates, we have 
\be
{\bm\ve}^r = -{\bm\ve}_u = - r \,dr\wg d\phi\,.
\ee
Hence, the pullback of $d{\bm\chi}$ on $\S$ is\footnote{The 2-form $dr\wg d\phi$ is singular at $r=0$ so we need to restrict the integration range to $r\geq \ve$ and take $\ve\to 0$ at the end. This is always what we will be doing implicitly.}
\be
d{\bm\chi}|_\S = 2 r \xi^a \d E_{a r} dr\wg d\phi\,.
\ee
From \eqref{dchideltaE}, we obtain\footnote{ We thank Hongliang Jiang for pointing out a mistake in the previous version of this formula.}
\be\label{firsteqproof3d}
\int_{-\lph}^{\lph}d\phi\int_0^{+\infty} dr \le(r^2\r{sin}\,\phi\, \d E_{rr} + r\le(\r{cos}\,\phi- \r{cos}(\lph) \ri) \d E_{r\phi}\ri)=0 ~.
\ee
Expanding  this equation at small $\l_\phi$ implies that
\bea\label{expsmalllphi}
\int_0^{+\infty} dr \,( r^2 \p_\phi \d E_{rr}(0,r, 0) + r \d E_{r\phi}(0,r,0))=0~.\-
\eea
\vspace{-1.1cm}
\paragraph{Rotations and time translations.} We can consider new configurations obtained by performing rotations. They are the same as the reference configuration but centered at $\phi=\phi_0$. The new RT suraces are obtained as the image under the bulk isometries
\be
\phi\rightarrow \phi+\phi_0.
\ee
The Jacobian of this transformation is simply the identity. Therefore, following the logic exposed in the previous section, we obtain that the vanishing of the functional \eqref{expsmalllphi} but applied to the image of $\d E_{ab}$ under this isometry:
\be
\int_0^{+\infty} dr \,( r^2 \p_\phi \d E_{rr}(0,r, \phi_0) + r \d E_{r\phi}(0,r,\phi_0))=0,
\ee
for any angle $\phi_0$. We can do the same with translation $u\ra u + u_0$ in retarded time $u$, to obtain
\be\label{Co}
\int_0^{+\infty} dr \,( r^2 \p_\phi \d E_{rr}(u_0,r, \phi_0) + r \d E_{r\phi}(u_0,r,\phi_0))=0.
\ee

\paragraph{light sheaf deformation.} 

We consider the same boundary interval $A$ as in the reference configuration \eqref{boundaryAref}. The latter followed the prescription in which the light rays $\g_+$ and $\g_-$ intersect the spatial origin $r=0$. This is not the most general prescription, as discussed in Sec.\ \ref{3dgenpresc}. Here, we will use a more general prescriptions to derive more constraints on $\d E_{ab}$. An alternative proof of this step is presented in the App.\ \ref{app:alt3d}.

We consider a more general light sheaf for the interval $A$. We take the parametrization \eqref{3dlightsheet} where we set $\l_u= Y_-=0 $ and $Y_+=Y$. The two light rays intersect the bifurcation surface at $\tilde{x}=0$ and $\tilde{x}=-Y\,\cos\,(\frac{\ell_\phi}{2})$. The first law tells us that for any $Y$, we have
\be
\int_{\S_{Y}} d{\bm\chi} = 0 ~,
\label{ConstrY}
\ee
where the surface $\S_{Y}$ depends on $Y$ and can be chosen to be any surface such that $\partial \S_{Y}=\tilde{A}\cup A$. In particular, one can choose $\S_{Y}=\S_{\{Y=0\}}\cup N_Y$, where $N_Y$ is the strip created by the union of all the half light rays $\g_+$ given in \eqref{3dlightsheet} where the parameter $Y_+$ goes from $0$ to $Y$. From \eqref{ConstrY}, it then follows that for any $Y$, we have
\be
\int_{N_{Y}} d{\bm\chi} = 0~.
\label{ConstrYY}
\ee
We now take the derivative with respect to $Y$ and evaluate at $Y=0$. The integral reduces to an integral over the $Y_+=0$  light ray and the integrand is contracted with $\partial_{\tilde{x}}$ as the effect of changing $Y_+$ is to translate the light ray in the $\tilde{x}$-direction. At the end, we get
\be
\int_{\gamma_+} \, \p_{\tilde{x}}\cdot d{\bm\chi} = 0.
\ee
where $\gamma_+$ is the usual light ray from the origin to the point $(u,\phi) = (0,\lph)$. Converting the vector to Bondi coordinates, we obtain
\be
\p_{\tilde{x}} = {1\/\r{sin}(\lph)}\le[\le(\r{cos}(\lph) -\r{cos}\,\phi \ri)\p_u + \r{cos}\,\phi\,\p_r - {\r{sin}\,\phi\/r} \p_\phi \ri] ~.
\ee
The integral is evaluated at $\phi = \lph$ where the expressions for $\p_\tx$ and for the bulk modular flow \eqref{3dbulkmodref} simplify to 
\be
\p_\tx =\r{cot}(\lph) \p_r - {1\/r} \p_\phi,\qq  \xi_A =   2\pi r\,\p_r
\ee
The pullback on $\g_+$ only keeps the $dr$ component so in the expression \eqref{dchideltaE} for $d{\bm\chi}$, we only have a contribution from ${\bm\ve}^r = - r dr\wg d\phi$. As a result, $\p_\tx\cdot d{\bm\chi}|_{\g_+} = -4 \pi r\, \d E_{rr} \,dr $ and we obtain
\be\label{3dintErr}
\int_0^{+\infty} dr \,r\,\d E_{rr}(u_0,r,\phi_0)  = 0~,
\ee
where as above, we have used rotations and time translations to make this expression valid for any $u_0$ and $\phi_0$.

\paragraph{Radial translations.} Let's consider a new configuration which is obtained by translating the reference configuration by a distance $r_0$ in the direction $\phi_0$ of the light ray on which \eqref{3dintErr} is integrated. In Cartesian coordinates, such a translation is given by
\be
t \rightarrow t+  r_0, \qq x\rightarrow  x+r_0 \,\r{cos}\,\phi_0, \qq y\rightarrow  y + r_0 \,\r{sin}\,\phi_0~.
\ee
These configurations are illustrated in Fig.\ \ref{fig:newRT}. We can apply the reasoning presented in Sec.\ \ref{genstrat} for these new configurations. In Bondi coordinates, the transformation becomes
\bea
u  &\rightarrow & r+r_0+u -\sqrt{r^2+2\, r \, r_0\, \cos (\phi -\phi_0)+r_0^2} ~,\\
r &\rightarrow & \sqrt{r^2+2\, r \, r_0\, \cos (\phi -\phi_0)+r_0^2}~, \\
\phi &\rightarrow & \arctan({r
   \,\sin (\phi )+r_0\, \sin (\phi_0) \/ r\,\cos (\phi )+r_0 \,\cos (\phi_0)} ) .
\eea
The constraint \eqref{Co} applied to the image of $\d E_{ab}$ under this isometry gives the new constraint
\be
\int_{r_0}^{+\infty} dr\,(r-r_0)\,\d E_{rr}(u_0,r,\phi_0) = 0 ~,
\ee
where we have also performed the change of variable $r\to r-r_0$ in the integral. 
Taking two derivatives with respect to $r_0$ shows that
\be
\d E_{rr}(u_0,r_0,\phi_0) = 0~,
\ee
for any value of $u_0,r_0,\phi_0$. From this, the equation \eqref{Co} simplifies to
\be
\int_0^{+\infty} dr \, r\, \d E_{r\phi}(u_0,r,\phi_0)=0~.
\ee
We use the same radial translation on this equation to obtain the constraint
\be
\int_{r_0}^{+\infty} dr \, {(r-r_0)^2\/r} \d E_{r\phi}(u_0,r,\phi_0)~.
\ee
Taking three derivatives with respect to $r_0$ implies that
\be
\d E_{r\phi}(u_0,r_0,\phi_0) = 0~,
\ee
which is true for any value of $u_0,r_0,\phi_0$. Hence, we have shown that
\be\label{resradtrans}
\d E_{rr } =\d E_{r\phi}=0~,
\ee
everywhere in the bulk.

\paragraph{General translations.}  We  consider a general bulk translation $\d x^\mu = v^\mu$. This generates a new family of configurations, illustrated in Fig.\ \ref{fig:newRT}. Acting with the infinitesimal translation on $\d E_{r\phi}=0$ leads to
\be
(v_y\r{cos}\,\phi - v_x\,\r{sin}\,\phi) (r^2 \d E_{u r} +\d E_{\phi\phi}) = 0\,,
\ee
which implies that
\be
\d E_{\phi \phi} =  -r^2 \d E_{u r} \,, 
\ee
everywhere in the bulk.

\paragraph{Conservation equation.} We now consider the conservation equation
\be
\n_a(\d E^{ab})=0~,
\ee
which is always satisfied by the equations of motion. Here, $\n_a$ is the derivative with respect to the background Minkowski spacetime. We will use this equation together with an additional holographic input to cancel the remaining components. Indeed one should remember that in AdS, the proof requires a holographic input that is the conservation and the tracelessness of the boundary stress tensor. In a radial Hamiltonian perspective, they correspond to initial conditions on the boundary surface. In the flat case, similar initial conditions are required. We will show in the next section how to make sense of a boundary "stress tensor" and derive its constraint equations using a flat limit in AdS.

For $b = u$, the conservation equation implies
\be
\p_r(\d E_{ur}) =0\,,
\ee
which leads to $\d E_{ur} = C_0(u,\phi)$ and $\d E_{\phi\phi} = -r^2 C_0(u,\phi)$. We expect that the trace conditions \eqref{Trace1} and \eqref{Trace2} imply that $C_0=0$ although we have not been able to show it conclusively.\footnote{This would be done by turning on an off-shell perturbation in the Bondi gauge such that \eqref{Trace1} and \eqref{Trace2} are violated which would allows us to identify the corresponding components of Einstein equations. We leave this analysis for future work.} Assuming that this is the case, we obtain
\be
\d E_{ur} = \d E_{\phi\phi} =0\,,
\ee
everywhere in the bulk. The conservation equation for $b = \phi$ then gives
\be
  \d E_{u\phi} + r \p_r(\d E_{u\phi})  =0\,.
\ee
The solution of this equation is
\be
\d E_{u\phi} = {C_2(u,\phi)\/r}\,.
\ee
In the next section, we show that the equation $C_2 = 0$ is precisely the conservation equation \eqref{Cons2} of the boundary stress tensor, so we have $\d E_{u\phi}=0$. Finally, the component with $b=r$ gives
\be
\d E_{uu} + r\p_r(\d E_{u u}) =0 \,,
\ee
with solution
\be
\d E_{uu} = {C_1(u,\phi)\/r}\,.
\ee
The equation $C_1=0$ is the other conservation equation \eqref{Cons1} of the boundary stress tensor, so we have $\d E_{uu} = 0$. Hence, we have shown that all the components of the linearized gravitational equation vanish.

\section{Holographic stress tensor in flat spacetime}\label{holst}

In AdS, the boundary is a timelike hypersurface which allows for the definition of a non-degenerate boundary metric whose dual operator is the boundary stress tensor. In flat space, things are more subtle, because the metric becomes degenerate on the boundary (its determinant vanishes). This is simply because $\mathcal{I}^+$ is a null hypersurface. To have  a good understanding of the flat case, it is  helpful to start from its AdS counterpart and perform a flat limit sending the AdS radius to infinity, we will see that this amounts to perform a Carrollian limit on the boundary (or ultra-relativistic limit). We will show that the induced geometry on a null hypersurface contains more than a degenerate metric and that additional geometrical objects appear naturally when performing the flat limit. The concept of boundary stress tensor will also have to be modified.

\subsection{AdS$_3$ in Bondi gauge}

We consider the following metric, written in Bondi gauge:
\begin{equation}
ds^2=\frac{\tilde{V}}{r}e^{2\tilde{\beta}}du^2-2e^{2\tilde{\beta}}dudr+r^2e^{2\tilde{\varphi}}(d\phi-\tilde{U}du)^2,
\label{BondiGauge}
\end{equation}
We are going to consider small perturbations around global AdS, the most generic perturbation in Bondi gauge is given by
\begin{equation}
\tilde{\beta}=\lambda\beta,\quad \tilde{V}=-r\left(1+\frac{r^2}{\l^2}\right)+\lambda V,\quad \tilde{U}=\lambda U, \quad\tilde{\varphi}=\lambda\varphi,
\end{equation}
where $\lambda$ is a small parameter. From now on, all the expressions will be linearized in $\lambda$. Solving the $(r,r)$, $(r,u)$, $(r,\phi)$ and $(\phi,\phi)$-components of the linearized Einstein equations, with negative cosmological constant, gives
\begin{equation}
\begin{split}
\beta &=\beta_0(u,\phi),\\
U &= -\frac{N(u,\phi)}{r^2}+U_0(u,\phi)+\frac{2\partial_\phi\beta_0}{r},\\
V &=rM(u,\phi)+r\left(-\frac{2r^2\beta_0}{\l^2}-2r\left(\partial_\phi U_0+\partial_u\varphi\right)\right).
\end{split}
\end{equation} 
The flat limit was considered for the case $\beta_0=U_0=0$ in \cite{Barnich:2012aw}. There are two residual equations, given by the $(u,u)$ and $(u, \phi)$-components of Einstein equations
\begin{equation}
\begin{split}
&\partial_u M= 2 \partial_\phi U_0+2
   \partial_\phi^2 U_0-2 \partial_u\beta_0-4 \partial_u\partial_\phi^2 \beta_0+2 \partial_u\varphi+2 \partial_u\partial_\phi^2\varphi+2\l^{-2}\partial_\phi N\,,\\
&\partial_u N=\frac{1}{2}\partial_\phi M-\partial_\phi\beta_0\,.
\end{split}
\end{equation}
The latter can be understood as the conservation of a boundary stress tensor
\begin{equation}
\nabla_{\mu}T^{\mu\nu}=0\,,
\end{equation}
where $\mu=\{u,\phi\}$. The boundary metric and the stress tensor are given by
\begin{equation}\arraycolsep=5pt\def\arraystretch{1.2}
g_{\mu\nu}=\begin{pmatrix}
-\frac{1+4\lambda \beta_0}{\l^2}& -\lambda U_0\\
-\lambda U_0 & 1+2\lambda \varphi
\end{pmatrix}, \quad T^{\mu\nu}=\frac{1}{8G}\tau^{\mu\nu},
\end{equation}
where
\begin{equation}
\begin{split}
\tau^{uu}&=-\l^3\left(-1+\lambda\left( M + 6\beta_0+4\partial_\phi^2\beta_0\right)\right),\\
\tau^{u\phi}&=\l \lambda\left(2N+\l^2\left(U_0+2\partial_\phi^2U_0+2\partial_u\partial_\phi\varphi\right)\right),\\
\tau^{\phi\phi}&=-\l\left(-1+\lambda\left(M+ 2\beta_0+2\varphi+2\l^2\partial_u\partial_\phi U_0+2\l^2\partial_u^2\varphi\right)\right).
\end{split}
\end{equation}
This stress tensor can be obtained, for example,  through the Brown and York procedure. It is well-known that the boundary theory is a 2d CFT whose central charge is given by \cite{Brown:1986nw}
\begin{equation}
c=\frac{3 \l}{2G}\,,
\end{equation}
and this is confirmed by computing the anomalous trace of the stress tensor 
\begin{equation}
T^{\mu}_{~\mu}=-\frac{c}{12}R=-\frac{\l }{8G}R\,,
\label{TraceAnomaly}
\end{equation}
where $R$ is the scalar curvature of the boundary metric.

\subsection{Flat limit and Carrollian geometry}\label{Flat Limit}

We  have now all the ingredients to perform the flat limit. In the bulk, the $\l\rightarrow \infty$ limit of the metric is given by  another metric in the Bondi gauge \eqref{BondiGauge} but whose defining functions are
\begin{equation}
\tilde{\beta}=\lambda\beta,\quad \tilde{V}=-r+\lambda V,\quad \tilde{U}=\lambda U, \quad\tilde{\varphi}=\lambda\varphi,
\end{equation}
we notice that this is now a perturbation around Minkowski. Solving the $(r,r)$, $(r,u)$, $(r,\phi)$ and $(\phi,\phi)$-components of the linearized Einstein equations, this time without cosmological constant, gives
\begin{equation}
\begin{split}
\beta &=\beta_0(u,\phi),\\
U &= -\frac{N(u,\phi)}{r^2}+U_0(u,\phi)+\frac{2\partial_\phi\beta_0}{r},\\
V &=rM(u,\phi)+r\left(-2r\left(\partial_\phi U_0+\partial_u\varphi\right)\right).
\end{split}
\end{equation} 
The two residual equations, the $(u,u)$ and $(u, \phi)$-components of Einstein equations, are
\begin{eqnarray}
\label{Cons1}
\partial_u M &=& 2 \partial_\phi U_0+2
   \partial_\phi^2 U_0-2 \partial_u\beta_0-4 \partial_u\partial_\phi^2 \beta_0+2 \partial_u\varphi+2 \partial_u\partial_\phi^2\varphi,\\
\label{Cons2}
\partial_u N &=& \frac{1}{2}\partial_\phi M-\partial_\phi\beta_0.
\end{eqnarray}
To be more precise, we have that the $(u,u)$ and $(u, \phi)$-components of the linearized Einstein equations scale with $r$ as 
\begin{equation}\label{flatlimitconst}
\delta E_{uu}=\frac{C_1(u,\phi)}{r} \quad \text{and}\quad  \delta E_{u\phi}=\frac{C_2(u,\phi)}{r},
\end{equation}
such that $C_1=0\Leftrightarrow $ \eqref{Cons1} and $C_2=0\Leftrightarrow$ \eqref{Cons2}. These conditions are the holographic input we need for the proof of Sec.\ \ref{3dproof}. The difference with the AdS case is that we cannot recast these two conservation equations as the divergence of a boundary energy--momentum tensor for the simple reason that there is no non-degenerate boundary metric that allows us to build the usual covariant derivative. In the following we will show how to obtain the right geometrical structure to describe the boundary geometry.

To perform the limit on the boundary,  it is useful to decompose the boundary metric and energy--momentum tensor with respect to their scaling with $\l$. We start with the metric
\begin{equation}
g_{\mu\nu}=h_{\mu\nu}-\l^{-2}n_{\mu}n_{\nu},
\end{equation}
where 
\begin{equation}\arraycolsep=5pt\def\arraystretch{1.2}
n_\mu=\begin{pmatrix}
1+2\lambda\beta_0\\
0
\end{pmatrix},\quad
h_{\mu\nu}=\begin{pmatrix}
0 & -\lambda  U_0\\
-\lambda U_0 & 1+2\lambda\varphi
\end{pmatrix}.
\end{equation}
The inverse metric is
\begin{equation}
g^{\mu\nu}=-\l^2v^{\mu}v^{\nu}+h^{\mu\nu},
\end{equation}
where
\begin{equation}\arraycolsep=5pt\def\arraystretch{1.2}
v^\mu=\begin{pmatrix}
1-2\lambda\beta_0\\
\lambda U_0
\end{pmatrix},\quad
h^{\mu\nu}=\begin{pmatrix}
0 & 0\\
0 & 1-2\lambda\varphi
\end{pmatrix}.
\end{equation}
This decomposition allows us to define properly the geometry on the null infinity. It will be composed of a degenerate metric $h_{\mu\nu}$ (which induces a real metric on the boundary circle) whose kernel is given by the vector field $v^\mu$ which represents the time direction, a temporal one-form $n_{\mu}$ and the pseudo-inverse metric $h^{\mu\nu}$ (indeed, as $h_{\mu\nu}$ is degenerate, it does not enjoy a true inverse). These are the ingredients of a Carrollian geometry \cite{Bekaert:2015xua, Hartong:2015xda}. One can check that they satisfy the following relations
\begin{equation}
h_{\mu\nu}v^\nu=0,\quad h^{\mu\nu}n_\nu=0,\quad v^\mu n_\mu=1 \quad \text{and}\quad h^{\mu\sigma}h_{\sigma\nu}=\delta^\mu_\nu-v^\mu n_\nu\,,
\end{equation}
at first order in $\la$. These can be taken as the defining relations of a Carrollian geometry. We will also make use of the scalings of the Christoffel symbols with $\l$:
\begin{equation}
\Gamma^{\mu}_{\nu\rho}=\l^2X^{\mu}_{\nu\rho}+Y^{\mu}_{\nu\rho}+\l^{-2}Z^{\mu}_{\nu\rho},
\end{equation} 
where $X^{\mu}_{\nu\rho}$, $Y^{\mu}_{\nu\rho}$ and $Z^{\mu}_{\nu\rho}$ can be written in terms of the Carrollian geometry as\footnote{One can check that $Y^{\mu}_{\nu\rho}$ is a torsionless "compatible" Carrollian connection \cite{Bekaert:2015xua}, which means that it parallel transports $v^{\mu}$ and $h_{\mu\nu}$.}
\begin{eqnarray}
X^{\mu}_{\nu\rho} &=&-\frac{1}{2}v^\nu v^\sigma(\partial_\nu h_{\sigma\rho}+\partial_\rho h_{\sigma\nu}-\partial_\sigma h_{\nu\rho}),\\
Y^{\mu}_{\nu\rho} &=& \gamma^{\mu}_{\nu\rho}+v^{\mu}v^{\sigma}\left(n_{(\nu}\partial_{\rho)}n_\sigma-(\partial_\sigma n_{(\nu})n_{\rho)}\right)+v^\mu\partial_{(\nu}n_{\rho)},\\
Z^{\mu}_{\nu\rho} &=& h^{\mu\sigma}\left((\partial_\sigma n_{(\nu})n_{\rho )}-n_{(\nu}\partial_{\rho)}n_\sigma\right),
\end{eqnarray}
where $\gamma^{\mu}_{\nu\rho}=\frac{1}{2}h^{\mu\sigma}(\partial_\nu h_{\sigma\rho}+\partial_\rho h_{\sigma\nu}-\partial_\sigma h_{\nu\rho})$ is the Levi-Civita of the pseudo metric $h_{\mu\nu}$.

The boundary energy--momentum tensor scales with $\l$ as
\begin{equation}
T^{\mu\nu}=\l^3T_1^{\mu\nu}+\l\, T_0^{\mu\nu},
\end{equation}
so the boundary dynamical data decomposes in two pieces, $T_0^{\mu\nu}$ and $T_1^{\mu\nu}$, defined on $\cI^+$. For the perturbation in Bondi gauge, they are given by
\arraycolsep=6pt\def\arraystretch{1.2}
\begin{eqnarray}
T_0^{\mu\nu}&=&\frac{1}{8G}\begin{pmatrix}
0 & 2\lambda N \\
2\lambda N & 1-\lambda(M+2\beta_0+2\varphi)
\end{pmatrix}\,,\\
T_1^{\mu\nu}&=&{1\/8G}\begin{pmatrix}
1-\lambda(M+6\beta_0+4\partial_\phi^2\beta_0)& \lambda(U_0+2\partial_\phi^2U_0+2\partial_\phi\partial_u\varphi)\\
\lambda(U_0+2\partial_\phi^2U_0+2\partial_\phi\partial_u\varphi) & -\lambda(2\partial_u\partial_\phi U_0+2\partial_u^2\varphi)
\end{pmatrix}.
\end{eqnarray}
We can now take the $\l\rightarrow\infty$ limit of the conservation equations. We obtain the two following conservation laws, a scalar one and a vector one 
\begin{eqnarray}
n_{\sigma}\left(\partial_\mu T_1^{\mu \sigma}+Y^{\mu}_{\mu\rho}T_1^{\rho \sigma}+Y^{\sigma}_{\mu\rho}T_1^{\mu\rho } +X^{\mu}_{\mu\rho}T_0^{\rho \sigma}+X^{\sigma}_{\mu\rho}T_0^{\mu\rho }\right)&=&0\,,\\
h_{\nu\sigma}\left(\partial_\mu T_0^{\mu \sigma}+Y^{\mu}_{\mu\rho}T_0^{\rho \sigma}+Y^{\sigma}_{\mu\rho}T_0^{\mu\rho } +Z^{\mu}_{\mu\rho}T_1^{\rho \sigma}+Z^{\sigma}_{\mu\rho}T_1^{\mu\rho }\right)&=&0\,.
\end{eqnarray}
In three dimensions, the vector conservation corresponds only to one equation since its projection on $v^\mu$ vanishes by definition. These two equations are the analog of the conservation of the stress tensor in AdS$_3$ and reproduce perfectly the two equations \eqref{Cons1} and \eqref{Cons2}. They are the holographic input that we need in the proof in Sec.\ \ref{3dproof} to cancel the integration constants $C_1$ and $C_2$.

There is also a Carrollian equivalent of the relation between the trace of $T^{\mu\nu}$ and the scalar curvature. It is obtained simply by taking the $\l\rightarrow\infty$ of the formula \eqref{TraceAnomaly} which splits into two equations:
\begin{eqnarray}
\label{Trace1} h_{\mu\nu}T_0^{\mu\nu}-n_{\mu}n_\nu T_1^{\mu\nu}&=&-\frac{R_0}{8G}\,,\\
\label{Trace2}
h_{\mu\nu}T_1^{\mu\nu}&=&-\frac{R_1}{8G}\,,
\end{eqnarray}
where $R_0$ and $R_1$ are two Carrollian scalar curvatures defined as
\begin{eqnarray}\nt
R_0 &=& R_Y-2v^\mu v^\nu\left(\partial_{[\alpha} Z^\alpha_{\nu]\mu }+Y^\beta_{\mu[\nu}Z^\alpha_{\alpha]\beta}+Z^\beta_{\mu[\nu}Y^\alpha_{\alpha]\beta}\right)+2h^{\mu\nu}\left(Z^\beta_{\mu[\nu}X^\alpha_{\alpha]\beta}+X^\beta_{\mu[\nu}Z^\alpha_{\alpha]\beta}\right),\\
\hspace{-0.8cm} R_1 &=&-2v^\mu v^\nu\left(\partial_{[\alpha}Y^{\alpha}_{\nu]\mu}+Y^\beta_{\mu[\nu}Y^{\alpha}_{\alpha]\beta}+Z^\beta_{\mu[\nu}X^{\alpha}_{\alpha]\beta}+X^\beta_{\mu[\nu}Z^{\alpha}_{\alpha]\beta}\right)\\ \notag
\hspace{-0.8cm} &&+2h^{\mu\nu}\left(\partial_{[\alpha}X^{\alpha}_{\nu]\mu}+Y^\beta_{\mu[\nu}X^{\alpha}_{\alpha]\beta}+X^\beta_{\mu[\nu}Y^{\alpha}_{\alpha]\beta}\right),
\end{eqnarray}
and $R_Y$ is the scalar curvature associated with $Y^{\mu}_{\nu\rho}$:
\be
R_Y=h^{\mu\nu}\left(\partial_\alpha Y^{\alpha}_{\nu\mu}-\partial_\nu Y^{\alpha}_{\alpha\mu}+Y^{\beta}_{\mu\nu}Y^{\alpha}_{\alpha\beta}-Y^{\beta}_{\mu\alpha}Y^{\alpha}_{\nu\beta}\right).
\ee
Equations \eqref{Trace1} and \eqref{Trace2} are the third holographic input that we have to impose for the proof in Sec.\ \ref{3dproof}. They are the equivalent of the tracelessness condition for the holographic stress tensor in AdS, that one has to impose on top of its conservation. For the Bondi perturbation, $R_0$ and $R_1$ are given by
\begin{eqnarray}
R_0 &=& -4\partial_\phi^2\beta_0\,,\\
R_1 &=& 2(\partial_\phi\partial_u U_0+\partial_u^2\varphi)\,.
\end{eqnarray}

Finally, we can focus on the case $\beta_0=U_0=\varphi=0$, which is the space of solutions considered in Sec.\ \ref{First Law} (see \cite{Barnich:2010eb}). The two cuvature elements $R_0$ and $R_1$ vanish, therefore it corresponds to a "flat" Carrollian geometry on the boundary (we also have that $X^{\mu}_{\nu\rho}$, $Y^{\mu}_{\nu\rho}$ and $Z^{\mu}_{\nu\rho}$ vanish). Moreover, the two pieces of boundary dynamical data simplify to
\begin{eqnarray}
T_0^{\mu\nu}&=&\frac{1}{8G}\begin{pmatrix}
0 & 2\lambda N \\
2\lambda N & 1-\lambda M
\end{pmatrix}\,,\\
T_1^{\mu\nu}&=&{1\/8G}\begin{pmatrix}
1-\lambda M& 0\\
0 & 0
\end{pmatrix},
\end{eqnarray}
and their two conservation laws become
\begin{eqnarray}
\partial_u M &=& 0\, ,\\
\partial_\phi M &=& 2\partial_u N\, .
\end{eqnarray}
The solutions are given by $M=\Theta(\phi)$ and $N=\frac{u}{2}\partial_\phi \Theta+\Xi(\phi)$. One can check that with these defining functions, together with $\beta_0=U_0=\varphi=0$, the line element \eqref{BondiGauge} becomes \eqref{pert3d}:
\be
ds^2=\eta_{ab} dx^a dx^b +\la\left(\Theta(\phi)\, du^2+2\le(\Xi(\phi)+\frac{u}{2}\partial_{\phi}\Theta(\phi)\ri)du d\phi\right)+O(\lambda^2),
\ee
which is the metric perturbation we have used for exact on-shell computations.

\section{Generalization to 4d}\label{4dproof}

In this section, we give the Ryu-Takayanagi prescription in 4d that follows from the assumptions given in Sec.\ \ref{assumptions}. We find a Rindler transformation and describe the corresponding entangling regions and RT surfaces. We show that the general RT prescription depends on the choice of an infalling light sheaf, i.e. a choice of bulk light rays which intersects $\cI^+$ at the boundary $\p A$ of the entangling region. Using these RT surfaces, we show that the gravitational equations of motion are equivalent to the first law of entanglement, assuming that the constraints on the boundary stress tensor imply the vanishing of $\d E_{ua}$ at infinity. Our proof is valid for any theory of gravity, including higher-derivative terms.

\subsection{Ryu-Takayanagi prescription in 4d Minkowski}

\paragraph{Rindler transformation.} We describe a transformation which satisfies the assumptions of the generalized Rindler method. It maps the coordinates $(u,\t,\phi)$ on $\cI^+$ into the coordinates  $(\tau,\rho,\eta)$ according to
\bea\label{rindler4d}
u \= {\tau\/\r{cosh}\,\rho} \,,\-
\t \=  \r{arctan}\le(\r{sinh}\,\rho\ri)+{\pi\/2} \,,\-
\phi \= \eta\,.
\eea
This can be compared with the 3d case \eqref{rindler3d}. It is in fact a BMS$_4$ superrotation, which maps the round sphere into a conformally flat space
\be
d\t^2 +\r{sin}^2\t\,d\phi^2 = {1\/\r{cosh}^2 \rho}(d\rho^2+d\eta^2)\,.
\ee
It is a Rindler transformation because the space that we obtain has a thermal identification
\be
\rho \sim \rho+2\pi i\,.
\ee
The modular flow $\z_A$ is the generator of this thermal circle, given by 
\be
\z_A = 2\pi \p_\rho = -2\pi\le( u \,\r{cos}\,\t\,\p_u + \r{sin}\,\t\,\p_\t\ri)\,.
\ee
This vector belongs to the BMS$_4$ algebra and hence annihilates the vacuum, as required for a boundary modular flow. To obtain the bulk modular flow, we can look for a Killing of 4d Minkowski which asymptotes to $\z_A$. We obtain
\be\label{4dmodflow}
\xi_A =  -2\pi \le( u \,\r{cos}\,\t\,\p_u - (r+u) \,\r{cos}\,\t \,\p_r +{(r+u)\/r} \r{sin}\,\t\,\p_\t\ri)\,.
\ee
Note that this is much simpler than trying to find the gravitational solution which is dual to a thermal state, i.e. the flat space analog of the hyperbolic black hole, which is what we do in App. \ref{appbulkrindler}.

\begin{figure}
\begin{center}
	\begin{tabular}{cc}
		\subf{\vspace{0.5cm}\includegraphics[width=6cm]{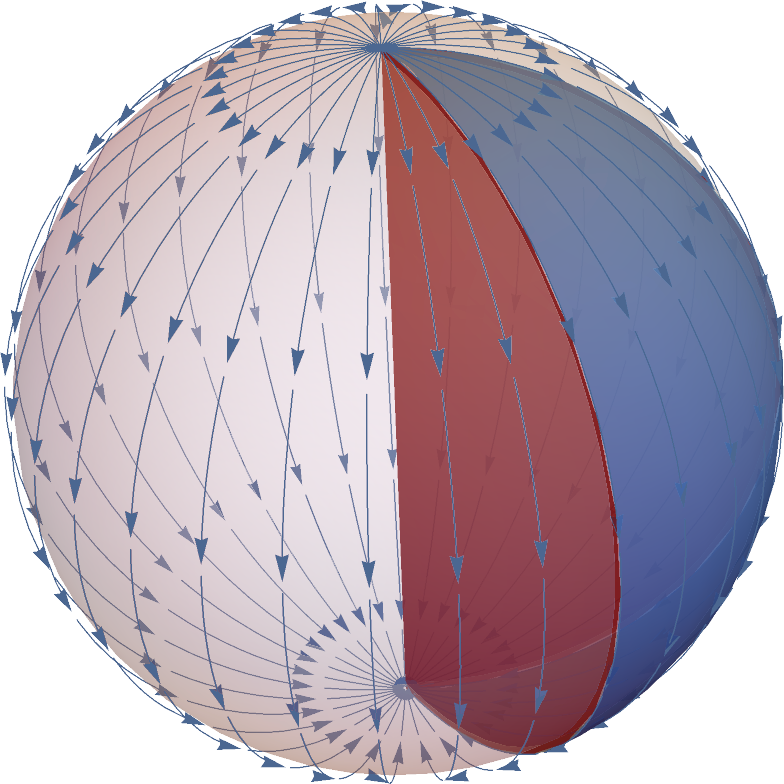}}{\vspace{0.5cm}"Watermelon slice" with $\t_0=0$}&
		\subf{ \hspace{1cm}\vspace{0.5cm}\includegraphics[width=6cm]{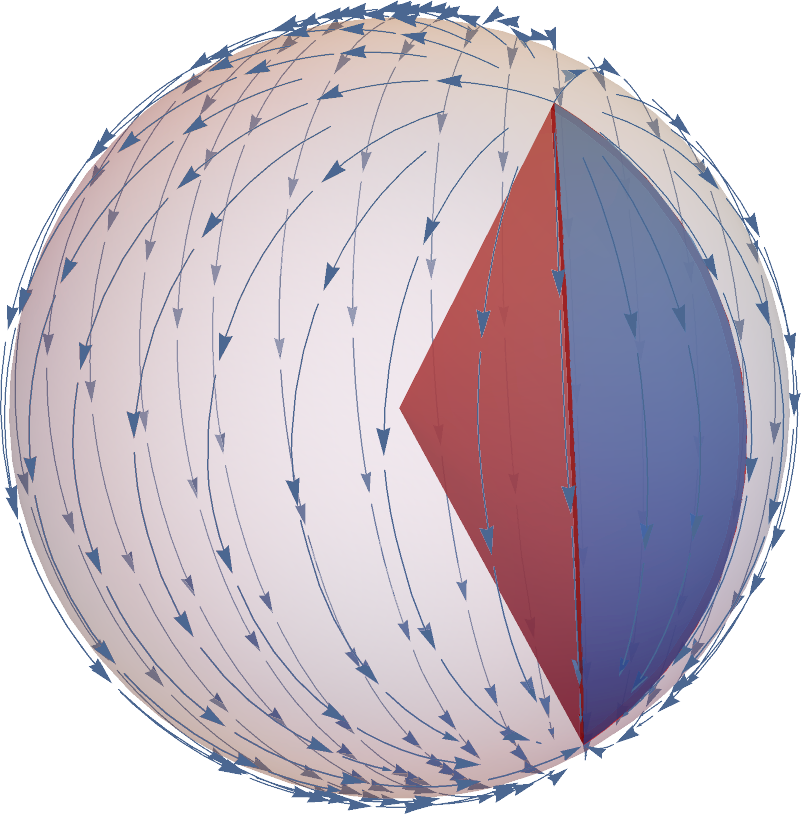}}{Deformed watermelon with $\t_0> 0$\vspace{1cm}} \\
		\subf{ \vspace{0.5cm}\includegraphics[width=6cm]{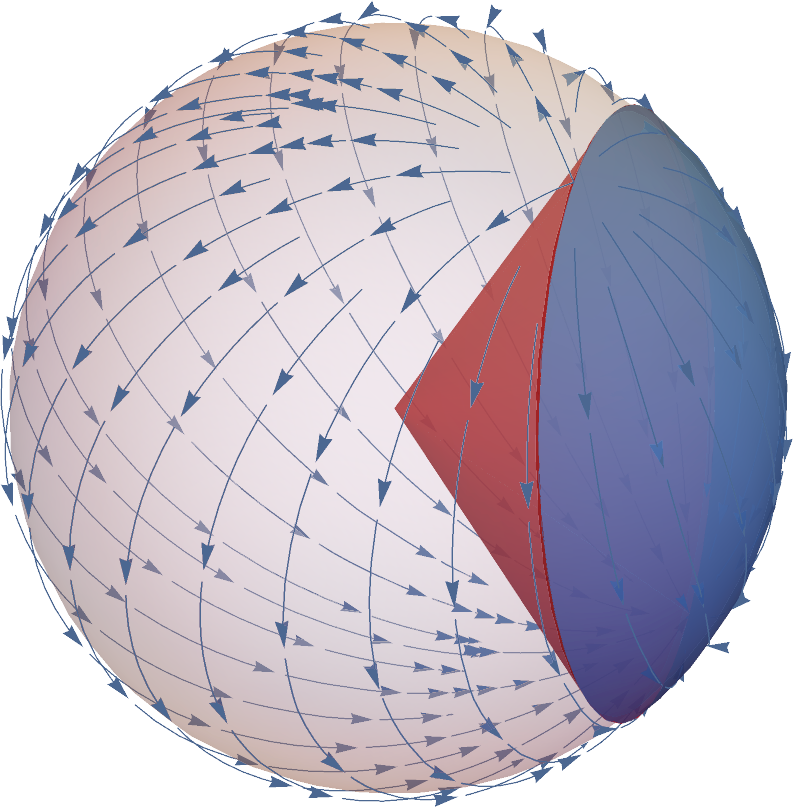}}{Disk at $\l_\phi = \pi$} &
		\subf{ \hspace{1cm}\vspace{0.5cm}\includegraphics[width=6cm]{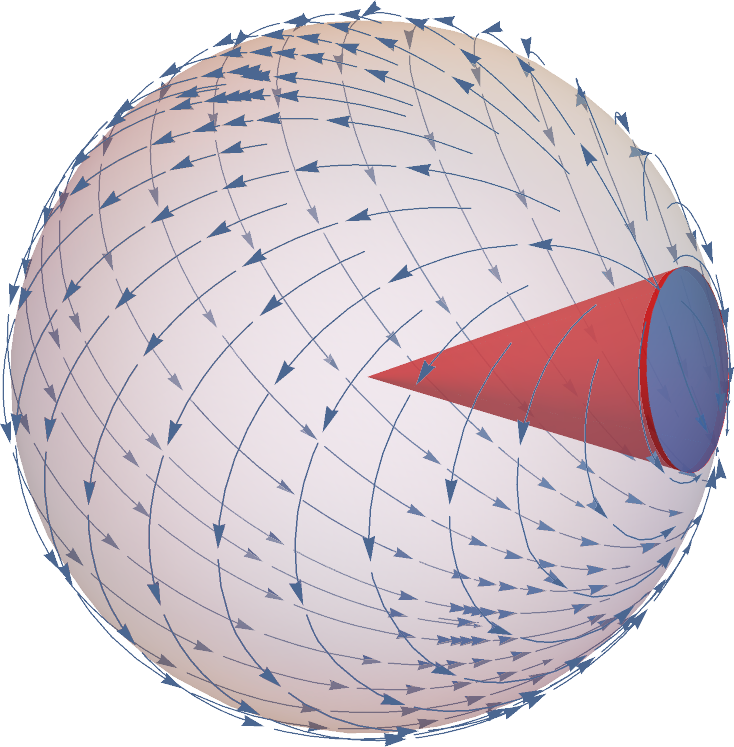}}{Smaller disk at $\l_\phi = \pi$} 
	\end{tabular}
	\put(-105, 115){{\color{red}{$\wt{A}$}}}
\put(0, 140){{\color{bluemathematica}{$A$}}}
		\put(-320, 125){{\color{red}{$\wt{A}$}}}
		\put(-211, 150){{\color{bluemathematica}{$A$}}}
	\put(-315, -123){{\color{red}{$\wt{A}$}}}
\put(-210, -80){{\color{bluemathematica}{$A$}}}
\put(-85, -120){{\color{red}{$\wt{A}$}}}
\put(0, -100){{\color{bluemathematica}{$A$}}}
\end{center}\
\caption{Examples of entangling regions (in blue) and associated RT surfaces (in red) for 4d Minkowski on the constant $u=0$ slice. They are associated to the modular flow \eqref{4dbdymodularflow} and its bulk extension \eqref{4dbulkmodularflow}.}\label{4dFig}
\end{figure}
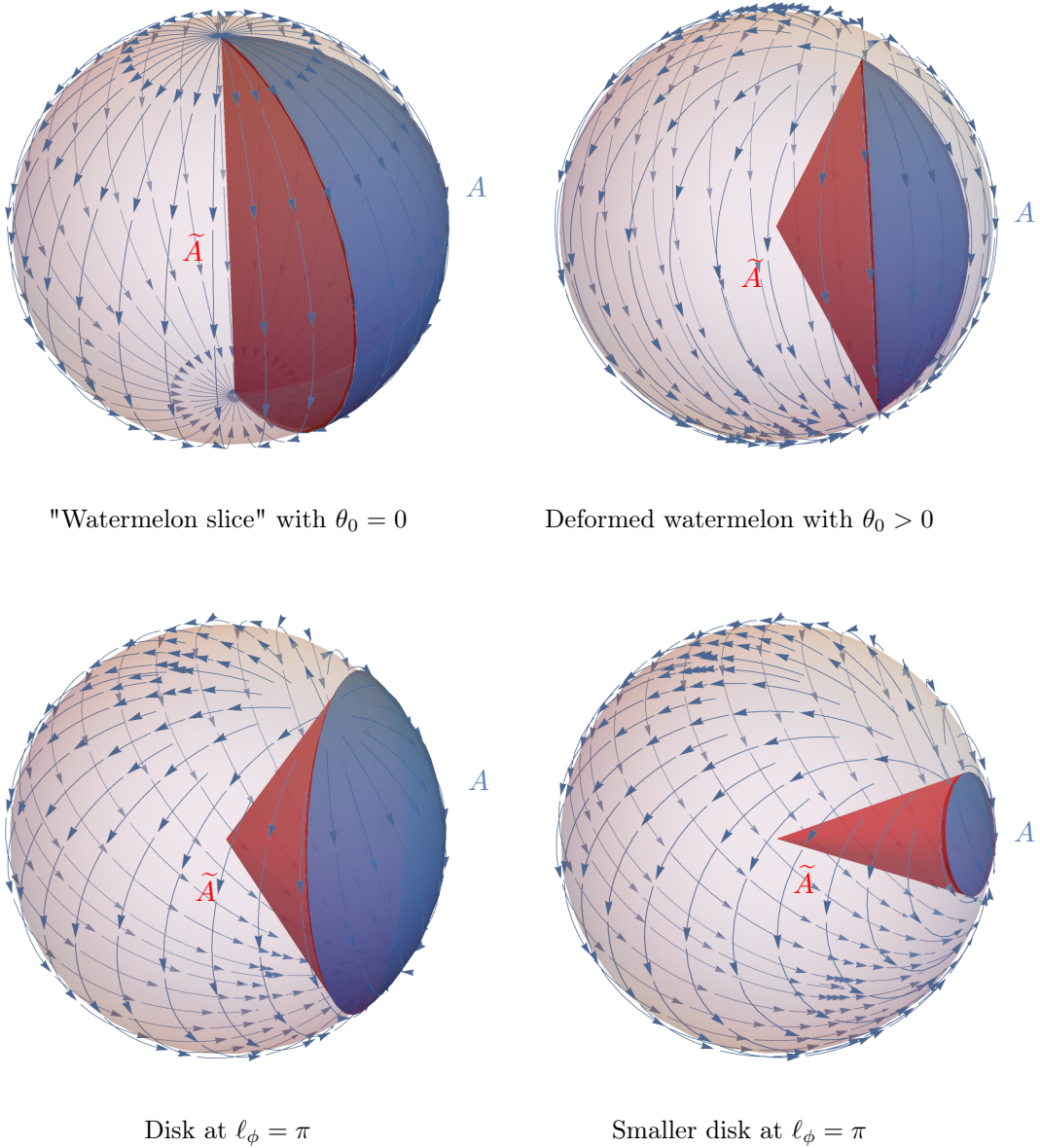

\paragraph{Watermelons.} We focus on entangling regions that lie on the slice $u=0$ as other configurations can be obtained by acting with bulk isometries. The entangling regions $A$ are given by patches on the sphere at infinity that are invariant under the flow. They are "watermelon slices" whose boundaries follow the flow and with width $\l_\phi$. They can be parametrized as
\be
-\lph\leq \phi \leq \lph, \qq 0\leq \t\leq \pi\,,
\ee
and are represented in Fig.\ \ref{4dFig}. The domain of dependence $\cD$ and its boundary $\p \cD$ can be checked to be invariant under the flow.\footnote{The boundary $\p A$ is not fixed pointwise by the flow, which is different from the  AdS case or in 3d Minkowski. This is inevitable for 4d Minkowski because there is no conformal Killing on the sphere which admits a one-dimensional set of fixed points \cite{2011JGP....61..589B}.}

\paragraph{Generalized Rindler transformations.}

When the sphere is written in complex coordinates
\be
z = e^{i\phi} \r{cot}(\tfrac\t2),\qq \bar{z} = e^{-i\phi} \r{cot}(\tfrac\t2)\,,
\ee
we observe that the Rindler transformation \eqref{rindler4d} can be written as
\be\label{Rindlerprev}
z\to e^{-w},\qq \bar{z}\to e^{-\bar{w}},
\ee
where $w= \rho - i \eta,\,\bar{w} = \rho+i \eta$. This suggests a way to obtain more general Rindler transformations, obtained by acting with a Möbius transformation on the sphere. Let's consider the following transformation
\bea\label{newRindler}
u&\ra& {\r{cos}\,\t_0\/\r{cosh}\,\rho\,+ \r{cos}\,\eta\,\r{sin}\,\t_0}\tau~,\-
z &\ra& {\r{sin}\,\t_0 + e^{w} (1+\r{cos}\,\t_0)\/\r{sin}\,\t_0 \,e^{w}+  (1+\r{cos}\,\t_0)},\-
\bar{z} &\ra& {\r{sin}\,\t_0 + e^{\bar{w}} (1+\r{cos}\,\t_0)\/\r{sin}\,\t_0 \,e^{\bar{w}}+  (1+\r{cos}\,\t_0)}\,,
\eea
which is a BMS$_4$ transformation. The boundary modular flow is the vector $2\pi\p_\rho$ given by
\be\label{4dbdymodularflow}
\z_A = -{2\pi\/\r{cos}\,\t_0} \le( u\,\r{cos}\,\t\,\p_u  +k\ri)\,,
\ee
where $k$ is a conformal Killing of the sphere given by
\be\label{confKilling}
k = (\r{sin}\,\t - \r{sin}\,\t_0\,\r{cos}\,\phi)\, \p_\t +\r{sin}\,\t_0 \,\r{cot}\,\t\,\r{sin}\,\phi\,\p_\phi ~.
\ee 
The bulk modular flow is
\be\label{4dbulkmodularflow}
\xi_A = {2\pi\/\r{cos}\,\t_0} \le( (u+r)\,\r{cos}\,\t\,\p_r - {u\/r} \r{sin}\,\t\,\p_\t \ri) + \z_A\,.
\ee
It is obtained as the Killing vector of 4d Minkowski spacetime which matches with $\z_A$ on the boundary. The transformation described in \eqref{newRindler} has also the thermal identification $\rho\sim \rho+2\pi i$. It is a one-parameter generalization of the previous Rindler transformation \eqref{Rindlerprev}, obtained by considering a more general conformal Killing $k$ of the sphere. 

\paragraph{Generalized watermelons.}

To understand the entangling regions associated to this modular flow, we should look at regions on $S^2$ that are preserved under $k$. There are two fixed points given by
\be
P_-: (\t,\phi) = (\t_0,0),\quad P_+: (\t,\phi) = (\pi-\t_0,0)\,.
\ee
The vector field $k$ is a flow from $P_-$ to $P_+$. The entangling regions are deformed "watermelons slices" whose boundaries are tangent to this flow, as depicted in Fig.\ \ref{4dFig}. The domain of dependence $\cD$ and its boundary $\p \cD$ can be checked to be invariant under the flow. An entangling region $A$ can be parametrized by
\be\label{4dAparam}
- \l(\t)\leq \phi \leq \l(\t), \qq \t_0\leq \t\leq \pi-\t_0\,,
\ee
where $\l(\t)$ satisfies the condition
\be
\l'(\t) = \le. ({k^\phi/k^\t})\ri|_{\phi= \l(\t)}  \,,
\ee
which ensures that the boundary $\p A$ is tangent to the vector field $k$. This makes sure that $A$ and $\p A$ are preserved under the modular flow. Explicitly, we obtain
\be\label{defltheta}
\r{tan}\,\l(\t) = {\r{cos}(2\t_0)-\r{cos}(2\t)\/ 2\,\r{sin}\,\t\,\le(\r{cot}(\lph)\,\r{sin}\,\t+\r{sin}\,\t_0\,\sqrt{1+\r{cot}^2(\lph) - {\r{sin}^2\t_0\/\r{sin}^2\t}} \ri)}\,,
\ee
where $\l_\phi$ parametrizes the width of the entangling region. For $\t_0 = 0$, we have $\l(\t)=\l_\phi/2$. At small $\l_\phi$, we have
\be\label{lthetasmall}
\l(\t) = \le( 1 - {\r{sin}\,\t_0\/\r{sin}\,\t}\ri) \,{\l_\phi\/2} + O(\l_\phi^2)\,.
\ee
At the special value $\l_\phi = {\pi\ov 2}$, the watermelon becomes a disk on the sphere. This is illustrated in Fig.\ \ref{4dFig}. The opening angle of the disk is $\pi-2\t_0$.

\paragraph{Ryu-Takayanagi surfaces.} The entangling regions described above are the generalization of the 3d story with $\l_u = 0$. The bulk modular flow \eqref{4dbulkmodularflow} is very similar to the bulk modular flow in three dimensions \eqref{3dbulkmodularflow}. The RT surfaces associated to the above regions are easy to describe, they lie on the slice $u=0$ and are the union of all light rays starting at the origin and ending on $\p A$. We illustrate this prescription in Fig.\ \ref{4dFig} by representing the sphere at infinity on the slice $u=0$. The entangling regions $A$ are in blue and the RT surfaces $\wt{A}$ are in red. We also represent the boundary modular flow on the sphere.  The entanglement entropy of the region $A$ is then given by
\be
S_A = \int_{\wt{A}} {\rm\textbf{Q}}[\xi_A]~.
\ee
For Einstein gravity in the Minkowski vacuum, the areas of all these RT surfaces vanish because they have a null tangent vector everywhere. 

\paragraph{Perturbations.} As an illustration, we can consider on-shell perturbations of 4d Minkowski in the Bondi gauge. The flat metric is given by
\be
ds^2 = - du^2 - 2 du dr + r^2 \g_{ij}dx^i dx^j,\qq \g_{ij}dx^i dx^j = d\t^2 + \r{sin}^2\t\,d\phi^2\,,
\ee
we consider the linearized on-shell perturbations studied in \cite{Campoleoni:2017qot} with $C_{ij}=0$, which corresponds to setting the gravitational wave aspect to zero. Asymptotically, the perturbation reads
\be
h_{uu} = {2\/r} \cM(x^i)+ O(r^{-2}),\qq h_{ui} = {1\/r}\cN_i(x^i) + O(r^{-2}),\qq h_{ij} = O(1)\,.
\label{pert4dh}
\ee
The subleading pieces in $r$ should not contribute to the charges at infinity. This allows us to compute $\d E_A$ in a similar way as in the previous section. We obtain on a slice $u=0$
\be
\d E _A= {3\/8 \,\r{cos}\,\t_0} \int_A d\t d\phi\le[ (\r{cos}\,\phi\,\r{sin}\,\t_0 - \r{sin}\,\t) \cN_\t(\t,\phi) - \r{cot}\,\t\,\r{sin}\,\phi\,\r{sin}\,\t_0\,\cN_\phi(\t,\phi) \ri],
\ee
which can be written in terms of the boundary modular flow \eqref{4dbdymodularflow} as
\be
\d E_A={3\/16\pi} \int_A d\t d\phi\,\z_A^i\cN_i(\t,\phi)\,.
\label{dEA4d}
\ee 
Exactly as in the 3d case, the entropy has to be computed using the refined prescription \eqref{refinedpresc} where we regulate the corner of the RT surface. The fact that $\d E_A = S_A$, which has to be positive, gives some constraints on the perturbations that can be described by a quantum system on $\cI^+$ satisfying our assumptions, similar to the discussion in Sec.\ \ref{constraints}. These constraints impose the functions $\cN_i$ in the perturbation to be such that \eqref{dEA4d} is positive for a given region $A$. This selects a subspace $\cH_\text{code}$ on which $K_A$ is bounded from below and this makes the density operator $e^{-K_A}$ is well-defined.

\subsection{General 4d prescription}\label{4dgenpr}

In this section, we discuss the general RT prescription in 4d, in the same spirit as the 3d discussion of Sec.\ \ref{3dgenpresc}. Given a boundary entangling region, we will describe the most general choice of light sheaf that satisfies the requirements to give a good RT configuration. That is, the light sheaf must connect $\p A$ to the Rindler bifurcation surface and the modular flow must be tangent to it. As explained in the 3d case, the first condition ensures that we can define an RT surface (as a portion of the Rindler bifurcation surface) and the second condition is required to have a well-defined first law.

\paragraph{Modular flow for non-zero $\l_u$.}

In Cartesian coordinates $(t,x,y,z)$, the bulk modular flow given \eqref{4dmodflow} takes the following form
\be
\xi_A = {2\pi\/\r{cos}\,\t_0} \le[ z\,\p_t +z \,\r{sin}\,\t_0\,\p_x+\le( t-x\,\r{sin}\,\t_0\ri)\p_z\ri]~.
\ee
We note that this it is similar to the 3d bulk modular flow at $\l_u = 0$. This suggests the following generalization for $\l_u\neq 0$ in 4d, obtained by performing a bulk  translation 
\be
z\ra z + {\l_u\/2\,\r{cos}\,\t_0}~,
\ee
which leads to
\be\label{4dbulkmodflow}
\xi_A = {2\pi\/\r{cos}\,\t_0} \le[ \le( z +{\l_u\/2\,\r{cos}\,\t_0}\ri)\p_t +\le( z \,\r{sin}\,\t_0\,+{\l_u \,\r{tan}\,\t_0\/2}\ri)\p_x+\le( t-x\,\r{sin}\,\t_0\ri)\p_z\ri]~.
\ee
Going back to Bondi coordinates $(u,r,\t,\phi)$ and taking the limit $r\to+\infty$, we obtain the corresponding 4d boundary modular flow, which reads
\bea
\z_A\= {2\pi\/\r{cos}\,\t_0} \le[ \le(- u\,\r{cos}\,\t+ {\l_u\/2\,\r{cos}\,\t_0} \le( 1 - \r{sin}\,\t_0\, \r{sin}\,\t\,\r{cos}\,\phi \ri)\ri) \p_u\ri. \-
&&\hspace{1.5cm} \le. \vphantom{\l_u\/2\,\r{cos}\,\t_0} + \le(\r{sin}\,\t_0\,\r{cos}\,\phi - \r{sin}\,\t \ri)\p_\t  + \r{sin}\,\t_0\,\r{cot}\,\t\,\r{sin}\,\phi\,\p_\phi \ri]~.
\eea
One can check that this modular flow follows from a generalized Rindler transform, which is the previous Rindler transform \eqref{newRindler} with a different transformation for $u$
\bea\label{4dRindlunonzero}
u&\ra& {\r{cos}\,\t_0\/\r{cosh}\,\rho\,+ \r{cos}\,\eta\,\r{sin}\,\t_0}\le(\tau +{\l_u\/2\,\r{cos}\,\t_0}\r{sinh}\,\rho \ri), \-
z &\ra& {\r{sin}\,\t_0 + e^{w} (1+\r{cos}\,\t_0)\/\r{sin}\,\t_0 \,e^{w}+  (1+\r{cos}\,\t_0)},\-
\bar{z} &\ra& {\r{sin}\,\t_0 + e^{\bar{w}} (1+\r{cos}\,\t_0)\/\r{sin}\,\t_0 \,e^{\bar{w}}+  (1+\r{cos}\,\t_0)}\,,
\eea
and which remains a BMS$_4$ transformation. The generator of the thermal circle $2\pi\p_\rho$ reproduces the boundary modular flow given above. This was guaranteed to work because, as in 3d, the case $\l_u\neq 0 $ is simply the image of the case $\l_u = 0$ by a bulk translation, which becomes on the boundary
\be
u\ra u +{\l_u\/2\,\r{cos}\,\t_0} \r{cos}\,\t~.
\ee 
On the boundary, this bulk translation changes the shape of the region $A$ which is the same as before but with an extension in $u$:
\be\label{4dAu}
u  ={\l_u\/2\,\r{cos}\,\t_0} \r{cos}\,\t ~, \qq \theta\in [\theta_0, \pi-\theta_0]~.
\ee
Similarly to 3d, the bulk modular flow \eqref{4dbulkmodflow} is simply a boost. This can be seen explicitly by defining new coordinates
\be\label{newcoord4d}
\tilde{t}  = {1\/\r{cos}\,\t_0} t -\r{tan}\,\t_0\,x,\qq \tilde{x}  = {1\/\r{cos}\,\t_0}\,x-\r{tan}\,\t_0\,t  ,\qq \tilde{z} = z+{\l_u\/2\,\r{cos}\,\t_0}~,
\ee
in which the modular flows is given by
\be
\xi_A = 2\pi \le( \tilde{z} \,\p_{\tilde{t}} + \tilde{t} \,\p_{\tilde{z}}\ri)~.
\ee
In App.\ \ref{appbulkrindler} we show that, exactly like in the 3d case, there exists a change of coordinates in the bulk defined on the exterior of a Rindler horizon that maps to the transformation \eqref{4dRindlunonzero} on the boundary.

\paragraph{RT prescription.} In 4d, the prescription where we impose that the light rays pass through the origin $r=0$  is inconsistent in the case $\l_u\neq 0$ because most light rays won't have an intersection with the bifurcation surface. Instead, we should consider the most general light sheaf which satisfies the requirements necessary for a good RT configuration, as was done in Sec.\ \ref{3dgenpresc} for the 3d case. We will take all these choices of light sheaf to be equally physical, reflecting a choice of UV cutoff in the putative dual theory.

\begin{figure}
	\centering
	\includegraphics[width=10cm]{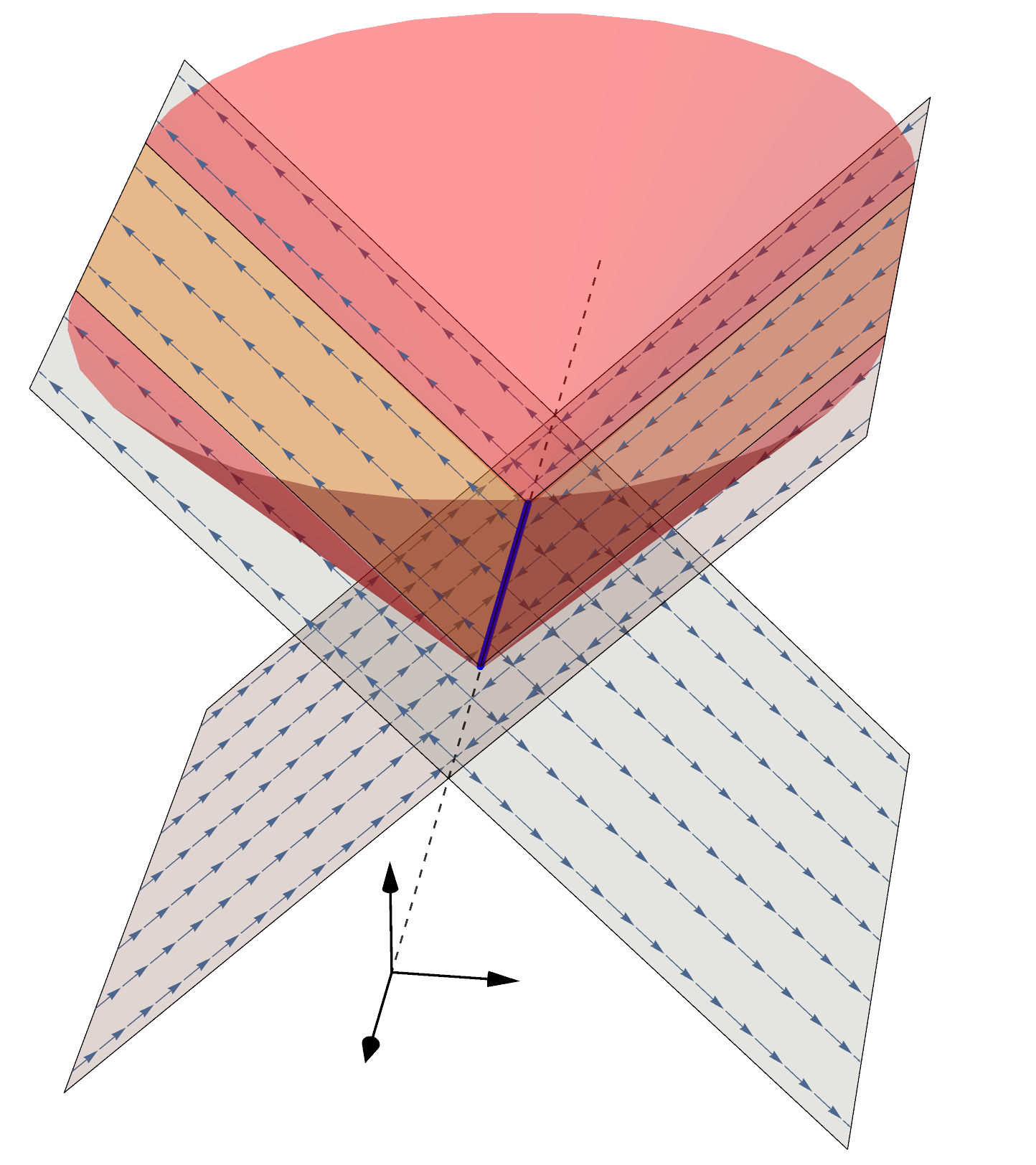}
		\put(-198,18){$(\tx,\ty)$}
		\put(-188,72){$\tit$}
		\put(-138,47){$\tz$}
		\put(-140, 325){{\color{red}{$\g_+$}}}
\put(-290, 225){{\color{red}{$\g_-$}}}
\put(-139, 160){{\color{blue}{$\g$}}}
\put(-275, 260){{\color{orange}{$\g_S$}}}
\put(-33, 246){{\color{orange}{$\g_N$}}}
	\caption{Ryu-Takayanagi configuration in coordinates $(\tit,\tx,\ty,\tz)$ in which the bulk modular flow is a boost. The RT surface $\wt{A}$ is given by the union of the light sheaf $\g_-\cup\g_N\cup \g_S \cup\g_+$ with the surface $\g$ on the Rindler bifurcation surface $(\tx,\ty)$. See Fig.\ \ref{fig:4dxtyt} for an illustration of $\g$ in the $(\tx,\ty)$-plane. The modular flow is tangent to the light sheafs $\g_N,\g_S$ because they are portions of the Rindler horizons and to $\g_-,\g_+$\ because they are half-cones whose transverse sections are hyperbolas which are tangent to the boost. }\label{4drindler}
\end{figure}

The boundary of $A$ on $\cI^+$ has two pieces $\p A= B_+\cup B_-$ which can be parametrized as
\begin{align}
B_+ &: \quad  \phi = \l(\t), \qq & \t_0  \leq\t\leq\pi-\t_0 ~,\- 
B_- &: \quad  \phi = -\l(\t),  \qq & \t_0 \leq\t\leq\pi-\t_0  ~.
\end{align}
where $\l(\t)$ is defined in \eqref{defltheta}, while their extension in the $u-$direction is given by \eqref{4dAu}. The most general light rays that arrive at a point $(\t,\phi) = (\t,\pm\l(\t))$ on $\cI^+$ can be parametrized as follows in Cartesian coordinates
\be\label{4dgamma}
\g_+(\t) : \begin{cases} t=s + T_+ (\t)\\
x =  s\,\r{sin}\,\t\,\r{cos}\,\l(\t) + X_+(\t)\\ 
y = s\,\r{sin}\,\t\,\r{sin}\,\l(\t) + Y_+(\t)\\
z = s\,\r{cos}\,\t + Z_+(\t) \end{cases},\qq \g_-(\t) : \begin{cases} t=s +T_- (\t)\\
x =  s\,\r{sin}\,\t\,\r{cos}\,\l(\t) + X_-(\t)\\ 
y = s\,\r{sin}\,\t\,\r{sin}\,\l(\t) + Y_-(\t)\\
z = s\,\r{cos}\,\t + Z_-(\t) \end{cases}~,
\ee
and the arbitrary functions $T_\pm(\t),X_\pm(\t),Y_\pm(\t),Z_\pm(\t)$ reflect the ambiguity in choosing these light rays. This ambiguity will be partially fixed by imposing the necessary requirements. Firstly, the light rays $\g_+(\t)$ and $\g_-(\t)$ should intersect $\cI^+$ at $\p A$, so that the value of $u$ at infinity is given by \eqref{4dAu}.  Then we should impose that all these light rays intersect the bifurcation surface of the Rindler horizon associated with the bulk modular flow, i.e. $\tilde{t}=\tilde{z}=0$. To do this, we impose that after transforming \eqref{4dgamma} to the new Cartesian coordinates \eqref{newcoord4d}, $\tz$ and $\tit$ become proportional. This also imposes the relation
\be
\tz = f(\t) \,\tit,\qq f(\t) = {\r{cos}\,\t\,\r{cos}\,\t_0\/ 1 - \r{cos}\,\l(\t)\,\r{sin}\,\t\,\r{tan}\,\t_0}~.
\ee
Denoting the two light sheafs
\bea
\g_+ \= \{\g_+(\t) \mid \t\in[\t_0,\pi-\t_0] \} ~,\-
\g_- \= \{\g_-(\t) \mid \t\in[\t_0,\pi-\t_0] \}~,
\eea
we see that $\g_+$ and $\g_-$ span over the quadrant $\tit\geq |\tz|$ because the function $f(\t)$ is a bijection between the interval $[\t_0, \pi-\t_0]$ and the interval $[-1,1]$.  To find the region $\g$, which is a 2d surface in 4d, we should consider the intersection of $\g_\pm$ with the bifurcation surface, which is the plane $(\tx,\ty)$. From the explicit parametrization, we find that the intersection of $\g_\pm $ with this plane is restricted to the lines
\be\label{lines}
\tx \pm \r{cos}\,\t_0\,\r{tan}(\lph)\, \ty = 0~.
\ee
Lastly, we should impose that the modular flow is tangent to the light sheaf $ \g_+\cup \g_-$ which is required to have a well-defined first law. This is necessary because we need $\xi_A\cdot \Theta$ to vanish when integrated on the light sheaf, see the paragraph below for more details. To do this, we consider the two tangent vectors
\be
{\p x^\mu\/\p\t} \p_\mu,\qq {\p x^\mu\/\p s} \p_\mu~,
\ee
and we require that the modular flow $\xi_A$ can be written as a linear combination of those. For the light sheaf $\g_+$, we find that this is only possible if the light sheaf $\g_+$ intersects the bifurcation surface at a single point $P_+$. That is, we need all the light rays in $\g_+$ to converge to the same point $P_+$ on the bifurcation surface. We have a similar condition on $\g_-$ which should intersect the bifurcation surface at a single point $P_-$. These points cannot be arbitrary in the plane $(\tx,\ty)$ since they have to belong to the lines given in \eqref{lines}. Importantly, $P_+$ and $P_-$ don't have to be the same. Enforcing all these constraints, we are able to fix the functions $T_\pm(\t),X_\pm(\t),Y_\pm(\t),Z_\pm(\t)$ and we can write the following simpler parametrization for the light sheafs
\bea
\g_\pm(\t) : \begin{cases} t=\mp \tilde{y}_\pm \,\r{sin}\,\t_0\,\r{tan}(\lph) + s ,\\
x=  \mp \tilde{y}_\pm\,\r{tan}(\lph) + s\,\r{sin}\,\t\,\r{cos}\,\l(\t)\\
y = \tilde{y}_\pm \pm s\,\r{sin}\,\t\,\r{sin}\,\l(\t)\\
z = -{\l_u\/2\,\r{cos}\,\t_0} + s\,\r{cos}\,\t  
\end{cases}.
\eea
In this parametrization, the light sheafs $\g_\pm$ intersect the bifurcation surface $\tit=\tz=0$ at $P_+$ and $P_-$ whose coordinates are given by
\be
P_\pm:\quad \tit=\tz= 0,\qq \ty =\ty_\pm,\qq \tx = \mp \r{cos}\,\t_0\,\r{tan}(\lph) \ty_\pm~.
\ee
The simplest choice is to take $P_- =P_+$. The RT configuration that we obtain is the one described in the previous section (up to a bulk isometry) and the RT surface has a conical shape. We can also have configurations where $P_-$ and $P_+$ are separated. In this case, we should add additional light rays to close the light sheaf. To do this, we define new light sheafs $\g_N$ and $\g_S$ consisting of light rays that go from the two poles of $\p A$ given by
\be
N :(\t,\phi) = (\t_0,0),\qq S:(\t,\phi) = (\pi-\t_0,0)~,
\ee
and intersect the bifurcation surface. It turns out that it is possible to make such a light ray intersect an arbitrary point on the bifurcation surface. For example, a parametrization of $\g_N$ and $\g_S$ can be given as
\bea \label{gammaN}
\g_N(v) : \begin{cases} \tit=  s\,\r{cos}\,\t_0  ,\\
\tx=  X_N(v) \\
\ty = Y_N(v)\\
\tz =  s\,\r{cos}\,\t_0  
\end{cases},\qq \g_S(v) : \begin{cases} \tit=  s \,\r{cos}\,\t_0 ,\\
\tx=  X_S(v) \\
\ty = Y_S(v)\\
\tz =  -s\,\r{cos}\,\t_0  ~,
\end{cases}
\eea
where $v$ parametrizes the different light rays in the light sheafs $\g_N$ and $\g_S$. These light sheafs satisfy our requirements: they intersect $\cI_+$ at the two poles $N$ and $S$ (with the required value of $u$) and the bulk modular flow is tangent to them. The intersection of $\g_{N}$ with the bifurcation surface is at $s=0$ and gives a curve $C_N: (\tx,\ty) = (X_N(v),Y_N(v))$ parametrized by $v$. Similarly, $\g_S$ intersects the bifurcation surface at the curve $C_S: (\tx,\ty) = (X_S(v),Y_S(v))$. Both of those curves must connect $P_+$ to $P_-$. The total light sheaf is given by $\g_+\cup \g_N\cup \g_S \cup \g_-$. This configuration is illustrated in Fig.\ \ref{4drindler}.

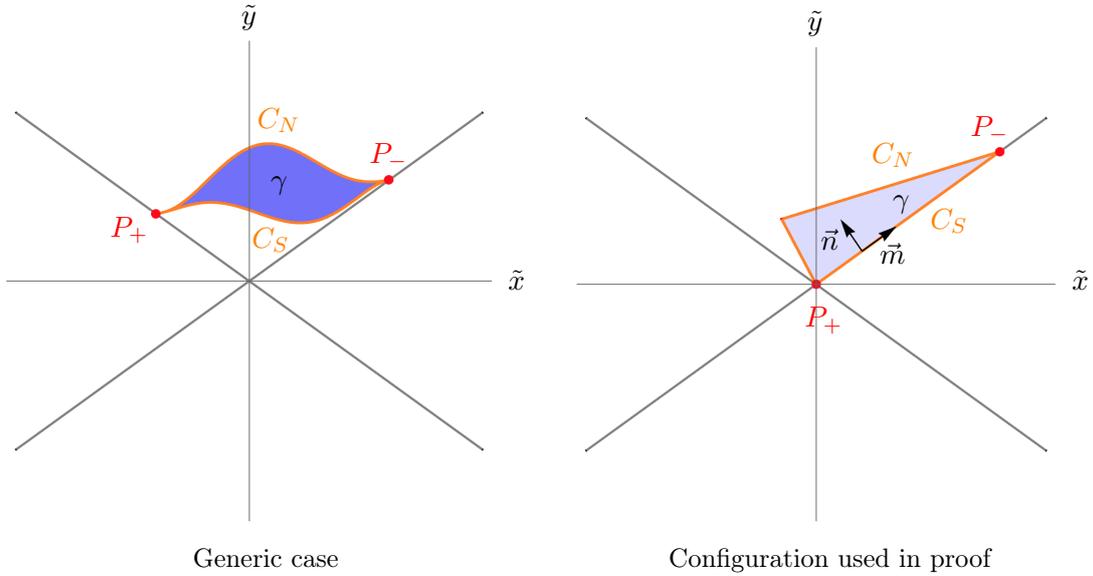
\begin{figure}
\centering
	\begin{tabular}{cc}
		\subf{\includegraphics[width=7cm]{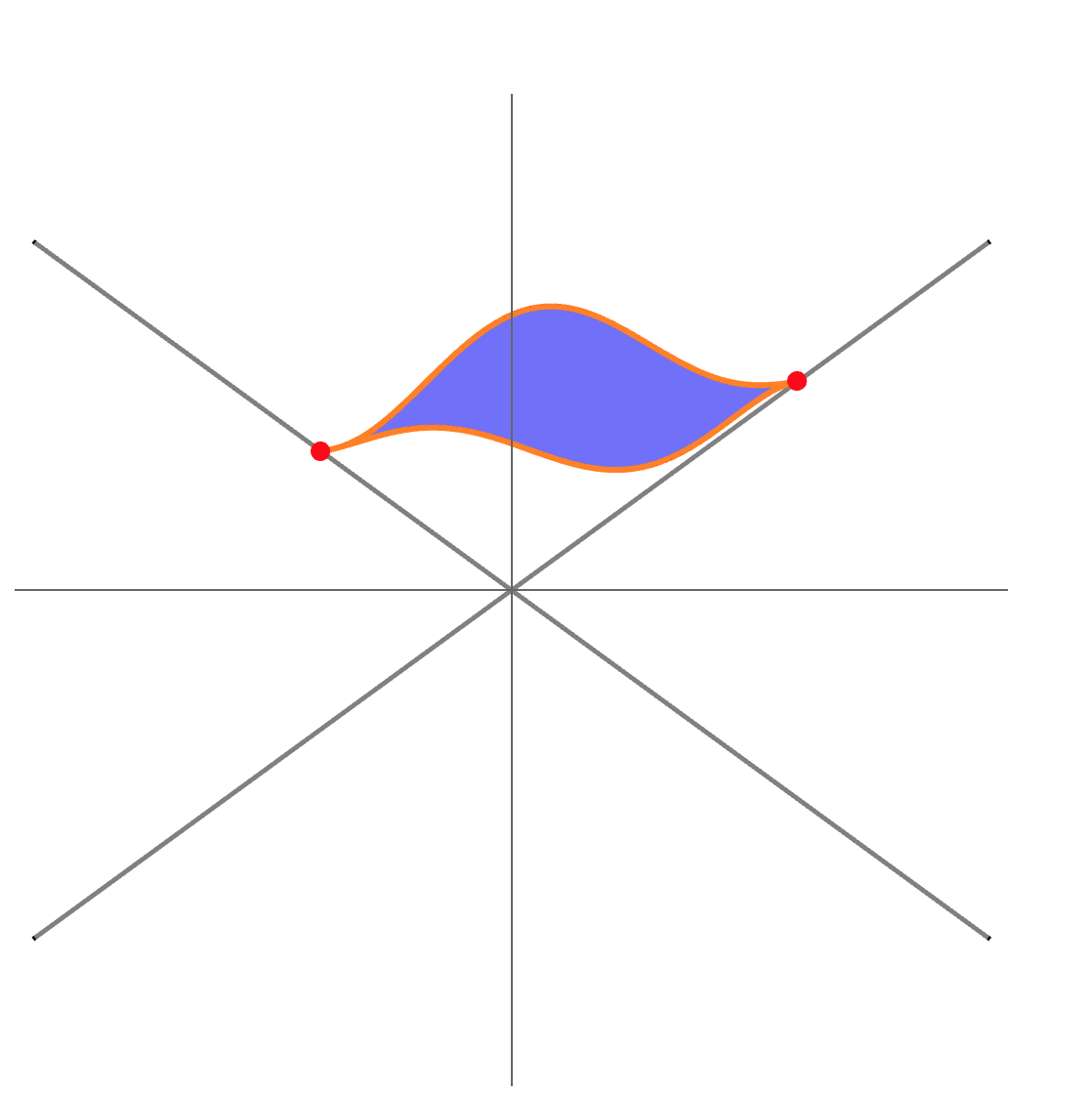}}{Generic case} &
		\subf{\includegraphics[width=7cm]{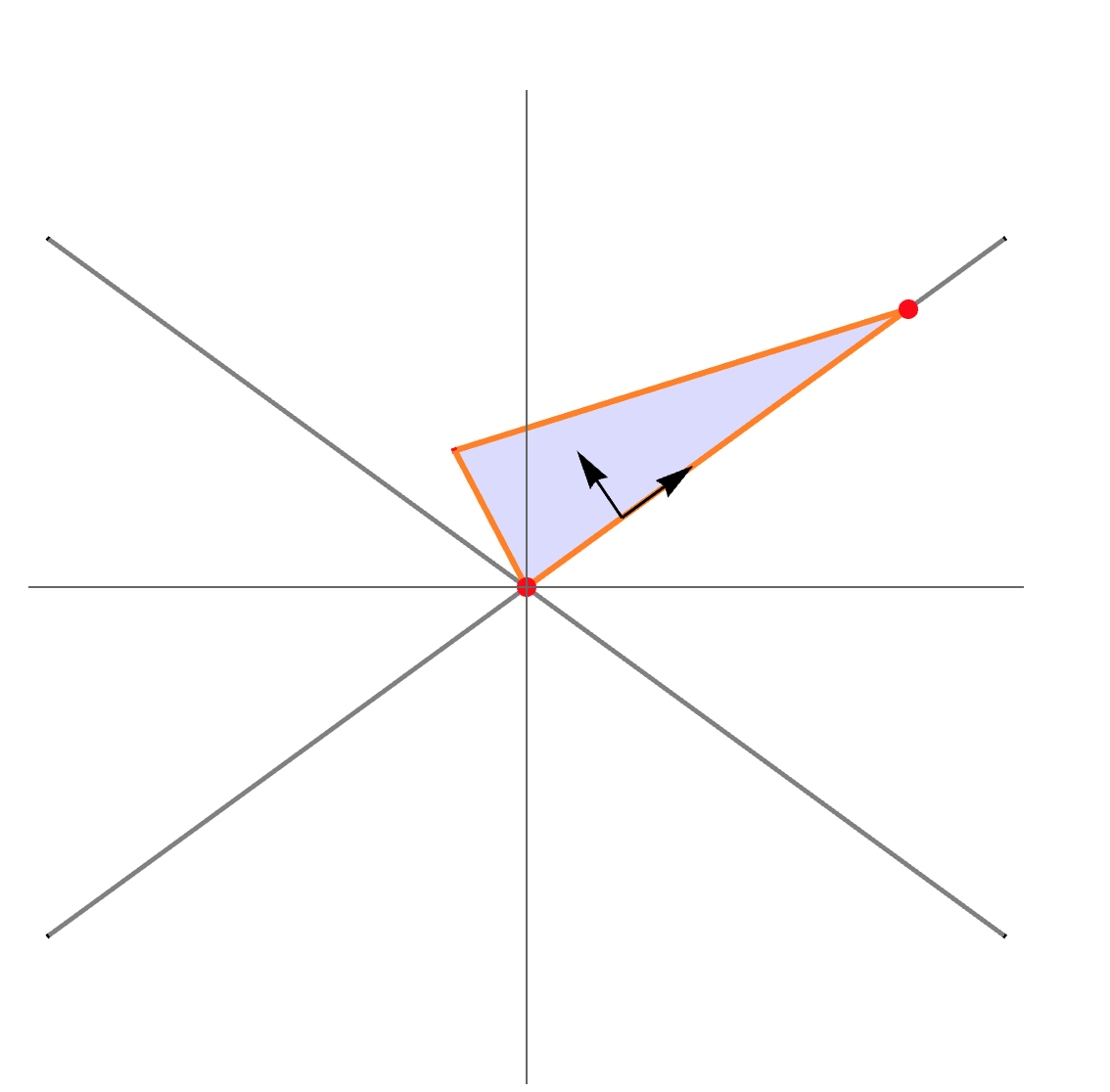}}{Configuration used in proof}
			\end{tabular}
\put(-375, 22){{\color{red}{$P_+$}}}
\put(-278, 50){{\color{red}{$P_-$}}}
\put(-320, 63){{\color{orange}{$C_N$}}}
\put(-322, 18){{\color{orange}{$C_S$}}}
\put(-315, 40){{\color{black}{$\g$}}}
\put(-115, -12){{\color{red}{$P_+$}}}
\put(-53, 60){{\color{red}{$P_-$}}}
\put(-90, 50){{\color{orange}{$C_N$}}}
\put(-68, 25){{\color{orange}{$C_S$}}}
\put(-82, 33){{\color{black}{$\g$}}}
\put(-109, 17){{\color{black}{$\vec{n}$}}}
\put(-87, 12){{\color{black}{$\vec{m}$}}}
\put(-15, 2){{\color{black}{$\tilde{x}$}}}
\put(-114, 100){{\color{black}{$\tilde{y}$}}}
\put(-226, 2){{\color{black}{$\tilde{x}$}}}
\put(-326, 102){{\color{black}{$\tilde{y}$}}}
\caption{Ryu-Takayanagi configuration in the Rindler bifurcation surface $(\tx,\ty)$. This surface intersects the light sheafs $\g_+$ and $\g_-$ at the points $P_+$ and $P_-$, which are restricted to lie on the lines \eqref{lines} (in gray). The light sheafs $\g_N$ and $\g_S$ intersect the bifurcation surface at the curves $C_N$ and $C_S$ (in orange). These light sheafs are represented in Fig.\ \ref{4drindler}. }\label{fig:4dxtyt}{}
\end{figure}

The surface $\g$ is the portion of the bifurcation surface which is in the interior of the contour formed by $C_N$ and $C_S$. It is depicted in the plane $(\tx,\ty)$ in Fig.\ \ref{fig:4dxtyt}. The RT surface $\wt{A}$ is the union of the total light sheaf with $\g$. The entanglement entropy of $A$ is given by
\be
S_A = \int_{\wt{A}} {\rm\textbf{Q}}[\xi_A]~.
\ee
In Einstein gravity, the integration of Wald's functional on the light sheaf vanishes so the entanglement entropy of $A$ is given by the area of the region $\g$
\be
S_A =  {\r{Area}(\g)\/4G}~.
\ee
The possible regions $\g$ can be obtained by the following procedure: put two points $P_\pm$ on the two lines \eqref{lines} (depicted in grey in Fig.\ \ref{fig:4dxtyt}). Then, connect them by two arbitrary curves $C_N$ and $C_S$ so that their union has a well-defined interior. This interior is the region $\g$ and the entropy is given by the area of $\g$ (in Einstein gravity). We see that as in 3d, the entropy is sensitive to the choice of light sheaf, which should reflect a choice of UV cutoff in the putative dual field theory. 

\paragraph{First law of entanglement.} We have the following definitions
\be
\d S_A = \int_{\wt{A}} \d {\rm\textbf{Q}}[\xi_A], \qq \d E_A = \int_A {\bm\chi}~,
\ee
the first law states that these two expressions are equal on-shell. The 3d derivation of Sec.\ \ref{First Law} can be carried out in 4d.  In this derivation, the first law follows from the fact that
\be
\int_{\wt{A}} \xi_A\cdot {\bm\Theta}[\d g] = 0~,
\ee
which holds whenever $\xi_A$ is tangent to $\wt{A}$. This is the case here since $\xi_A$ vanishes on $\g$ and is tangent to the light sheaf (this was one of our requirements). As a result, all the RT surfaces described here  satisfy a first law for perturbations.

\subsection{Linearized gravitational equations}

In this section, we prove that the four-dimensional linearized gravitational equations follow from the first law of entanglement. The proof is very similar to the three-dimensional case described in Sec.\ \ref{3dproof}, to which we refer for more details.

\paragraph{Reference configuration.} We consider a watermelon $A$ at $u=0$ with $\ell_u=0$. The first law of entanglement gives the equation
\be
\int_{\t_0}^{\pi-\t_0} d\t\int_{-\l(\t)}^{\l(\t)} d\phi \int_0^{+\infty} dr \,\xi^a \d E_{ab }{\bm\ve}^b = 0\,,
\ee
where $\ve_a = {1\/6} \ve_{abcd}dx^b\wg dx^c\wg dx^d$ and $\l(\t)$ is defined in \eqref{defltheta} and contains the parameter $\l_\phi$ which parametrizes the width of $A$. The dependence on $\l_\phi$ enters in a complicated fashion. However, we can differentiate with respect to $\l_\phi$ at $\l_\phi = 0$, where we can use the expansion \eqref{lthetasmall}. This leads to
\be
\int_{\t_0}^{\pi-\t_0} d\t \le( 1 - {\r{sin}\,\t_0\/\r{sin}\,\t}\ri)\int_0^{+\infty} dr \,\xi^a \d E_{ab }{\bm\ve}^b = 0\,,
\ee
where the LHS is evaluated at $\phi=0$. In Bondi coordinates, we have 
\be
{\bm\ve}^r = -{\bm\ve}_u = - r^2\r{sin}\,\t\,dr\wg d\t\wg d\phi \,.
\ee
The bulk modular flow \eqref{4dbulkmodularflow} evaluated at $u=0$ and $\phi=0$ is given by
\be\label{4dmodflowuzerophizero}
\xi_A = {2\pi\/\r{cos}\,\t_0} \le( r\,\r{cos}\,\t\,\p_r - (\r{sin}\,\t - \r{sin}\,\t_0)\, \p_\t\ri) \,,
\ee
so the integral becomes
\bea\label{4dfirstexpr}
\hspace{-1cm}0 \= \int_{\t_0}^{\pi-\t_0} d\t\le( \r{sin}\,\t - \r{sin}\,\t_0\ri) \int_0^{+\infty} dr \le( -r^3\r{cos}\,\t\,\d E_{r r} + r^2(\r{sin}\,\t - \r{sin}\,\t_0)\,\d E_{r\t}\ri)\,.
\eea
The expansion around $\t_0={\pi\/2}$ implies that
\be\label{4drefexp}
\int_0^{+\infty} dr\le. \le( r^3 \p_\t \d E_{rr} +  2 \,r^2 \d E_{r\t}\ri)\ri|_{(u,\t,\phi)=( 0,\tfrac\pi2,0)}=0~.
\ee

\paragraph{Rotations and time translations.} As in the 3d case, we can consider new configurations obtained by performing rotations. They are the same as the reference configuration but centered at $\phi=\phi_0$.  We can also consider a translation $u\ra u + u_0$ in retarded time $u$. The Jacobians of these transformations are the identity which implies that the expression \eqref{4drefexp} becomes
\be\label{4drefexpall}
\int_0^{+\infty} dr\le. \le( r^3 \p_\t \d E_{rr} +  2 \,r^2 \d E_{r\t}\ri)\ri|_{(u,\t,\phi)=( u_0,\tfrac\pi2,\phi_0)}=0~,
\ee
for any $u_0$ and $\phi_0$.

\paragraph{light sheaf deformation.} 

We consider the same boundary region $A$ but with the more general configuration described in Sec.\ \ref{4dgenpr}. For the proof, we consider the configuration depicted on the right of Fig.\ \ref{fig:4dxtyt}. We put $P_+$ at the origin and $P_-$ at a distance $\l$ from $P_+$ on one of the axis and we connect them by the two curves $C_N$ and $C_S$, as represented on the figure. The configuration is parametrized by the length $\l$ of the segment $[P_+P_-]$ and the overture angle $\a$ at $P_-$. The first law of entanglement gives
\be\label{IhL}
I(\a,\l) = \int_{\S(\a,\l)} d{\bm\chi} = 0~,
\ee
where $\S$, the interior of the RT surface, depends on these two parameters $\a$ and $\l$. Let's denote by $n$ the vector normal to the segment $C_S$ 
\be
n= \r{cos}(\lph) \p_\tx + \r{cos}\,\t_0\,\r{sin}(\lph) \p_\ty~,
\ee
Taking the derivative of \eqref{IhL} with respect to $\a$ and evaluating at $\a=0$, we obtain
\be
\int_0^{+\infty} ds \int_{0}^{\ell} dv \le( n\cdot d{\bm\chi}\ri) = 0~,
\ee
where we have used the fact that $\g_N$ \eqref{gammaN} can be parametrized by $s$ and $v$. The vector tangent to $C_S$ is given by
\be
m=  \r{cos}(\lph) \p_\ty-\r{cos}\,\t_0\,\r{sin}(\lph) \p_\tx~.
\ee
We can now take the derivative with respect to $\l$ and evaluate at $\l=0$. 
\be
\int_0^{+\infty} dr\, (m\cdot(n\cdot d{\bm\chi}))|_{(\t,\phi)=(\t_0,0)} = 0
\ee
We have reduced the integral to a light ray going from the origin to the point of $\p A$ with $(\t,\phi) = (\t_0,0)$ (and $u=0$ as we are considering a region $A$ with $\ell_u=0$). We can then go to Bondi coordinates. From the change of coordinates, we can compute
\bea
m \= -\r{sin}\,\t_0\,\r{sin}(\lph) \,\p_r - {\r{cos}\,\t_0\,\r{sin}(\lph)\/r} \p_\t + {\r{cos}(\lph)\/r\,\r{sin}\,\t_0} \p_\phi~,\-
n \= \r{cos}(\lph)\,\r{tan}\,\t_0\,\p_r + {\r{cos}(\lph)\/r} \p_\t + {\r{cot}\,\t_0\,\r{sin}(\lph)\/r} \p_\phi~,
\eea
when evaluated at $\t=\t_0$ and $\phi=0$.  In the definition \eqref{dchideltaE} of $d{\bm\chi}$, the non-trivial contribution comes from
\be
{\bm\ve}^r = - {\bm\ve}_u = -r^2\,\r{sin}\,\t\, dr\wg d\t\wg d\phi~.
\ee
Hence, we obtain that
\be
m\cdot(n\cdot d{\bm\chi})|_{\g_N} = 2\le(1-  \r{sin}^2\t_0\,\r{sin}^2(\lph)\ri)\,\xi_A^a \d E_{ar} dr~.
\ee
The bulk modular flow at $(u,\t,\phi)=(0,\t_0,0)$ is simply given by
\be
\xi_A = 2\pi r \,\p_r
\ee
Hence, we obtain
\bea
&& \int_0^{+\infty}dr \,r \d E_{rr}(0,r,\t_0,0) = 0~,
\eea
As previously, we can act with rotations and time translations to show that we have
\bea\label{4ddERR}
&& \int_0^{+\infty}dr \,r \d E_{rr}(u_0,r,\t_0,\phi_0) = 0~,
\eea
for arbitrary $u_0,\t_0,\phi_0$.

\paragraph{Radial translations.} Let's consider a new configuration which is obtained by translating the reference configuration by a distance $r_0$ in the direction $(\theta_0, \phi_0)$ of the light ray on which \eqref{4ddERR} is integrated. In Cartesian coordinates, such a translation is given by
\be
t \rightarrow t+  r_0, \qq x\rightarrow  x+r_0 \,\r{cos}\,\t_0\,\r{cos}\,\phi_0, \qq y\rightarrow  y + r_0 \,\r{cos}\,\t_0\,\r{sin}\,\phi_0,\qq z\ra z+ r_0\,\r{sin}\,\t_0~.
\ee
This leads to the new constraint
\be
\int_{r_0}^{+\infty} dr\,(r-r_0)\,\d E_{rr}(u_0,r,\theta_0,\phi_0) = 0 ~,
\ee
where we have also performed the change of variable $r\to r-r_0$ in the integral. 
Taking two derivatives with respect to $r_0$ shows that
\be
\d E_{rr}(u_0,r_0,\theta_0,\phi_0) = 0~,
\ee
for any value of $u_0,r_0,\theta_0,\phi_0$. From this, the equation \eqref{4drefexpall} simplifies to
\be
\int_0^{+\infty} dr \, r^2\, \d E_{r\t}(u_0,r,\tfrac\pi2, \phi_0)=0~.
\ee
We use the same radial translation on this equation to obtain the constraint
\be
\int_{r_0}^{+\infty} dr \, {(r-r_0)^2\/r} \d E_{r\t}(u_0,r,\tfrac\pi2,\phi_0)~.
\ee
Taking three derivatives with respect to $r_0$ implies that
\be\label{dErthetaint}
\d E_{r\t}(u_0,r_0,\tfrac\pi2,\phi_0) = 0~,
\ee
which is true for any value of $u_0,r_0,\phi_0$.

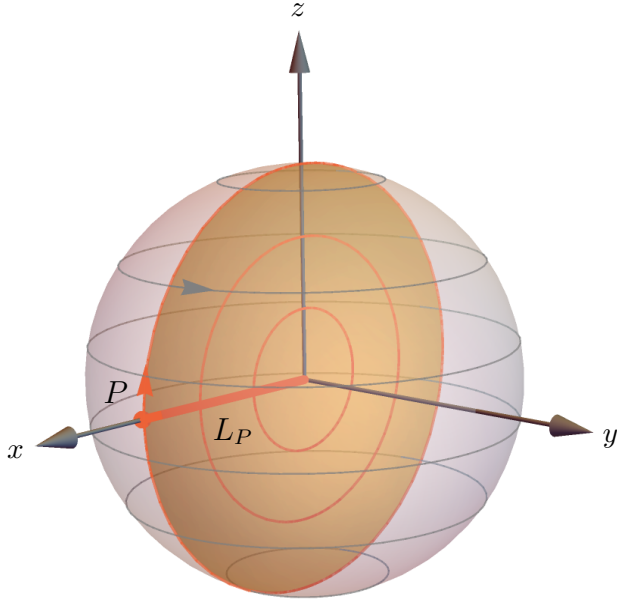
\begin{figure}
	\centering
	\includegraphics[width=8cm]{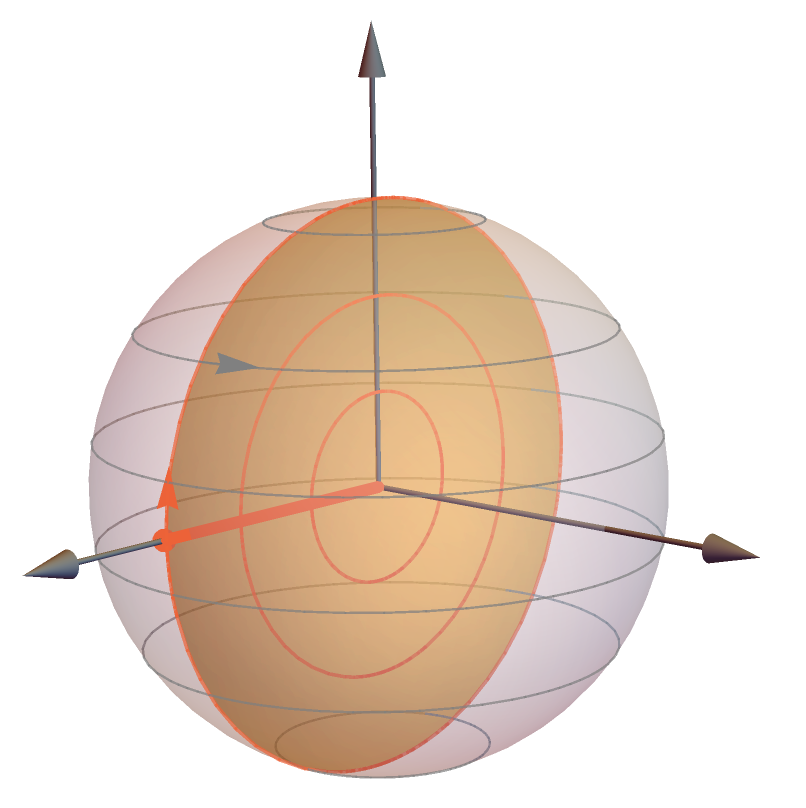}
	\put(-9,67){$y$}
	\put(-232,61){$x$}
		\put(-126,228){$z$}
	\put(-196,82){\large$P$}
	\put(-155,68){\large$L_P$}
	\caption{Illustration of the proof that $\d E_{r\t} = 0$. After showing that $\d E_{r\t} $ vanishes on the orange line $L_P$, we use the $(x,z)$-rotation (orange arrow) to show that $\d E_{r\t} = 0$ in the orange disk. With the $(x,y)$-rotation (gray arrow), we can show that $\d E_{r\t} = 0$ everywhere in the ball. These transformations all have diagonal Jacobians where they are evaluated so they don't mix $\d E_{r\t}$ with other components.}\label{fig:illustr}

\end{figure}

\paragraph{Vanishing of $\d E_{r\t}$ everywhere.} The equation \eqref{dErthetaint} at $\phi_0 = 0$ shows that $\d E_{r\t}$ vanishes on the semi-infinite line $L_P$ given by $(\t,\phi)= (\tfrac\pi2,0)$. Let's now consider rotations in the plane $(x,z)$. Under such rotations, $L_{P}$ covers the full disk in the $y=0$ plane, shown in orange in Fig.\ \ref{fig:illustr}. The Jacobian of this transformation, when evaluated at $\phi=0$, is diagonal in Bondi coordinates because it simply corresponds to a shift in $\t$. It is given explicitly by
\be\arraycolsep=5pt\def\arraystretch{1.0}\label{Jacobxz}
{\p x^a\/ \p\tilde{x}^b} = \bpm 1 &  &  & \\ & 1 & & \\ & & 1  & \\ & & & \r{cos}\,\a - \r{cot}\,\t\,\r{sin}\,\a\epm\,,
\ee
so we obtain $\d E_{r\t}=0$ when evaluated on this disk. For any point on this disk, we can then consider a rotation in the $(x,y)$-plane, whose Jacobian is the identity. This shows that $\d E_{r\t}=0$ vanishes everywhere inside the ball. This implies that
\be
\d E_{r\t} = 0\,,
\ee
everywhere in the bulk. This procedure is illustrated in Fig.\ \ref{fig:illustr}.

\paragraph{Boosts and rotations.}  We now act with boosts and rotations on the previous configurations to generate more constraints on $\d E_{ab}$. Transforming the equation $\d E_{r\t}=0$ under the infinitesimal $(x,z)$-rotation, the $(t,y)$-boost and the $(t,x)$-boost, we obtain
\be
\d E_{r\phi} = \d E_{\t\phi}= \d E_{rr}= 0,\qq \d E_{\t\t} = -r^2 \d E_{u r}\,.
\ee
Then, the image of $\d E_{\t\phi}= 0$ under the $(x,z)$-rotation implies that
\be
\d E_{\phi\phi}=-r^2\r{sin}^2\t\, \d E_{u r}\,.
\ee
\paragraph{Conservation equation.} As in 3d, we consider the  conservation equation 
\be\label{4dconserveq}
\n_a (\d E^{ab}) = 0\,,
\ee
which is always satisfied by the equations of motion. For $b = r$, this implies that
\be
\p_r(\d E_{ur}) =0\,,
\ee
which leads to 
\be
\d E_{ur} = C_0(u,\phi),\qq \d E_{\t\t} = -r^2 C_0(u,\phi),\qq \d E_{\phi\phi} = -r^2 \r{sin}^2\t\,C_0(u,\phi)\,.
\ee
We expect that an analysis similar to the 3d one in Sec.\ \ref{holst} can be performed in 4d and that it will lead to a trace condition and three conservation equations for the holographic stress tensor in 4d Minkowski. A proof of this statement will require a detailed analysis of the flat limit of perturbed AdS$_4$ in Bondi gauge, which we leave for future work. From now on, we will assume that these boundary conditions ensure the vanishing of the components $\d E_{ua }$ at leading asymptotic order. The trace condition, similar to \eqref{Trace1} and \eqref{Trace2} in 3d, should imply that $C_0=0$, leading to
\be
\d E_{ur} = \d E_{\t\t}=\d E_{\phi\phi} =0\,,
\ee
everywhere in the bulk. The conservation equation \eqref{4dconserveq} for $b = \t,\phi$ gives
\be
  \p_r(\d E_{u\t}) +{2\/r}\d E_{u\t}  =0, \qq \p_r(\d E_{u\phi})+{2\/r}  \d E_{u\phi}  =0\,.
\ee
The solutions of these equations are
\be
\d E_{u\t} = {C_1(u,\t)\/r^2} ,\qq \d E_{u\phi} = {C_2(u,\phi)\/r^2} \,.
\ee
We expect that the conservation of the boundary stress tensor implies that $C_1=C_2=0$, leading to $\d E_{u\t}=\d E_{u\phi}=0$. Finally, the conservation equation \eqref{4dconserveq} for $b=u$ gives
\be
\p_r(\d E_{u u})  + {2\/r} \d E_{uu}=0 \,.
\ee
which is solved by 
\be
\d E_{uu} = {C_3(u,\phi)\/r^2}\,,
\ee
and $C_3=0$ is expected to follow from the conservation of the boundary stress tensor. Thus, we have shown that all the components of the linearized gravitational equation vanish.

\section{Conclusion}

In this paper, we have considered holographic entanglement entropy in asymptotically flat spacetimes. Under some general assumptions on the dual field theory, an analog of the Ryu-Takayanagi formula was obtained in \cite{Jiang:2017ecm} to compute the entanglement entropies of 3d Minkowski spacetime. We have refined and generalized this prescription and showed that it satisfies a first law when perturbations are considered. Using this RT prescription, we have shown that the first law of entanglement is equivalent to the linearized gravitational equations of motion. We have also extended all these results to 4d.

This result could have also been phrased  purely in classical gravity, although it is natural to motivate it from the perspective of holography. It will be important to understand better the dual field theory, and try to prove the assumptions detailed in Sec.\ \ref{assumptions}. Some recent progress in this direction include \cite{Barnich:2014kra, Oblak:2015sea, Campoleoni:2016vsh, Oblak:2016eij, Bagchi:2019xfx,Ball:2019atb,Himwich:2019dug,Donnay:2018neh,Hijano:2018nhq, Hijano:2019qmi}.

Another line of research would be to push further the consequences of the RT prescription described here. One could hope to get some hints on the microscopic definition of the dual field theory, or show that one of the assumptions was incorrect. An important feature of our analysis is the importance of the choice of an infalling light sheaf. We believe that this is a hint towards the UV structure of the dual theory, which we hope to investigate in future work. The RT formula in AdS has given rise to a wealth of results connecting quantum information to the emergence of spacetime. It would be interesting to investigate these ideas in asymptotically flat spacetimes, using the RT prescription described here.

\section*{Acknowledgments}

It is a pleasure to thank Jan de Boer, Samuel Guérin, Hongliang Jiang, Erik Mefford, Kevin Morand, Romain Ruzziconi and Wei Song for useful discussions. We are grateful for the hospitality of Sylvia and Melba Huang  and to the 2019 Amsterdam string theory summer workshop where this work was completed. This work was supported in part by the $\Delta$-ITP consortium, a program of the NWO that is funded by the Dutch Ministry of Education, Culture and Science (OCW) and the ANR-16-CE31-0004 contract Black-dS-String.

\appendix

\section{Bulk Rindler transformation}\label{appbulkrindler}

In this appendix, we describe the bulk extension of the generalized Rindler transform \eqref{rindler3d} on the boundary. The image of Minkowski spacetime under this bulk transformation turns out to be the upper wedge of a Rindler spacetime. 

\paragraph{Bulk Rindler transformation in 3d.} We describe the change of coordinates that brings the metric in Bondi coordinates to the upper wedge of a Rindler spacetime. The Cartesian coordinates are related to Bondi coordinates using
\begin{equation}
t=u+r, \qq x=r\,\cos\,\phi, \qq y=r\,\sin\,\phi,
\end{equation}
and the coordinates in which the modular flow is a boost are
\be
\tilde{t} = {t\/\r{sin}(\lph)} -\r{cot}(\lph)\,x, \qq \tilde{x} = {x\/\r{sin}(\lph)}- \r{cot}(\lph) \,t ,\qq \tilde{y} = y + {\l_u\/2\,\r{sin}(\lph)}.
\ee
We define new coordinates $(\tilde{\tau},\rho)$ satisfying
\begin{equation}
\tilde{t}=e^{\tilde{\tau}}\cosh\,\rho,\quad \tilde{y}=e^{\tilde{\tau}}\sinh\,\rho.
\end{equation}
These coordinates only cover the upper wedge $\tilde{t}^2-\tilde{y}^2>0$. In these coordinates the bulk metric and modular flow are given by
\be
\xi_A =2\pi \partial_\rho,\quad ds^2=e^{2\tilde{\tau}}(-d\tilde{\tau}^2+d\rho^2)+d\tilde{x}^2.
\ee
We recognize the Rindler metric and the bulk modular flow generates the (spacelike) Rindler evolution. The Rindler horizon is situated at $\tilde{\tau}=-\infty$. To obtain the bulk extension of the generalized Rindler transform, consider the new coordinates $\{\tau, \tilde{x}, \rho \}$ satisfying
\begin{equation}
\tau=e^{\tilde{\tau}}-\tilde{x},
\end{equation}
defined only for $\tau>-\tilde{x}$. The metric becomes
\begin{equation}
ds^2=-d\tau^2-2d\tau d\tilde{x}+(\tau+\tilde{x})^2d\rho^2,
\end{equation}
and the bulk modular flow is still $\xi_A=2\pi\partial_\rho$. The Rindler horizon is at $\tau=-\tilde{x}$. Finally, the bulk transformation is obtained by writing the new coordinates in terms of Bondi coordinates $\{u, r, \phi\}$:
\begin{eqnarray}\nt
\tau &=&-\tilde{x}+\le[ {1\/\r{sin}^2(\lph)} \left(r+u-r\,
   \cos \left(\frac{\ell_\phi }{2}\right)
   \cos \,\phi \,\right)^2-\frac{1}{4} \left({\l_u\/\,\r{sin}(\lph)} +2 r\, \sin \,\phi\,\right)^2\ri]^{1/2},\\\nt
\tilde{x} &=&  \frac{r\, \cos \,\phi}{\r{sin}(\lph)}
   -\r{cot}(\lph)\, (r+u) ,\\
\rho &=& \text{arccoth} \left(\frac{r+u-r \,\r{cos}(\lph)
\, \r{cos} \,\phi}{{\l_u\/2}+ r\, \r{sin}(\lph)\,\r{sin}\,\phi}\right).
\end{eqnarray}
This coordinate system allows us to perform an asymptotic limit $r\rightarrow\infty$, which gives
\begin{eqnarray}
\tau &=& \frac{2 u\, \r{sin}(\lph)- \l_u\,\r{sin}\,\phi}{2\, \r{cos}\,\phi  -2 \,\r{cos}(\lph)},\\
   \rho &=& \text{arccoth} \left(\frac{ 1- \r{cos}(\lph)\,\r{cos}\,\phi}{  \r{sin}(\lph)\,\r{sin}\,\phi}\right).
\end{eqnarray}
One can check that this is exactly the inverse of the boundary generalized Rindler transformation \eqref{rindler3d}, reproduced below
\bea
u \= {\r{sin}(\lph)\,\/\r{cosh}\,\rho + \r{cos}(\lph)}\left(\tau+\frac{\ell_u}{2\,\sin(\frac{\ell_\phi}{2})}\sinh \rho\right) \,, \-
\phi \=  \r{arctan}\le({\r{sin}(\lph)\,\r{sinh}\,\rho\/1+\r{cos}(\lph)\,\r{cosh}\,\rho}\ri) \,.
\eea

\paragraph{Bulk Rindler transformation in 4d.} The same procedure can be carried out in 4d. Again, consider the bulk transformation from Bondi coordinates to Rindler coordinates in the upper wedge:
\begin{equation}
t=u+r, \quad x=r\,\sin\,\theta\,\cos\,\phi, \quad y=r\,\sin\,\theta\,\sin\,\phi,\quad z=r\,\cos\,\theta,
\end{equation}
followed by
\begin{equation}
\tilde{t}=\frac{t}{\cos\,\theta_0}-\tan\,\theta_0\,x, \quad \tilde{x}=\frac{x}{\cos\,\theta_0}-\tan\,\theta_0\,t, \quad \tilde{y}=y, \quad \tilde{z}=z+\frac{\ell_u}{2\,\cos\,\theta_0},
\end{equation}
and then
\begin{equation}
\tilde{t}=e^{\tilde{\tau}}\cosh\,\rho,\quad \tilde{z}=e^{\tilde{\tau}}\sinh\,\rho,\quad \tilde{x}=\mu\, \cos\,\eta,\quad \tilde{y}=\mu\,\sin\,\eta,
\end{equation}
where the last two spacelike coordinates are mapped to polar coordinates: $\mu\in [0,\infty[$ and $\eta\in [0,2\pi[$. In these coordinates, the metric and the bulk modular flow become
\begin{equation}
\xi_A=2\pi\partial_\rho,\quad ds^2=e^{2\tilde{\tau}}(-d\tilde{\tau}^2+d\rho^2)+d\mu^2+\mu^2d\eta^2.
\end{equation}
Exactly like in 3d, we recognize the Rindler metric and the bulk modular flow generates the (spacelike) Rindler evolution. The Rindler horizon is at $\tilde{\tau}=-\infty$. To obtain the bulk extension of the boundary generalized Rindler transform, we consider the new coordinates $\{\tau, \mu, \rho, \eta\}$, such that
\begin{equation}
\tau=e^{\tilde{\tau}}-\mu,
\end{equation}
defined only for $\tau>-\mu$. The metric becomes
\begin{equation}
ds^2=-d\tau^2-2d\tau d\mu+(\tau+\mu)^2d\rho^2+\mu^2d\eta^2,
\end{equation}
while the bulk modular flow is still given by $\xi_A=2\pi\partial_\rho$. The new radial coordinate is $\mu$ and by taking the limit $\mu\rightarrow \infty$ we confirm that the boundary metric is indeed the degenerate flat metric $d\rho^2+d\eta^2$. The Rindler horizon is at $\tau=-\mu$. Finally, the bulk transformation is obtained by writing the new coordinates in Bondi coordinates:
\begin{eqnarray}
\tau &=& \sqrt{\left(\frac{r+u}{\r{cos}\,\t_0}-r\, \r{sin}\,\t\,\r{tan}\,\t_0\,\r{cos}\,\phi\right)^2-\frac{1}{4}
   \left(\frac{\ell_u}{\r{cos} \,\theta_0}+2 r\, \r{cos}\,\t\right)^2}\\
   && - \sqrt{r^2 \,\r{sin}^2\t \,\r{sin}^2\phi +\left(\frac{r\, \r{sin}\,\t\,\r{cos}\,\phi }{\r{cos}\,\t_0}-\r{tan}\,\t_0 \,
   (r+u)\right)^2}   ,\\
\mu &=&  \sqrt{r^2 \,\r{sin}^2\t\,\r{sin}^2\phi+\left(\frac{r \,\r{sin}\,\t\,\r{cos}\,\phi}{\r{cos}\,\t_0}-\r{tan}\,\t_0\, (r+u)\right)^2}  , \\
\rho &=& \text{arctanh}\left(\frac{{\l_u\/2} + r\,\r{cos}\,\t_0\,\r{cos}\,\t }{(r+u)-r\,\r{sin} \,\t_0\,\r{sin}\,\t\,\r{cos}\,\phi }\right)   ,\\
\eta &=&  \text{arctan}\left(\frac{r\,\r{cos}\,\t_0\,\r{sin}\,\t\,\r{sin}\,\phi}{r\,\r{sin}\,\t\,\r{cos}\,\phi-\r{sin}\,\t_0\, (r+u)}\right)    .
\end{eqnarray}
This allows us to perform the asymptotic limit $r\rightarrow\infty$ which gives
\begin{eqnarray}
\tau &=&  { u \,\r{cos}\,\t_0 - {\l_u\/2}\,\r{cos}\,\t \/  \sqrt{ (\r{sin}\,\t_0\,\r{sin}\,\t\,\r{cos}\,\phi -1)^2 -\r{cos}^2\t_0\,\r{cos}^2\t }} ~,\\
\rho \= \r{arctanh}\le( {\r{cos}\,\t_0\,\r{cos}\,\t\/1-\r{sin}\,\t_0\,\r{sin}\,\t\,\r{cos}\,\phi}\ri)~, \- 
\eta \= \r{arctan}\le({\r{cos}\,\t_0\,\r{sin}\,\t\,\r{sin}\,\phi \/\r{sin}\,\t\,\r{cos}\,\phi - \r{sin}\,\t_0} \ri)~.
\end{eqnarray}
One can check that this is precisely the inverse of the boundary generalized Rindler transformation \eqref{4dRindlunonzero}, reproduced below
\begin{eqnarray}
u&\ra& {\r{cos}\,\t_0\/\r{cosh}\,\rho\,+ \r{cos}\,\eta\,\r{sin}\,\t_0}\le(\tau +{\l_u\/2\,\r{cos}\,\t_0}\r{sinh}\,\rho \ri), \-
z &\ra& {\r{sin}\,\t_0 + e^{w} (1+\r{cos}\,\t_0)\/\r{sin}\,\t_0 \,e^{w}+  (1+\r{cos}\,\t_0)},\-
\bar{z} &\ra& {\r{sin}\,\t_0 + e^{\bar{w}} (1+\r{cos}\,\t_0)\/\r{sin}\,\t_0 \,e^{\bar{w}}+  (1+\r{cos}\,\t_0)}\,,
\end{eqnarray}
where  $z=e^{i\phi}\cot(\frac{\theta}{2})$ and $w=\rho-i\eta$.

\section{Precisions on the general strategy}\label{precisionstrat}

In this appendix, we make precise the general strategy explained in Sec.\ \ref{genstrat}. Let
$g:M\rightarrow M$ be a bulk isometry, $i:\Sigma\rightarrow M$ the original RT surface and $i_g=g \circ i:\Sigma\rightarrow M$ the image of this surface through isometry. The original RT surface is associated to a bulk modular flow $\xi$ to which corresponds a two-form $d{\bm\chi}[\xi]$. The pullback of this two-form on $\Sigma$ is
\begin{eqnarray}\label{plop}
i^*(d{\bm\chi}[\xi])=\xi^a(i(\sigma))\delta E_{ab}(i(\sigma))\frac{1}{2}\varepsilon^b_{cd}(i(\sigma))\frac{\partial x^c}{\partial \sigma^\alpha}\frac{\partial x^d}{\partial \sigma^\beta}d\sigma^\alpha\wedge d\sigma^\beta,
\end{eqnarray}
where $\sigma$ stands for the coordinates on the two-dimensional manifold $\Sigma$. Suppose that from the vanishing of the integral of this two-form on $\Sigma$, we have been able to derive that some functional of $\delta E_{ab}$ vanishes at $i(\sigma)$, 
\be\label{plap}
\cF\left[\delta E_{ab}(i(\sigma))\right] =0 ~,
\ee
for some $\bar{a},\bar{b}$. We can now consider another surface, $(g\circ i) (\Sigma)$ in $M$ and we call its associated bulk modular flow $\xi_g$. We should consider the pullback on the corresponding two-form $d{\bm\chi}[\xi_g]$ because
\begin{eqnarray}
\int_{(g\,\circ\, i)(\Sigma)}d{\bm\chi}[\xi_g]=\int_{\Sigma}i^*_g d{\bm\chi}[\xi_g]~.
\end{eqnarray}
The pullback is given by
\begin{equation}
i^*_g(d{\bm\chi}[\xi_g])=\xi_g^a(g\circ i(\sigma))\delta E_{ab}(g\circ i(\sigma))\frac{1}{2}\varepsilon^b_{cd}(g\circ i(\sigma))\frac{\partial g^c}{\partial x^e}\frac{\partial g^d}{\partial x^f}\frac{\partial x^e}{\partial \sigma^\alpha}\frac{\partial x^f}{\partial \sigma^\beta}d\sigma^\alpha\wedge d\sigma^\beta.
\end{equation}
Now we can insert the identity matrix $\delta^a_b=\frac{\partial g^a}{\partial x^c}\frac{\partial x^c}{\partial g^b}$ to impose the equality of two $b$-index, leading to
\begin{eqnarray}
i^*_g(d{\bm\chi}[\xi_g])\=\xi_g^a(g\circ i(\sigma))\delta E_{ab}(g\circ i(\sigma))
\frac{\partial g^b}{\partial x^g}\left(\frac{\partial x^g}{\partial g^h}
\frac{1}{2}\varepsilon^h_{cd}(g\circ i(\sigma))\frac{\partial g^c}{\partial x^e}\frac{\partial g^d}{\partial x^f}\right)\-
&& \times \frac{\partial x^e}{\partial \sigma^\alpha}\frac{\partial x^f}{\partial \sigma^\beta}d\sigma^\alpha\wedge d\sigma^\beta.
\end{eqnarray}
Now we can use the fact that $g$ is an isometry, while $\varepsilon^h_{cd}  $ is the volume form to obtain than the parenthesis is actually $\frac{1}{2}\varepsilon^g_{ef}( i(\sigma))$. Moreover we know that the modular flow for the image surface is the image of the modular flow of the initial surface under the $g$-transformation: $\xi_g^a(g\circ i(\sigma))=\frac{\partial g^a}{\partial x^b}\xi^b(i(\sigma))$. Finally, we obtain
\begin{equation}
i^*_g(d{\bm\chi}[\xi_g])=\xi^i( i(\sigma))\left(\frac{\partial g^a}{\partial x^i}\frac{\partial g^b}{\partial x^g}\delta E_{ab}(g\circ i(\sigma))\right)
\frac{1}{2}\varepsilon^g_{ef}( i(\sigma))\frac{\partial x^e}{\partial \sigma^\alpha}\frac{\partial x^f}{\partial \sigma^\beta}d\sigma^\alpha\wedge d\sigma^\beta,
\end{equation}
which, is exactly \eqref{plop} with the replacement 
\begin{equation}
\delta E_{ab}(i(\sigma))\rightarrow \frac{\partial g^c}{\partial x^a}\frac{\partial g^d}{\partial x^b}\delta E_{cd}(g\circ i(\sigma))~,
\end{equation} 
which implies that \eqref{plap} ensures that 
\begin{eqnarray}\label{plip}
\cF\left[\frac{\partial g^c}{\partial x^{a}}\frac{\partial g^d}{\partial x^{b}}\delta E_{cd}(g\circ i(\sigma))\right]=0\,.
\end{eqnarray}
For example, if we can show that some components of $\d E_{ab}$ vanish using a set of RT surfaces, we immediately obtain that other components, obtained by applying bulk isometries according to \eqref{plip}, will also vanish.

\section{Alternative proof in 3d}\label{app:alt3d}

In this appendix, we provide an alternative to the step in the 3d proof of Sec.\ \ref{sec:3dproof} where we used the light sheaf deformation. Here, we insist on doing this step using only RT configurations where the light rays $\g_+$ and $\g_-$ pass through the spatial origin $r=0$. We will consider such configurations with $\l_u\neq 0 $ described in \eqref{3dreview} which is the prescription used in \cite{Jiang:2017ecm}. Although a better and equivalent\footnote{This is because all the configurations described in Sec.\ \ref{3dgenpresc} can be transformed with a bulk translation to a configuration where the two light rays pass through the line $r=0$.} derivation is presented in the main text, it is instructive to perform this step as presented here.

We should note that if we consider only the surfaces with $\l_u = 0$ (and with light rays passing through $r=0$), together with their image under bulk isometries, then the first law does \emph{not} imply the gravitational equations: these surfaces don't provide  enough constraints. Indeed, the only constraint that we obtain is
\be
\d E_{r\phi} + r\p_r \d E_{r\phi} - r\p_\phi \d E_{rr} =0~, 
\ee
and its image under bulk isometries. This does not imply that $\d E_{ab} = 0$ as it's possible to find explicit counterexamples.

Hence, we need to consider RT surfaces with $\l_u\neq 0 $ (still requiring that the light rays pass through $r=0$). The computation becomes simpler in the limit of small $\l_u$. More precisely, we consider 
\be
\l_u =\la \,\ve^2 ,\qq\l_\phi = \ve~,
\ee
where we take $\ve$ to be small. We would like to compute
\be
I = \int_\S \xi^a \d E_{ab}{\bm\ve}^b
\ee
in an expansion around $\ve=0$. The first law of entanglement will constrain $\d E_{ab}$ to be such that $I=0$. It turns out that $\lim_{\ve\to 0} I = 0$ for any perturbation, so we don't get any constraint at zero order in $\ve$.  To compute $I$ at first order in $\ve$, it is enough to consider the surface $\S$ at first order in $\ve$\footnote{This can be justified as follows. Denoting $i_\ve:S\ra M$ the embedding of $\S_\ve$ in $M$, we have
\be
I = \int_{\S_\ve} \xi_\ve^a \d E_{ab} {\bm\ve}^b = {1\/2}\int _S \xi_\ve^a(i_\ve(\s)) \d E_{ab}(i_\ve(\s)) \ve_{cd}^b(i_\ve(\s)) (J_\ve)^c_{~\a} (J_\ve)^d_{~\b}d\s^\a\wg d\s^\b\,
\ee
where $(J_\ve)^c_{~\a}$ is the Jacobian of the embedding. This shows that, to compute the leading non-trivial term of $I$, it is enough to take $i_\ve$ at first order in $\ve$, which corresponds  to taking the surface $\S_\ve$ at first order in $\ve$.}. The configuration simplifies because the points $B_+$ and $B_-$ are at $u=O(\ve^2)$. Hence, we have
\be
B_+:\quad (u,\phi) = \le(0, {\ve\/2}\ri),\qq B_-:\quad (u,\phi) = \le(0, -{\ve\/2}\ri)~,
\ee
to first order in $\ve$. We  also have the following parametrization for the light rays
\bea
\g_+ && :\quad (t,x,y) = \le(- 2\eta+s , -2\eta+s, {\ve\/2}(-2\eta+s)\ri), \qq s\geq 0\-
\g_- && :\quad (t,x,y) = \le( 2\eta+s , 2\eta+s, -{\ve\/2}(2\eta+s)\ri), \qq s\geq 0 ~,
\eea
where we only kept the terms at first order in $\ve$. The curve $\gamma$ is simply a straight line connecting the two points
\begin{equation}
P_+:\quad (t,x,y)=(-2\eta, -2\eta, -\eta \epsilon),\quad P_-:\quad (t,x,y)=(2\eta, 2\eta, -\eta \epsilon).
\end{equation}
We can show that $\g_-$ stays at $u=0$ everywhere and that $\g_+$ is at $u=0 $ for $s\geq  2\eta$, which corresponds to all its points before it crosses the origin. Let's call $\tilde{\g}_-$ the segment that connects the origin to $P_-$, which , which is in the continuation of $\g_-$ past $P_-$. The plane surface bounded by $\g_-, \tilde{\g}_-$ and $\g_+$ (up to the origin) lies on the constant slice $u=0$. It has the same shape as the RT surface for $\l_u=0$ depicted in Fig.\ \ref{fig:3dpresc}.

The additional piece consists in another triangle, bounded by $\g$, $\tilde{\g}_-$ and $\tilde{\g}_+$, where $\tilde{\g}_+$ is the piece of $\g_+$ connecting the origin to $P_+$. This is the triangle $T=P_-P_+ O$. Let's introduce coordinates
\be
x_+ = t+x, \qq x_- = t-x
\ee
In these coordinates, we have (at first order)
\bea
P_+ && :\quad  (x_+,x_-,y) = (-4 \la,0, -\la\ve)\-
P_- && :\quad  (x_+,x_-,y) = (4 \la,0, -\la\ve)
\eea
We see that the triangle $T = P_+ P_- O$ can be parametrized as follows
\be
x_-= 0,\qq  -\la \ve\leq y \leq 0, \qq |x_+| \leq -{4y \/\ve}
\ee
The integration over the triangle is
\be
I =\int_{-\la \ve}^0 d y \int_{4 y/\ve}^{-4y/\ve} d x_+\,F(x_+,x_-,y)~,
\ee
where $F$ is the appropriate integrand. We can redefine $y= \eta \ve \tilde{y}$ so that it becomes
\be
I =\la\ve \int_{-1}^0 d\tilde{y}\int_{4 \la \tilde{y}}^{-4 \la \tilde{y}} dx_+ F(x_+,x_-,\la\ve y)~.
\ee
We now come back to the full integral 
\be
I = \int_\S \xi^a \d E_{ab} {\bm\ve^b}\,,
\ee
which we want to evaluate at first order in $\varepsilon$. The integral splits in an integral over the pizza slice and an integral over the triangle
\be
I = I_P+I_T~.
\ee
The integral over the pizza slice is
\bea
I_P \= \int_{-\ve/2}^{\ve/2} d\phi\int_0^{+\infty}dr\,r\,\xi^a \d E_{ar}~.
\eea
The integral over the triangle is found by looking at the metric in the $(x_+,x_-,y)$ coordinates. We have $\p_\pm = {1\/2}(\p_t\pm\p_x)$ so that
\be
g_\mn = {1\/2} \bpm 0 & -1 & 0 \\ -1 & 0 &0 \\ 0 & 0 & 1 \epm,\qq g^\mn = 2 \bpm 0 & -1 & 0 \\ -1 & 0 &0 \\ 0 & 0 & 1 \epm\,.
\ee
The volume form on the triangle is
\be
{\bm \ve}^{x_+} = -2  {\bm \ve}_{x_-} =  dy\wg dx_+\,
\ee
this implies that
\bea
I_T \= \int_{-\eta \ve}^0 d y \int_{4 y/\ve}^{-4y/\ve} d x^+ \, \xi^a \d E_{a x_+}~.
\eea
Both integrals $I_P$ and $I_T$ can be computed explicitly at first order in $\varepsilon$. We now take derivatives of the result with respect to $\eta$. The first law gives $I=0$ so for any $\eta$ we have
\be
\p_\eta^3 I|_{\eta=0} = 0.
\ee
On the other hand, one find that
\bea
\p_\eta^3 I_P |_{\eta=0}  \= O(\ve^2)~, \-
\p_\eta^3  I_T |_{\eta=0}\= - 16 \pi \ve\le( \d E_{rr}(0,0,0) - 2 \d E_{ur}(0,0,0) + 2 \d E_{uu}(0,0,0) \ri) + O(\ve^2)~,
\eea
which provides the new constraint
\be\label{meq}
\d E_{rr}(0,0,0) - 2 \d E_{ur}(0,0,0) + 2 \d E_{uu}(0,0,0)=0~.
\ee
Following the general strategy, we obtain a new constraint by acting with the translation
\be
\tilde{t} = t+  r_0, \qq \tilde{x} = x+r_0 \,\r{cos}\,\phi_0, \qq \tilde{y} = y + r_0 \,\r{sin}\,\phi_0~.
\ee
Evaluating the result at $\phi=\phi_0$, we obtain
\be
\d E_{rr}(0,r_0,\phi_0) - 2 \d E_{ur}(0,r_0,\phi_0) + 2 \d E_{uu}(0,r_0,\phi_0)=0~,
\ee
for any $r_0,\phi_0$. We can then consider time translations to show that this relation holds at any $u$. Finally, acting with a boost in the $(t,x)$-plane and evaluating at $\phi=0$ leads to
\be
\d E_{rr}(u_0,r_0,\phi_0) = 0~,
\ee
for any $u_0,r_0,\phi_0$. The rest of the proof follows.

\bibliography{bibliography}

\end{document}